\documentclass[12pt]{article}
\pdfoutput=1
\usepackage{amsmath,amssymb,epsfig,ulem,color}
\usepackage[stable]{footmisc}
\usepackage[bookmarks=false]{hyperref}
\allowdisplaybreaks  

\renewcommand{\arraystretch}{1.2}

\def \refeq#1{(\ref{#1})}  
\def \refsec#1{Section~\ref{#1}}
\def \refapp#1{Appendix~\ref{#1}}
\def \reffig#1{Figure~\ref{#1}}
\def \reftab#1{Table~\ref{#1}}

\def \cLdB1{{{\cal L}_{\Delta B = 1}^{\rm EW}}} 

\def \Op{Q}

\def \One{\leavevmode\hbox{\small1\kern-3.6pt\normalsize1}} 

\def \MeV{{\rm \; MeV}}
\def \GeV{{\rm \; GeV}}
\def \TeV{{\rm \; TeV}}

\def \B0toK0mumu{{ \bar{B}^0 \to K^0 \bar{\mu} \mu}}

\def \beq{\begin{equation}}
\def \eeq{\end{equation}}
\def \bea{\begin{eqnarray}}
\def \eea{\end{eqnarray}}

\def\bm#1{\mbox{\boldmath$#1$\unboldmath}}

\def \ie{{\it i.e.}}

\begin{document}

\begin{titlepage}

\begin{flushright}
OUTP-11-49P \\
\end{flushright}

\vspace{15pt}
\begin{center}
  \Large\bf New Physics in $\bm{ \Gamma_{12}^s}$: 
  $\bm{(\bar s b) (\bar \tau \tau)}$ Operators
\end{center}

\vspace{5pt}
\begin{center}
  {\sc Christoph~Bobeth$^a$ and Ulrich~Haisch$^b$} \\
  \vspace{10pt} {\sl $^a$Institute for Advanced Study and Excellence 
    Cluster Universe \\ Technische Universit\"at M\"unchen
    \\ D-85748 Garching, Germany} \\
  \vspace{10pt} {\sl $^b$Rudolf Peierls Centre for Theoretical Physics \\
    University of Oxford\\
    OX1 3PN Oxford, United Kingdom}
\end{center}

\vspace{10pt}
\begin{abstract}\vspace{2pt} \noindent 
  Measurements performed at the Tevatron of both the like-sign dimuon charge
  asymmetry in $B_{d,s}$-meson samples and the mixing-induced CP asymmetry in
  $B_s \to J/\psi \phi$ depart from their standard model (SM) predictions. This
  could be an indication for new CP phases in $\Delta B = 2$ transitions,
  preferentially in $B_s$--$\bar B_s$ mixing. The experimental situation,
  however, remained inconclusive, as it favored values of the element
  $\Gamma_{12}^s$ of the decay matrix in the $B_s$-meson system that are notably
  different from the SM expectation, suggesting the presence of new physics in
  the $\Delta B = 1$ sector as well. The very recent LHCb measurements of $B_s
  \to J/\psi \phi$ and $B_s \to J/\psi f_0$, which do not find any evidence for
  a new-physics phase in the element $M_{12}^s$ of the mass matrix, point into
  this direction as well. In this article, we explore the potential size of
  non-standard effects in $\Gamma_{12}^s$ stemming from dimension-six operators
  with flavor content $(\bar s b) (\bar \tau \tau)$. We show that since the
  existing constraints imposed by tree- and loop-level mediated $B_{d,s}$-meson
  decays are quite loose, the presence of absorptive new physics of this type
  would lead to an improved global fit to the current data. The allowed effects
  are however far too small to provide a full explanation of the observed
  anomalies.  Our model-independent conclusions are finally contrasted with
  explicit analyses of the new-physics effects in $B_s$--$\bar B_s$ mixing that
  can arise from leptoquarks or $Z^\prime$ bosons.
\end{abstract}

\vfill
\end{titlepage}

\tableofcontents

%
\section{Introduction}
\label{sec:introduction}

The phenomenon of neutral $B_s$-meson mixing is encoded in the off-diagonal
elements $M_{12}^s$ and $\Gamma_{12}^s$ of the mass and decay rate matrix. These
two complex parameters can be fully determined by measuring the mass difference
$\Delta M_s = M_H^s - M_L^s$, the CP-violating phase $\phi_{J/\psi \phi}^s = -2
\beta_s$, the decay width difference $\Delta \Gamma_s = \Gamma_L^s -
\Gamma_H^s$, and the CP asymmetry $a_{fs}^s$ in flavor-specific decays.
  
The combined Tevatron and LHC determination of the mass difference
reads \cite{Abulencia:2006ze, LHCbnote50}
\begin{align}
  \label{eq:DMsexp}
  \Delta M_s & = (17.73 \pm 0.05) \, {\rm ps}^{-1} \,,
\end{align}
and agrees well with the corresponding standard model (SM) prediction
\cite{Lenz:2011ti}
\begin{align} 
  \label{eq:DMsSM}
  (\Delta M_s)_{\rm SM} & =  (17.3 \pm 2.6) \, {\rm ps}^{-1} \,.
\end{align}
Here the quoted errors correspond to 68\% confidence level (CL)
ranges.

The phase difference $\phi_{J/\psi\phi}^s$ between the $B_s$ mixing
and the $b \to s c \bar c$ decay amplitude and the width difference
$\Delta \Gamma_s$ can be simultaneously determined from an analysis of
the flavor-tagged time-dependent decay $B_s \to J/\psi \phi$. The
2010 HFAG average\footnote{Notice that only results published or
  accepted in a refereed journal by March 15, 2011 have been included
  in the average.} combines the results of CDF \cite{Aaltonen:2007he,
  Aaltonen:2007gf} and D{\O}~\cite{D0noteB58}, which are based on $1.35
\,{\rm fb}^{-1}$ and $2.8 \, {\rm fb}^{-1}$ of $p \bar p$ data at
$\sqrt{s} = 1.96 \TeV$, respectively.  At 68\%~CL they find
\cite{Asner:2010qj}
\begin{equation}
  \label{eq:phiDGold}
  \phi_{J/\psi\phi}^s  = \left (-44^{+17}_{-21} \right)^\circ \,, \qquad 
  \Delta \Gamma_s      = \left (0.154^{+0.054}_{-0.070} \right ) {\rm ps}^{-1} \,,
\end{equation}
as well as a symmetric solution located at $180^\circ - \phi_{J/\psi\phi}^s$ 
and $-\Delta \Gamma_s$.\footnote{The solution $180^\circ - \phi_{J/\psi\phi}^s$
and $-\Delta \Gamma_s$ has by now been excluded by LHCb at $4.7\sigma$ \cite{Aaij:2012eq}.}
The corresponding numbers in the SM are \cite{Lenz:2011ti, Lenz:2010gu}\footnote{A recent concise review of the theoretical uncertainties plaguing $(\Delta
  \Gamma_s)_{\rm SM}$ has been given in \cite{Lenz:2011zz}.}
\begin{equation}
  \label{eq:phiDGSM}
  (\phi_{J/\psi\phi}^s)_{\rm SM}  = 
    \arg \left[ \frac{\left (V_{ts}^\ast V_{tb} \right )^2}
                      {\left (V_{cs}^\ast V_{cb} \right )^2} \right] 
    = \left (-2.1 \pm 0.1 \right)^\circ \,, \qquad
  (\Delta \Gamma_s)_{\rm SM}  = \left (0.087 \pm 0.021 \right ) {\rm ps}^{-1} \,.
\end{equation}
They deviate from \refeq{eq:phiDGold} by $2.3 \sigma$ and $0.9
\sigma$, respectively. 

Meanwhile, CDF, D{\O}, and LHCb provided new measurements of $\phi_{J/\psi\phi}^s$
and $\Delta \Gamma_s$ with data from 2011 which have been combined in 
the new HFAG 2012 release \cite{Amhis:2012bh}. Both, the CDF \cite{:2012ie} 
and D{\O} \cite{Abazov:2011ry} collaborations, have upgraded their $B_s \to 
J/\psi \phi$ analysis to integrated luminosities of $9.6 \,{\rm fb}^{-1}$ and 
$8.0 \,{\rm fb}^{-1}$, whereas LHCb analyzed $B_s \to J/\psi \phi$ 
\cite{LHCbnote49} and $B_s \to J/\psi \pi^+\pi^-$ \cite{LHCb:2012ad} 
using $1 \,{\rm fb}^{-1}$ of data. The combined results show no evidence
for a new-physics phase in $B_s$--$\bar B_s$ mixing and a value for
$\Delta \Gamma_s$ that is above the SM expectation \refeq{eq:phiDGSM}, but compatible
within the uncertainties.\footnote{Very recently, the ATLAS collaboration 
has also presented measurements of $\phi_{J/\psi\phi}^s$ and $\Delta \Gamma_s$ 
\cite{:2012fu} based on  $4.9 \,{\rm fb}^{-1}$ of data. We add that the error
of $\Delta \Gamma_s = \left (0.053 \pm 0.023 \right ) {\rm ps}^{-1}$ reported by
ATLAS is small, and that the quoted central value is below the SM prediction.
These new results   have not yet been included in
the HFAG averages (\ref{eq:phiDGnew}) that we will use in our work. }  
The corresponding central values and 68\% CL error ranges are
\begin{equation} 
  \label{eq:phiDGnew}
  \phi_{J/\psi\phi}^s = \left (-2.5^{+ 5.2}_{- 4.9}\right)^\circ \,, \qquad 
  \Delta \Gamma_s    = \left (0.105 \pm 0.015 \right ) {\rm ps}^{-1} \,.
\end{equation}

The CP asymmetry in flavor-specific decays $a_{fs}^s$ can be extracted
from a measurement of the like-sign dimuon charge asymmetry $A_{\rm
  SL}^b$, which involves a sample that is almost evenly composed of
$B_d$ and $B_s$ mesons. Performing a weighted average of the results by CDF \cite{CDFnote9015} and 
D{\O}~\cite{Abazov:2010hv, Abazov:2010hj}, which are based on $1.6 \,{\rm fb}^{-1}$
and $6.1 \, {\rm fb}^{-1}$ of integrated luminosity (obtained before 2011), leads to
\begin{equation}
  \label{eq:ASLbold}
   A_{\rm SL}^b  = (-8.5 \pm 2.8) \cdot 10 ^{-3}\,,
\end{equation}
at 68\% CL. Utilizing the SM predictions for the individual
flavor-specific CP asymmetries \cite{Lenz:2011ti}
\begin{equation} 
  \label{eq:afsSM}
  (a_{fs}^d)_{\rm SM}  = -(4.1\pm 0.6) \cdot 10^{-4} \,, \qquad 
  (a_{fs}^s)_{\rm SM}  = (1.9\pm 0.3) \cdot 10^{-5} \,,
\end{equation}
one obtains
\begin{equation} 
  \label{eq:ASLbSMold}
  (A_{\rm SL}^b)_{\rm SM} = 
    (0.506 \pm 0.043) \hspace{0.5mm} (a_{fs}^d)_{\rm SM} 
  + (0.494 \pm 0.043) \hspace{0.5mm} (a_{fs}^s)_{\rm SM} 
  = (-2.0 \pm 0.4) \cdot 10^{-4} \,,
\end{equation}
which differs from \refeq{eq:ASLbold} by $3.0 \sigma$. Using the
measured value of the CP asymmetry in flavor-specific $B_d$ decays
\cite{Asner:2010qj},\footnote{The latest D{\O} measurement of $a_{fs}^d$
yielding $a_{fs}^d = \left (6.8 \pm 4.7 \right ) \cdot 10^{-3}$ 
\cite{:2012uia} is not considered here, because it is not incorporated
in the official HFAG average presently.}
\begin{equation}
  \label{eq:afsd}
  a_{fs}^d  = \left (-4.7 \pm 4.6 \right ) \cdot 10^{-3} \,,
\end{equation}
and \refeq{eq:ASLbold} in \refeq{eq:ASLbSMold}, one can also directly
derive a value of $a_{fs}^s$. One arrives at
\begin{align}
  \label{eq:afssold}
  a_{fs}^s & = (-1.2 \pm 0.7) \cdot 10^{-2} \,,
\end{align}
which compared to \refeq{eq:afsSM} represents a discrepancy of $1.7
\sigma$. Recently, the D{\O} collaboration has updated
its analysis of the like-sign dimuon charge asymmetry.  Employing 
$9.0 \, {\rm fb}^{-1}$ of data, they now find \cite{Abazov:2011yk}
\begin{align}
  \label{eq:ASLbnew}
   A_{\rm SL}^b & = (-7.87 \pm 1.72_{\rm stat.} \pm 0.93_{\rm syst.} ) \cdot 10 ^{-3}\,.
\end{align}
This new number should be confronted with the SM value
\begin{align}
  \label{eq:ASLbSMnew}
  (A_{\rm SL}^b)_{\rm SM} & 
  = (0.594 \pm 0.022) \hspace{0.5mm} (a_{fs}^d)_{\rm SM} 
  + (0.406 \pm 0.022) \hspace{0.5mm} (a_{fs}^s)_{\rm SM} 
  = (-2.4 \pm 0.4) \cdot 10^{-4} \,,
\end{align}
which corresponds to a tension with a statistical significance of
$3.9 \sigma$. This result has been combined by HFAG \cite{Amhis:2012bh}
with the latest $B$-factory results on $a_{fs}^d$ yielding
\beq
  \label{eq:afssnew}
  a_{fs}^d  = (-3.3 \pm 3.3) \cdot 10^{-3} \, , \qquad 
  a_{fs}^s  = (-1.05 \pm 0.64) \cdot 10^{-2} \, .
\eeq
Notice that the value for $a_{fs}^s$ quoted above differs from  $(a_{fs}^s)_{\rm SM}$ as given in  
\refeq{eq:afsSM} by $1.6 \sigma$ only. The average (\ref{eq:afssnew}) does not take into
account the latest LHCb measurement $a_{fs}^s = (-0.24 \pm 0.63) \cdot 10^{-2}$
\cite{LHCb:CONF-2012-022} which is in agreement with the SM within 
uncertainties. Incorporating the latter number into a naive weighted average, results in $a_{fs}^s = (-0.64 \pm 0.45) \cdot 10^{-2}$,  
which corresponds to  a  tension of $1.4 \sigma$. It should be noted, that the
combination  of $A_{\rm SL}^b$ and $a_{fs}^d$
in order to determine $a_{fs}^s$ accounts for contributions beyond the
SM which affect $a_{fs}^d$. When using instead the SM value of $a_{fs}^d$, 
the latest result of $A_{\rm SL}^b$ given in \refeq{eq:ASLbnew} implies 
$a_{fs}^s = (-1.88 \pm 0.49) \cdot 10^{-2}$, which deviates from the SM 
by $3.8\sigma$, {\ie}, a deviation of similar size as directly seen in $A_{\rm SL}^b$.

In view of the observed departures from the SM predictions, it is natural to ask
what kind of new physics is able to simultaneously explain the observed values
of $\Delta M_s$, $\phi_{J/\psi \phi}^s$, $\Delta \Gamma_s$, and $a_{fs}^s$.
Here we would like to address the found anomalies in a model-independent way by
parametrizing the off-diagonal elements of the mass and decay rate matrix as
follows
\begin{equation}
\begin{aligned} 
  \label{eq:M12G12paraNP}
  M_{12}^s = (M_{12}^s)_{\rm SM} + (M_{12}^s)_{\rm NP} & 
  = (M_{12}^s)_{\rm SM} \, R_{M} \, e^{i \phi_{M}} \,, \\[0.0cm]
 \Gamma_{12}^s = (\Gamma_{12}^s)_{\rm SM} + (\Gamma_{12}^s)_{\rm NP} &
  = (\Gamma_{12}^s)_{\rm SM} \, R_{\Gamma} \, e^{i \phi_{\Gamma}} \,.
\end{aligned}
\end{equation}
In the presence of generic new physics, the $B_s$-meson observables of interest
are then given to leading power in $|\Gamma_{12}^s|/|M_{12}^s|$ by
\begin{gather} 
  \Delta M_s = (\Delta M_s)_{\rm SM} \hspace{0.5mm} R_{M} \,, \qquad
  \phi_{J/\psi \phi}^s = (\phi_{J/\psi \phi}^s)_{\rm SM} + \phi_{M}
  \,, \nonumber \\[1mm] \Delta \Gamma_s = (\Delta \Gamma_s)_{\rm SM}
  \hspace{0.25mm} R_{\Gamma} \, \frac{\cos \left (\phi_{\rm SM}^s +
      \phi_{M} - \phi_{\Gamma} \right )} {\cos \phi_{\rm SM}^s }
  \approx (\Delta \Gamma_s)_{\rm SM} \hspace{0.5mm} R_{\Gamma} \, \cos
  \left ( \phi_{M} - \phi_{\Gamma} \right )
  \,, \label{eq:observablesNP} \\[2mm] a_{fs}^s = (a_{fs}^s)_{\rm
    SM}\, \frac{R_{\Gamma}}{R_{M}} \; \frac{\sin \left (\phi_{\rm
        SM}^s + \phi_{M} - \phi_{\Gamma} \right )}{\sin \phi_{\rm
      SM}^s } \approx (a_{fs}^s)_{\rm SM}\, \frac{R_{\Gamma}}{R_{M}}
  \, \frac{\sin \left (\phi_{M} - \phi_{\Gamma} \right )}{ \phi_{\rm
      SM}^s} \,. \nonumber
\end{gather}
Notice that in the case of $\Delta \Gamma_s$ and $a_{fs}^s$ the final results
have been obtained by an expansion in $\phi_{\rm SM}^s = \arg \left
  (-(M_{12}^s)_{\rm SM}/(\Gamma_{12}^s)_{\rm SM} \right )$.  Although this is an
excellent approximation, given that $\phi_{\rm SM}^s = (0.22 \pm 0.06)^\circ$
\cite{Lenz:2010gu}, we will employ the exact analytic expressions in our
numerical analysis.

\begin{figure}[!t]
\begin{center}
\mbox{\includegraphics[height=3in]{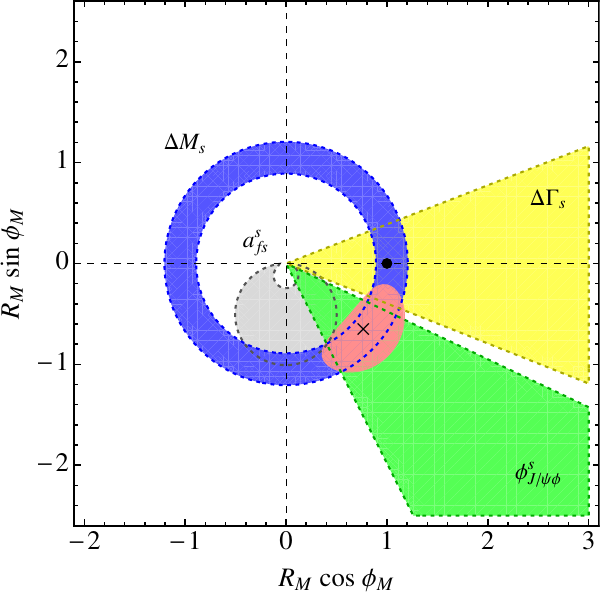}}
\quad
\mbox{\includegraphics[height=3in]{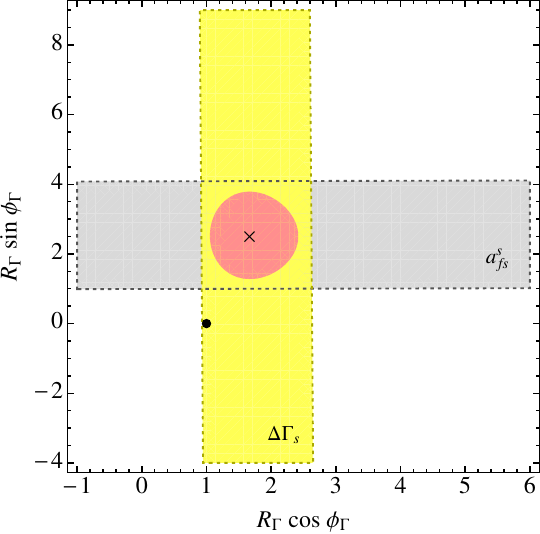}}
\end{center}
\vspace{-8mm}
\begin{center}
  \parbox{15.5cm}{\caption{\label{fig:M12G12old} Left (Right):
      Constraints on the parameter $R_M$ and $\phi_M$ ($R_\Gamma$ and
      $\phi_\Gamma$) from the global fit to the data set D1 in
      scenario S1 (S2). For the individual constraints the colored
      areas represent $68\%$ CL regions (${\rm dofs} = 1$), while for
      the combined fit to the $B_s$-meson mixing data the light red 
      area shows the $95\%$ probability region (${\rm dofs} = 2$). 
      In both panels the SM prediction (best-fit solution) is indicated
      by a dot (cross).}}
\end{center}
\end{figure}

The four real parameters $R_{M,\Gamma}$ and $\phi_{M,\Gamma}$ entering
\refeq{eq:M12G12paraNP} can be constrained by confronting the observed values of
$\Delta M_s$, $\phi_{J/\psi \phi}^s$, $\Delta \Gamma_s$, and $a_{fs}^s$ with
their SM predictions. In the following, we will analyse two different data
sets.\footnote{For simplicity we ignore the
  mirror solutions at $180^\circ - \phi_{J/\psi\phi}^s$ and $-\Delta
  \Gamma_s$ in our discussion hereafter.} The
first set (D1) consists out of \refeq{eq:DMsexp}, \refeq{eq:phiDGold}, and
\refeq{eq:afssold}, while in the second case (D2), we rely on \refeq{eq:DMsexp},
\refeq{eq:phiDGnew}, and \refeq{eq:afssnew}. We begin our analysis by asking how
well the SM hypothesis describes the two sets of data. In the case of D1, our
global fit returns $\chi^2= 9.1$ corresponding to {$1.9\sigma$ ($2.6\sigma$) 
for 4 (2) degrees of freedom (dofs), while for D2 we obtain 
$\chi^2 = 3.2$, which translates into $0.6 \sigma$ $(1.3 \sigma)$ if 4 (2) dofs are considered.  
One observes two features. First, the quality of the fit improves notably when going
from D1 to D2, which is due to the fact that the value of $\phi_{J/\psi\phi}^s$
in the new data set is fully consistent with the SM expectation and, second,
even the new data is not in perfect agreement with the SM hypothesis as a result
of the observed large negative value of $a_{fs}^s$ and, to a lesser extend, the
latest measurement of $\Delta \Gamma_s$, which resides above the SM
prediction. Let us also mention that using $A_{\rm SL}^b$ rather than $a_{fs}^s$
in the $\chi^2$ analysis (\ie, fixing $a_{fs}^d$ to its SM value and employing
\refeq{eq:ASLbold} instead of \refeq{eq:afssold} in the case of D1 and
\refeq{eq:ASLbnew} instead of \refeq{eq:afssnew} in the case of D2) would result
in notable worse fits.  One would find $\chi^2= 14.6$ ($\chi^2 = 15.1$)
for D1 (D2), corresponding to CL values at or above the $3\sigma$ level. The $\chi^2$
of the fit is in this case driven by $A_{\rm SL}^b$, which shows the largest
deviations from the SM. Notice that employing the poor measurement of $a^d_{fs}$
to extract $a^s_{fs}$ from $A^b_{\rm SL}$, as done in \refeq{eq:afssold}
(and to less extend for \refeq{eq:afssnew} by HFAG \cite{Amhis:2012bh}), while allowing 
to improve the fit to the $B_s$-meson data, comes of course with the price of a 
$1.0 \sigma$ tension in $a^d_{fs}$. This feature should be clearly kept in 
mind when interpreting the CL levels quoted here and below.

It is also instructive to determine the best-fit solution of the full
four-parameter fit to the data \refeq{eq:DMsexp}, \refeq{eq:phiDGold}, and
\refeq{eq:afssold} or \refeq{eq:DMsexp}, \refeq{eq:phiDGnew}, and
\refeq{eq:afssnew}.  In the case of D1, we find that $(R_M, \phi_M, R_\Gamma,
\phi_\Gamma) = (1.02, -44^\circ, 3.1, 13^\circ)$ leads to a vanishing $\chi^2$,
while for D2 the $\chi^2$ function equals zero at $(R_M, \phi_M, R_\Gamma,
\phi_\Gamma) = (1.02, -0.4^\circ, 2.5, 61^\circ)$. After marginalization the corresponding 
symmetrized 68\% CL parameter ranges are 
\beq \label{eq:D1CL68}
\begin{aligned}
  R_M & = 1.05 \pm 0.16 \,, & \qquad \phi_M & = (-46 \pm 19)^\circ \,, 
\\
  R_\Gamma & = 3.3 \pm 1.5 \,, & \qquad \phi_\Gamma & = (7 \pm 30)^\circ \,,
\end{aligned}
\eeq
and
\beq \label{eq:D2CL68}
\begin{aligned}
  R_M & = 1.05 \pm 0.16 \,, & \qquad \phi_M & = (-0.4 \pm 5.2)^\circ \,,
\\
  R_\Gamma & = 2.6 \pm 1.2 \,, & \qquad \phi_\Gamma & = (54 \pm 20)^\circ \,.
\end{aligned}
\eeq
We also note that the
combination of $(a_{fs}^d)_{\rm SM}$ with \refeq{eq:ASLbnew} in order to
determine $a_{fs}^s$ yields $(R_M, \phi_M, R_\Gamma,\phi_\Gamma) = 
(1.02, -0.4^\circ, 4.1, 73^\circ)$ and  
\beq \label{eq:D2:afsd:CL68}
\begin{aligned}
  R_M & = 1.05 \pm 0.16 \,, & \qquad \phi_M & = (-0.4 \pm 5.2)^\circ \,, 
\\
  R_\Gamma & = 4.2 \pm 1.2 \,, & \qquad \phi_\Gamma & = (72 \pm 9)^\circ \,.
\end{aligned}
\eeq 
Focusing on $R_\Gamma$ and $\phi_G$, we see that a good fit to the data
sets requires very large corrections to $\Gamma_{12}^s$, either without (D1) or
with (D2) a large weak phase. Even larger effects are required if one tries
to explain  \refeq{eq:ASLbnew} using $(a_{fs}^d)_{\rm SM}$ as an input
(a scenario that we will call D3 from hereon)}. Based on older data sets this
observation has been made before in \cite{Dighe:2010nj, Dobrescu:2010rh, Ligeti:2010ia,
  Bauer:2010dga} and triggered a considerable amount of theoretical
investigations \cite{Bai:2010kf, Alok:2010ij, Oh:2010vc, Kim:2010gx, Oh:2011nb,
  Dutta:2011kg, Dighe:2011du, Goertz:2011nx}.  The first publication that, to
the best of our knowledge, argued convincingly for the possibility to have
sizable new-physics effects in $\Gamma_{12}^s$ is the work \cite{Dighe:2007gt}.

\begin{figure}[!t]
\begin{center}
\mbox{\includegraphics[height=3.0in]{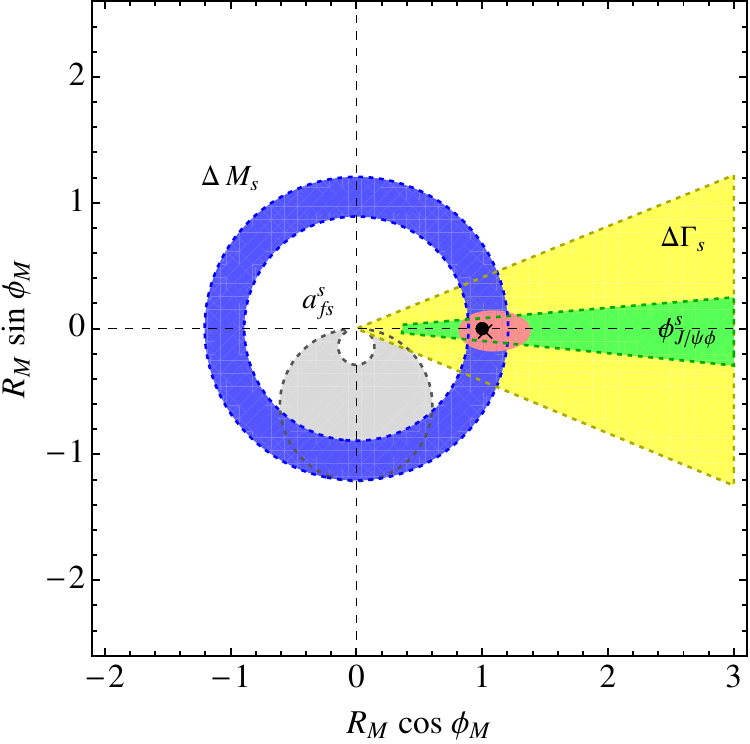}}

\vspace{4mm}

\mbox{\includegraphics[height=3.0in]{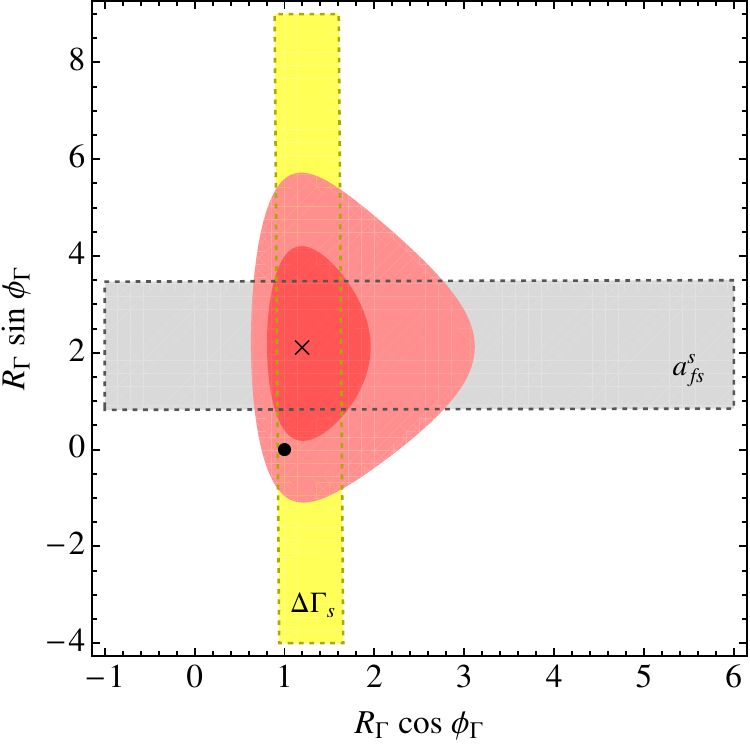}}
\qquad
\mbox{\includegraphics[height=3.0in]{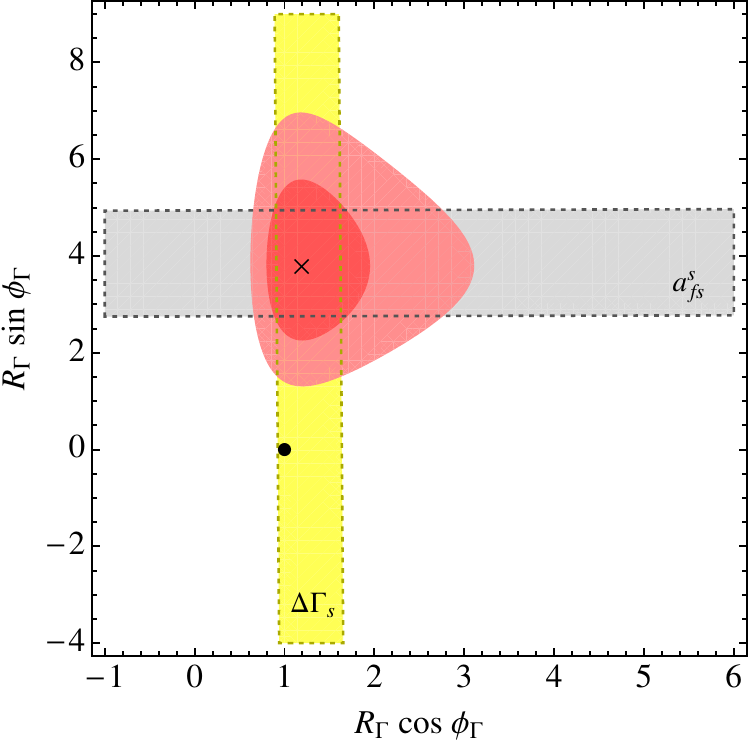}}
\end{center}
\vspace{-8mm}
\begin{center}
  \parbox{15.5cm}{\caption{\label{fig:M12G12new} 
  Upper (Lower left, right): Constraints on the parameter $R_M$ and 
  $\phi_M$ ($R_\Gamma$ and $\phi_\Gamma$) from the global fit to the
  data set D2 in scenario S1 (D2 in scenario S2, D3  in scenario S2). For
  the individual constraints the colored areas represent $68\%$ CL regions
  (${\rm dofs} = 1$), while for the combined fit to the $B_s$-meson mixing
  data the red (light red) area shows the 
  $68\%$ ($95\%$) probability region (${\rm dofs} = 2$). In all panels the
  SM prediction (best-fit solution) is indicated by a dot (cross).
  }}
\end{center}
\end{figure}

To further elucidate the latter point, we analyse two orthogonal hypothesis of
new physics in $B_s$--$\bar B_s$ oscillations, namely a scenario with
$(M_{12}^s)_{\rm NP} \neq 0$ and $(\Gamma_{12}^s)_{\rm NP} = 0$ and a scenario
with $(M_{12}^s)_{\rm NP} = 0$ and $(\Gamma_{12}^s)_{\rm NP} \neq 0$. We refer
to this two different scenarios as S1 and S2, respectively. The left (right)
panel in \reffig{fig:M12G12old} shows the results of our global fit for scenario
S1 (S2) in the $R_M \hspace{0.25mm} \cos \phi_M$--$R_M \hspace{0.25mm} \sin
\phi_M$ ($R_\Gamma \hspace{0.25mm} \cos \phi_\Gamma$--$R_\Gamma \hspace{0.25mm}
\sin \phi_\Gamma$) plane based on the data set D1. By inspection of the left
plot, we infer that in the scenario S1 the preferred 95\% CL parameter region,
is essentially determined by the interplay of $\Delta M_s$ and $\phi_{J/\psi
  \phi}^s$. The best-fit point is located at $(R_M, \phi_M) = (1.01,
-40^\circ)$. It has $\chi^2/{\rm dofs} = 3.1/2$ corresponding to $1.3
\sigma$,\footnote{This result agrees well with the global analysis performed in
  \cite{Lenz:2010gu}, but disagrees with the work \cite{Dighe:2011du}, which
  claims that the scenario S1 is disfavored by more than $3 \sigma$.}  which is
in almost equal parts due to $\Delta \Gamma_s$ and $a_{fs}^s$. This moderate
tension is also clearly visible in the figure.  Turning our attention to the
right panel, we observe that in the case of the hypothesis S2 only a small
region of parameter space has a CL that is better than 95\%. The best-fit
solution resides at $(R_\Gamma, \phi_\Gamma) = (3.0, 56^\circ)$ and has
$\chi^2/{\rm dofs} = 5.4/2$ corresponding to $1.8\sigma$. These numbers imply
that the data set D1 prefers the hypothesis S1 over S2.

By repeating the above statistical analysis for the new data set D2, we obtain
the results displayed in the upper and lower left panel in \reffig{fig:M12G12new}. 
From the upper panel, we glean
that in the scenario S1 the regions of all individual constraints apart from
$a_{fs}^s$ now overlap. Minimizing the $\chi^2$ function gives 
$(R_M, \phi_M) = (1.02, -1.0^\circ)$ and $\chi^2/{\rm dofs} = 3.2/2$ 
corresponding to $1.3\sigma$, which represents only a marginal improvement with
respect to the SM hypothesis. In consequence, the case for a non-zero non-standard 
contribution to $M_{12}^s$ is rather weak.  Comparing the right panel of
Figure~\ref{fig:M12G12old} with the lower left panel of Figure~\ref{fig:M12G12new}, we see that the 95\% CL
region is not only significantly larger in the scenario S2 than before, but that
now also a description of the data with a probability of better than 68\% is
possible. The best-fit point is located at $(R_\Gamma, \phi_\Gamma) = 
(2.4, 61^\circ)$. In fact, the latter parameters lead to an almost perfect fit with
$\chi^2/{\rm dofs} = 0.03/2$ corresponding to $0.02\sigma$. The new data set D2
hence statistically favors the new-physics scenario S2 over the hypothesis S1. 
We stress again that the best-fit points and CL values given in this and 
the last paragraph do depend on whether $a_{fs}^s$ or $A_{\rm SL}^b$ is part of
the $\chi^2$ function. In particular, even larger values for $R_\Gamma$ of around
$4$ would be needed to give a good fit in scenario S2, if \refeq{eq:ASLbold}
instead of \refeq{eq:afssold} or \refeq{eq:ASLbnew} instead of \refeq{eq:afssnew}
would been used. This feature is illustrated in the lower right panel in
\reffig{fig:M12G12new}, which shows the results of a fit to the data set D3. 

The above findings suggest that one hypothetical explanation of the
experimentally observed large negative values of $a_{fs}^s$ (or equivalent
$A_{\rm SL}^b$) consists in postulating new physics in $\Gamma_{12}^s$ that
changes the SM value by a factor of 3 or more. The goal of our work is to study
in detail whether or not and to which extend such an speculative option is
viable. While in principle any composite operator $(\bar s b) \hspace{0.25mm}
f$, with $f$ leading to an arbitrary flavor neutral final state of at least two
fields and total mass below the $B_s$-meson mass, can contribute to
$\Gamma_{12}^s$, the field content of $f$ is in practice very restricted, since
$B_s \to f$ and $B_d \to X_s\hspace{0.25mm} f$ decays to most final states
involving light particles are severely constrained. One notable exception is the
subclass of $B_s$- and $B_d$-meson decays to a pair of tau leptons
\cite{Dighe:2010nj, Bauer:2010dga, Alok:2010ij, Kim:2010gx, Dighe:2011du,
  Dighe:2007gt, Carpentier:2010ue}, to which we will devote our full attention
in this article. Specifically, we perform a thorough model-independent analysis
of the impact of the complete set of $(\bar s b) (\bar \tau \tau)$ operators on
$\Gamma_{12}^s$, taking into account all existing direct and indirect
constraints.  We find that the loose bounds on most of the Wilson coefficients
of the considered operators allow for enhancements of $\Gamma_{12}^s$ of maximal
$35\%$ compared to its SM value, but that the $300\%$ effects in $R_\Gamma$
needed to fully accommodate the data (see \refeq{eq:D1CL68} and
\refeq{eq:D2CL68}) can clearly not be obtained. The presence of $(\bar s b)
(\bar \tau \tau)$ operators alone can thus not reconcile the observed tensions
in the $B_{d,s}$-meson data.  Our work clarifies and extends the existing
analyses \cite{Dighe:2010nj, Alok:2010ij, Dighe:2007gt}. While most of our
discussion is based on an effective-field theory and thus general, we also
consider two explicit models of new physics that alter $\Gamma_{12}^s$, namely
leptoquarks and scenarios with new neutral gauge bosons.

This article is organized as follows. After introducing important definitions
and notations in \refsec{sec:preliminaries}, we study in \refsec{sec:direct} the
direct bounds on the Wilson coefficients of the $(\bar s b) (\bar \tau \tau)$
operators following from $B_s \to \tau^+ \tau^-$, $B \to X_s \tau^+ \tau^-$, and
$B^+ \to K^+ \tau^+ \tau^-$.  \refsec{sec:indirect} is devoted to a
comprehensive discussion of the indirect constraints arising from $b \to s
\gamma$, $b \to s \ell^+ \ell^-$ ($\ell = e, \mu$), and $b \to s \gamma
\gamma$. In \refsec{sec:numerics}, we calculate the effects of the full set of
$(\bar s b) (\bar \tau \tau)$ operators on $\Gamma_{12}^s$ and analyze their
numerical impact. Our model-independent findings are contrasted with two
explicit SM extensions in \refsec{sec:newphysics}.  A summary of our main
results and our conclusions are presented in \refsec{sec:summary}. A series of
appendices contains useful details concerning technical aspects of our
calculations.

%
\section{Preliminaries}
\label{sec:preliminaries}

In the SM, the effective $\Delta B = 1$ Lagrangian is given by
\beq \label{eq:Leff} 
{\cal L}_{\rm eff} = {\cal L}_{{\rm QCD} \hspace{0.25mm} \times
  \hspace{0.25mm} {\rm QED}} \left (u,d,s,c,b,e,\mu,\tau \right ) +
\frac{4 \hspace{0.25mm} G_F}{\sqrt{2}} \, V_{ts}^\ast V_{tb} \sum_i
C_i(\mu) \hspace{0.25mm} \Op_i \,,
\eeq 
and consists of products of Wilson coefficients $C_i$ and dimension-six
operators $\Op_i$. The Wilson coefficients $C_i$ and the matrix elements of the
operators $Q_i$ both depend on the renormalization scale $\mu$, which we set to
the bottom-quark mass $m_b$ throughout this article when evaluating the
$B$-meson observables of interest. In \refeq{eq:Leff} the Fermi constant $G_F$
and the leading CKM factor $V_{ts}^\ast V_{tb}$ have been extracted as a global
prefactor. Finally, the sum over $i$ comprises the current-current operators
($i=1,2$), the QCD-penguin operators ($i=3,4,5,6$), the electromagnetic and
chromomagnetic dipole operators ($i=7,8$), and the semileptonic operators
($i=9,10$).

For what concerns our work, the electromagnetic dipole and the
vector-like semileptonic operator are the most important ones. We
define them in the following way
\beq \label{eq:SMbasis}
\Op_7 = \frac{e}{(4 \pi)^2} \, m_b (\mu) \left ( \bar{s}\,
  \sigma^{\mu\nu} P_R\, b \right ) F_{\mu\nu} \,, \qquad \Op_9 =
\frac{e^2}{(4 \pi)^2} \left ( \bar{s}\, \gamma^\mu P_L\, b \right )
\left ( \bar{\ell}\, \gamma_\mu\, \ell \right ) \,,
\eeq
where $e$ is the electromagnetic coupling constant, $m_b(\mu)$ denotes the
$\overline{\rm MS}$ bottom-quark mass, $F_{\mu \nu}$ is the field strength
tensor associated to the photon, $P_{L,R} = (1 \mp \gamma_5)/2$ project onto
left- and right-handed chiral fields, and $\sigma^{\mu \nu} = i \left
  [\gamma^\mu, \gamma^\nu \right ]/2$. Within the SM one has numerically $C_{7,
  \rm SM} (m_b) \approx -0.3$ and $C_{9, \rm SM} (m_b) \approx 4.1$. The
chiral-flipped partners $\Op_7^\prime$ and $\Op_9^\prime$ of the electromagnetic
dipole and the vector-like semileptonic operator are obtained from
\refeq{eq:SMbasis} by simply replacing the chiral projectors $P_{L,R}$ through
$P_{R,L}$.

In what follows, we supplement the effective SM Lagrangian
\refeq{eq:Leff} by
\beq \label{eq:Leffsbtautau}
{\cal L}_{\rm eff}^\Lambda = \frac{1}{\Lambda^2} \sum_i
C_i^{\hspace{0.25mm} \Lambda} (\mu) \Op_i \,,
\eeq
where $\Lambda > m_t$ denotes the scale of new physics and the index
$i$ runs over the complete set of dimension-six operators with flavor
content $(\bar s b)(\bar \tau \tau)$, namely ($A, B = L, R$)
\beq \label{eq:Qsbtautau}
\begin{split}
  \Op_{S, AB} & = \left (\bar s \, P_A \, b \right ) \left (\bar \tau
    \, P_B \, \tau \right ) \,, \\
  \Op_{V, AB} & = \left (\bar s \, \gamma^\mu P_A \, b \right )
  \left (\bar \tau\, \gamma_\mu P_B \, \tau \right ) \,, \\
  \Op_{T, A} & = \left (\bar s \, \sigma^{\mu\nu} P_A \, b \right )
  \left (\bar \tau \, \sigma_{\mu \nu} P_A \,\tau \right ) \,.
\end{split}
\eeq 
As we will discuss in detail below, the Wilson coefficients of these
operators can be bounded by various direct and indirect constraints for any
given value of $\Lambda$.

For later convenience, we also introduce Wilson coefficients $C_i$
associated to the non-standard operators \refeq{eq:Qsbtautau} that are
normalized like the SM contributions \refeq{eq:Leff}. In terms of the
Wilson coefficients $C_i^\Lambda$ they are simply given by
\beq \label{eq:CiLambda}
C_i (\mu) = \frac{\sqrt{2}}{4 \hspace{0.25mm} G_F} \,
\frac{1}{V_{ts}^\ast V_{tb}} \, \frac{C_i^\Lambda (\mu)}{\Lambda^2}
\,.
\eeq
We now have enough definitions and notations in place to start the
discussion of the relevant restrictions.

%
\section{Direct Bounds on $\bm{ (\bar s b) (\bar \tau \tau)}$ Operators
  \label{sec:direct}
}

The ten operators entering \refeq{eq:Leffsbtautau} govern the purely leptonic
$B_s \to \tau^+ \tau^-$ decay, the inclusive semi-leptonic $B \to X_s
\tau^+\tau^-$ decay, and its exclusive counterpart $B^+ \to K^+ \tau^+\tau^-$,
making these channels potentially powerful constraints. In practice, however,
flavor-changing neutral current $B_{d,s}$ decays into final states involving
taus are experimentally still largely unexplored territory. Rather weak limits
on the branching ratios of the former two modes can be derived from the LEP
searches for $B$ decays with large missing energy \cite{Grossman:1996qj} and/or
the $B$-factory measurements of the ratio of the $B_s\hspace{0.25mm}$- and the
$B_d\hspace{0.25mm}$-meson lifetimes \cite{Dighe:2010nj}.

In the latter case, one finds by comparing the SM prediction
$\tau_{B_s}/\tau_{B_d} - 1 \in [-0.4, 0.0] \%$~\cite{Lenz:2011ti} with the
corresponding experimental result $\tau_{B_s}/\tau_{B_d} - 1 = (0.4 \pm 1.9)
\%$, which is based on the recent LHCb measurement $\Gamma_s = (0.656 \pm
0.012)\,\mbox{ps}^{-1}$ \cite{LHCbnote49} and $\tau_{B_d} = (1.519 \pm 0.007)\,
\mbox{ps}$~\cite{Nakamura:2010zzi}, good agreement compared to previous
measurements \cite{Asner:2010qj}.  Yet, following the arguments of
\cite{Dighe:2010nj}, values of
\beq \label{eq:directbound1}
{\cal B}(B_s \to \tau^+\tau^-) \, < \, 3\% \,,
\eeq
are still allowed at 90\% CL even after the first LHCb measurement.

Constraints also derive from the semileptonic branching ratios
\cite{Lenz:2010gu, Lenz:1997aa}.  For example, one can consider possible
contaminations of the decay samples $b \to u \ell \bar \nu_\ell$ due to one tau
in $B \to X_s \tau^+ \tau^-$ decaying leptonically and the other one
hadronically. If the second $\tau$ decays as $\tau^+ \to u (\bar{d},\,
\bar{s})\,\bar\nu_\tau$ such a decay chain results in the experimental signature
$B \to X_s\, \nu_\tau \bar\nu_\tau + (\pi^+, K^+)\, \ell \bar\nu_\ell$, which
contains at least one strangeness final-state meson.  Since this signal arises
in the SM as a sequence of tree-level decays, one has very naively
\beq
 {\cal B}(B \to X_s \tau^+\tau^-) \approx 
  \frac{{\cal B}(B \to X_s\, \nu_\tau \bar\nu_\tau + (\pi^+, K^+)\, \ell \bar\nu_\ell)}{
  {\cal B}(\tau \to \ell \bar\nu_\ell \nu_\tau) \, 
  {\cal B}(\tau^+ \to u (\bar{d},\, \bar{s})\,\bar\nu_\tau) } \,.
\eeq
Using now  \cite{Nakamura:2010zzi}
\beq
  {\cal B}(\tau \to \ell \bar\nu_\ell \nu_\tau)  \approx 36\% \,, \qquad 
  {\cal B}(\tau^+ \to u (\bar{d},\, \bar{s})\,\bar\nu_\tau)  \approx
    1 - {\cal B}(\tau \to \ell \bar\nu_\ell \nu_\tau) \approx 64\% \,,
\eeq
and taking into account the uncertainties of the relevant  Òsemileptonic decay
modesÓ of $B^{0/\pm}$-admixture \cite{Nakamura:2010zzi}
\beq 
\begin{split}
  {\cal B} (B  \to K^+ \, \ell \bar\nu_\ell + \mbox{anything}) & = 
  (6.2 \pm 0.5) \cdot 10^{-2} \,, \\
 {\cal  B} (B \to K^- \, \ell \bar\nu_\ell + \mbox{anything}) &= 
  (1.0 \pm 0.4) \cdot 10^{-2}\,, \\
  {\cal B} (B \to K^0/\bar{K}^0 \, \ell \bar\nu_\ell + \mbox{anything})  & = 
  (4.6 \pm 0.5) \cdot 10^{-2} \,.
\end{split}
\eeq  
suggests that branching ratios satisfying 
\beq \label{eq:directbound2}
{\cal B} (B \to X_s\tau^+\tau^-) \, \lesssim \, 2.5 \% \,, 
\eeq
would lead to no observable signal. Of course, a misidentification of $B \to
X_s\, \nu_\tau \bar\nu_\tau + (\pi^+, K^+)\, \ell \bar\nu_\ell$ as an exclusive
$B\to \pi\, \ell \bar\nu_\ell$ or an inclusive $B\to X_u\, \ell \bar\nu_\ell$
decay is also possible, {\it i.e.}, loosing the $X_s$ in the analysis. In order
to quantify the precise impact of misidentifications would obviously require a
sophisticated analysis, including all the relevant details of the experimental
analysis, which goes beyond the scope of this work. Let us add that the above
line of reasoning can also be applied to inclusive semileptonic $B$ decays, if
one knows the fraction $f_{\rm cut}$ of $B \to X_s\tau^+\tau^-$ events that
survive the experimental cuts imposed in the $B \to X \ell \nu_\ell$
analysis. Employing $({\cal B}_{\rm SL})_{\rm SM} = \left (11.7 \pm 1.4 \pm 1.0
\right ) \%$ \cite{Bagan:1994qw, Neubert:1996we} and $({\cal B}_{\rm SL})_{\rm
  exp} = \left (10.76 \pm 0.14 \right ) \%$ \cite{Nakamura:2010zzi} leads to the
limit
\beq \label{eq:BrSL}
  {\cal B} (B \to X_s\tau^+\tau^-) \, 
  \lesssim \, \frac{21 \%}{9.3 \hspace{0.25mm} f_{\rm cut} -1} \,.
\eeq
The assumption $f_{\rm cut} = 1$, {\it i.e.}, that all $B \to X_s\tau^+\tau^-$ events
represent a residual background, is quite unrealistic, but it yields the same
numerical value as in (\ref{eq:directbound2}). In practice $f_{\rm cut}$ can
be determined only with a dedicated detector simulation, and hence we expect a
significantly weaker constraint on $B \to X_s\tau^+\tau^-$ from $B \to X \ell
\nu_\ell$. We finally note that $B \to X_s \tau^+ \tau^-$ only starts to become
competitive with the other direct constraints, if its branching ratios can be
constrained to values below $1\%$ (this will become clear only later). Whether
\refeq{eq:directbound2} or the weaker limit of ${\cal O} (5\%)$ extracted in
\cite{Grossman:1996qj} is used, is hence irrelevant for (most of) the further
discussion.

Bounds of strength similar to those given in (\ref{eq:directbound1}) and
(\ref{eq:directbound2}) also follow from charm counting \cite{Kagan:1997qna,
  Kagan:1997sg}. Combining the latest BaBar result for the average of the $B^-$
and $B_d$ charm multiplicity $n_c = 1.20 \pm 0.06$ with ${\cal B} (B \to X_{s c
  \bar c}) = 0.24 \pm 0.02$~\cite{Aubert:2006mp}, yields ${\cal B} (B \to X_{\rm
  no \ charm}) = (4 \pm 5)\%$ at 68\% CL.  Subtracting from this number the SM
prediction for ${\cal B} (B \to X_{\rm no \ charm})$ of around 1.5\%
\cite{Lenz:1997aa}, suggests that \refeq{eq:directbound1} is a conservative
upper limit, that is consistent with charm counting.\footnote{The $B \to X_{\rm
    no \ charm}$ branching ratio has also been determined by DELPHI
  \cite{Abreu:1998xb}. The quoted 95\% CL bound of ${\cal B} (B \to X_{\rm no \
    charm}) < 3.7\%$ relies however on a Monte Carlo simulation to model the
  displaced vertex distribution for intermediate charm, which introduces an
  uncertainty that is hard to quantify.}
  
The unsatisfactory experimental situation concerning $B_{d,s}$-meson decays into
final states with tau pairs, has been improved recently by the BaBar
collaboration, which was able to set a first upper limit on the branching ratio
of $B^+ \to K^+ \tau^+ \tau^-$.  At 90\% CL this bound
is~\cite{Flood:2010zz}\footnote{The measurement is restricted to dilepton
  invariant masses of $14.23 \GeV^2 < q^2 <(M_{B^+} - M_{K^+})^2$.  This
  experimental cut is implemented in our calculation.}
\beq \label{eq:directbound3}
{\cal B} (B^+ \to K^+ \tau^+\tau^-) \, < \, 3.3 \cdot 10^{-3} \,.
\eeq

Although the limits \refeq{eq:directbound1} and \refeq{eq:directbound3} are many
orders of magnitude above the corresponding SM predictions, we will see below
that they provide currently the strongest constraints on the Wilson coefficients
of the scalar and vector operators of $(\bar s b) (\bar \tau \tau)$ type.  Let
us first have a look at $B_s \to \tau^+ \tau^-$ and $B \to X_s \tau^+
\tau^-$. In the case of the exclusive decay the relevant branching ratio reads
\cite{Bobeth:2002ch}
\beq \label{eq:Bstau2} 
{\cal B} (B_s \to \tau^+\tau^-) = {\cal N}_{B_s \to \tau^+\tau^-} \,
\left \{ \left(1 - \frac{4 m_\tau^2}{M_{B_s}^2} \right) \big |F_S \big
  |^2 + \left |F_P + \frac{2 \hspace{0.25mm}m_\tau}{M_{B_s}} \, F_A
  \right |^2 \right \} \,,
\eeq 
where
\beq \label{eq:NBstau2} 
{\cal N}_{B_s \to \tau^+\tau^-} = \frac{G_F^2 M_{B_s}^3 f_{B_s}^2
  \tau_{B_s}}{16 \pi} \, |V_{ts}^\ast V_{tb} |^2 \sqrt{1 -
  \frac{4\hspace{0.25mm} m_\tau^2}{M_{B_s}^2}} \; \approx \;  6.4 \cdot
10^{-2} \,,
\eeq 
and 
\beq \label{eq:FSPA}
\begin{split}
  & F_{S,P} = \frac{M_{B_s}}{m_b + m_s} \Big[
  C_{S,RR} \mp C_{S,LL} \pm C_{S,RL}- C_{S,LR}\Big]\,, \\
  & F_A = \frac{\alpha}{2 \pi} \, C_{10,\rm SM} + C_{V,LR} + C_{V,RL}
  - C_{V,LL}- C_{V,RR} \,.
\end{split}
\eeq
Here all Wilson coefficients are understood to be normalized as in
\refeq{eq:Leff} and \refeq{eq:CiLambda} and evaluated at $m_b$.  Notice also
that the SM contribution proportional to $C_{10, \rm SM} (m_b) \approx -4.2$ has
been kept separately in $F_A$. Compared to the new-physics contribution this
correction is suppressed by both a CKM and a loop factor, resulting in ${\cal B}
(B_s \to \tau^+\tau^-)_{\rm SM} \approx 8 \cdot 10^{-7}$. In order to obtain
this number, we have employed $G_F = 1.16637 \cdot 10^{-5} \GeV^{-2}$, $M_{B_s}
= 5.3663 \GeV$ \cite{Nakamura:2010zzi}, $f_{B_s} = 238.8 \MeV$
\cite{Laiho:2009eu}, $\tau_{B_s} = 1.477 \, {\rm ps}$ \cite{Asner:2010qj},
$|V_{ts}^\ast V_{tb}|^2 = 1.6 \cdot 10^{-3}$ \cite{Charles:2004jd}, $m_\tau
=1.777 \GeV$, and $\alpha = \alpha(M_Z) = 1/129$ \cite{Bobeth:2003at}.

Assuming the dominance of a single operator and neglecting the SM contribution
to \refeq{eq:FSPA}, we then find by combining \refeq{eq:directbound1} with
\refeq{eq:Bstau2} the following upper bounds on the magnitudes of the Wilson
coefficients
\beq \label{eq:Bstau2boundsmb}
|C_{S,AB} (m_b) | \, < \, 0.5 \,, \qquad 
|C_{V,AB} (m_b) | \, < \, 1.0 \,.
\eeq
Here $m_b = m_b^{\rm pole} = 4.8 \GeV$ and $m_s = 0.1\GeV$ have been used.

By means of the solutions to the LO renormalization group equations (RGEs)
presented in \refapp{app:mixing}, the results \refeq{eq:Bstau2boundsmb} can be
reinterpreted as limits on the matching corrections to the Wilson coefficients
at the scale $\Lambda$. Performing the evolution from $m_b$ ($m_t$) up to $m_t$
($\Lambda$) in a five-flavor (six-flavor) theory, we obtain in the case of the
scalar operators
\beq \label{eq:Bstau2boundsLambda}
|C_{S,AB} (\Lambda) | \, = \, \eta_5^{12/23} \, \eta_6^{4/7} \;
|C_{S,AB} (m_b) | \, < \, \eta_6^{4/7} \;   0.4 \,,
\eeq 
where $\eta_5 = \alpha_s(m_t)/\alpha_s(m_b) \approx 0.5$ and $\eta_6 = \alpha_s
(\Lambda)/\alpha_s (m_t) = [ 1.223 + 0.124 \ln \left (\Lambda/{\rm TeV} \right
)]^{-1}$. The bound on $C_{V,AB}$ is, on the other hand, scale independent,
because the vector operators $\Op_{V,AB}$ correspond to conserved currents. In
\reftab{tab:bounds1}, we summarize for convenience the currently available
direct bounds on the high-scale Wilson coefficients assuming a new-physics scale
of $\Lambda = 1 \TeV$.

\begin{table}
\begin{center}
\begin{tabular}{|c|ccc|}
  \hline
  $|C_i (1 \TeV)|$ &
  $B_s \to \tau^+\tau^-$ & $B \to X_s \tau^+\tau^-$ & $B^+ \to K^+ \tau^+\tau^-$
  \\
  \hline 
  $S,AB$ &  $0.3$  &  $1.9$  & $0.5$ 
  \\
  $V,AB$ & $1.0$ &  $1.5$  & $0.8$
  \\
  $T,A$ & --- &  $0.5$ & $0.5$
  \\
  \hline
\end{tabular}

\vspace{4mm}

\parbox{15.5cm}{\caption{\label{tab:bounds1} Direct upper bounds on
    the high-scale Wilson coefficients at $\Lambda = 1\TeV$.}}
\end{center}  
\end{table}

Unlike $B_s \to \tau^+ \tau^-$, the inclusive decay $B \to X_s \tau^+
\tau^-$ depends on all Wilson coefficients. Its differential branching
ratio might be split into the SM contribution, the interference of the
SM with the new-physics contributions, and the pure new-physics
corrections. Since we assume that the Wilson coefficients of the
operators \refeq{eq:Qsbtautau} are generated at tree level, we are
allowed to neglect both the SM and the interference terms, which are
suppressed relative to the new-physics effects by at least a factor of
order $\alpha/(4 \pi) \approx 6 \cdot 10^{-4}$. The relevant
contributions to the branching ratio differential in the dilepton
invariant mass, $s = q^2$, read \cite{Fukae:1998qy}
\beq \label{eq:dBrXsdq2}
\left ( \frac{d {\cal B} (B \to X_s \tau^+ \tau^-)}{ds} \right )_i =
{\cal N}_{B \to X_s \tau^+ \tau^-} \; |C_i|^2 \, M_i^{X_s} (s) \,.
\eeq
Here $i = S,AB, \hspace{0.5mm} V,AB, \hspace{0.5mm} T,A$ and
\beq \label{eq:NXs}
{\cal N}_{B \to X_s \tau^+ \tau^-} = \frac{3}{2m_b^8} \,
\frac{|V_{ts}^\ast V_{tb}|^2}{|V_{cb}|^2} \, \frac{{\cal B} (B \to X_c
  \ell \nu_\ell)_{\rm exp}}{f(z) \hspace{0.5mm} \kappa (z)} \, \approx
\, 1.1  \, \cdot 10^{-6} \,.
\eeq
The analytic expressions for the kinematic functions $M_i^{X_s} (s)$
can be found in \refapp{app:kinfun}. In order to obtain the numerical
result in \refeq{eq:NXs}, we have employed $|V_{ts}^\ast
V_{tb}|^2/|V_{cb}|^2 = 0.96$ \cite{Charles:2004jd}, ${\cal B} (B \to
X_c \ell \nu_\ell)_{\rm exp} = 10.23 \%$ \cite{Asner:2010qj}, and $z =
(m_c^{\rm pole}/m_b^{\rm pole})^2 = 0.084$. For this input the
phase-space factor and the NLO QCD corrections \cite{Nir:1989rm} of $B
\to X_c \ell \nu_\ell$ evaluate to $f (z) \approx 0.54$ and $\kappa
(z) \approx 0.88$.  The Wilson coefficients are bounded by integrating
the non-resonant branching ratio over the entire kinematical range $4
\hspace{0.25mm} m_\tau^2 < s < (m_b - m_s)^2$ and comparing the
obtained result to the experimental
extraction.\footnote{Experimentally the narrow $\psi^\prime$ resonance
  at $s \approx 13.7 \GeV^2$ has to be removed by making appropriate
  kinematic cuts in the invariant mass spectrum. In view of the
  poorness of the upper limit on ${\cal B} (B \to X_s \tau^+ \tau^-)$,
  it is at present immaterial if these cuts are imposed in the
  theoretical calculation.} 
Numerically, the integrations yield
\begin{align} 
  \label{eq:MXs}
  \int \! ds \, M_{S}^{X_s} (s) & \approx 2636 \,, &
  \int \! ds \, M_{V}^{X_s} (s) & \approx 10542 \,, & 
  \int \! ds \, M_{T}^{X_s} (s) & \approx 126505 \,.
\end{align}

From the bound ${\cal B} (B \to X_s \tau^+ \tau^-) \lesssim 2.5\%$,
which is five orders of magnitude above the SM 
expectation ${\cal B}(B \to X_s \tau^+ \tau^-)_{\rm SM} \approx 5 \cdot 10^{-7}$
\cite{Hewett:1995dk}, we then derive, by utilizing \refeq{eq:dBrXsdq2}
to \refeq{eq:MXs} and considering each type of Wilson coefficient
individually, the inequalities
\beq \label{eq:BXsboundsmb} 
|C_{S,AB} (m_b) | \, \lesssim \, 2.9 \,, \quad |C_{V,AB}
(m_b) | \, \lesssim \, 1.5 \,, \quad |C_{T,A} (m_b) | \,
\lesssim \, 0.4 \,.
\eeq
Comparing these results with \refeq{eq:Bstau2boundsmb}, one observes
that $B \to X_s \tau^+ \tau^-$ constrains the contributions arising
from scalar and vector operators much less severely than $B_s \to
\tau^+ \tau^-$.

It is again straightforward to find the corresponding bounds on the
high-scale Wilson coefficients. In the case of the scalar operators,
one proceeds in analogy to \refeq{eq:Bstau2boundsLambda}, the upper
bound on $C_{V,AB}$ is unchanged, and in the last case, we arrive at
\beq \label{eq:BXsboundsLambda}
|C_{T,A} (\Lambda) | \, = \, \eta_5^{-4/23} \, \eta_6^{-4/21} \,
|C_{T,A} (m_b) | \, \lesssim \, \eta_6^{-4/21} \; 0.5 \,.
\eeq 
The bound \refeq{eq:BXsboundsLambda}
and the limits on $C_{S,AB}$ and $C_{V,AB}$ stemming from $B \to X_s
\tau^+ \tau^-$ are collected in \reftab{tab:bounds1}.

Let us finally consider the bounds on the Wilson coefficients $C_i$
that follow from the upper limit \refeq{eq:directbound3}, which still
leaves ample room for new physics to enhance the branching ratio with
respect to the SM prediction of ${\cal B} (B^+ \to K^+ \tau^+
\tau^-)_{\rm SM} \approx 2 \cdot 10^{-7}$ \cite{Du:1993sh}. Keeping
again only the pure new-physics contributions, the individual
corrections to the differential branching ratio of $B^+ \to K^+ \tau^+
\tau^-$ take the form \cite{Bobeth:2007dw}
\beq \label{eq:dBrKdq2}
\left (\frac{d {\cal B} (B^+ \to K^+ \tau^+ \tau^-)}{ds} \right )_i =
{\cal N}_{B^+ \to K^+ \tau^+ \tau^-} \, |C_i|^2 \, M_i^{K^+} (s) \,,
\eeq
where 
\beq \label{eq:NKp}
{\cal N}_{B^+ \to K^+ \tau^+ \tau^-} = \frac{G_F^2 \hspace{0.25mm}
  |V_{ts}^\ast V_{tb}|^2 \hspace{0.25mm} \tau_{B^+}}{24 \pi^3
  M_{B^+}^3} \, \approx \, 5.0 \cdot 10^{-6} \,,
\eeq
corresponding to $\tau_{B^+} = 1.641 \, {\rm ps}$ \cite{Asner:2010qj}
and $M_{B^+} = 5.279 \GeV$ \cite{Nakamura:2010zzi}. The functions
$M_i^{K^+} (s)$ are collected in \refapp{app:kinfun}.  Integrating
them over $14.23 \GeV^2 < s < (M_{B^+} - M_{K^+})^2$ gives
\begin{align} 
  \label{eq:MKp}
  \int \! ds \, M_{S}^{K^+} (s) & \approx 1081 \,, &
  \int \! ds \, M_{V}^{K^+} (s) & \approx 1082 \,, &
  \int \! ds \, M_{T}^{K^+} (s) & \approx 3610 \,.
\end{align}

Inserting \refeq{eq:NKp} and \refeq{eq:MKp} into \refeq{eq:dBrKdq2},
and allowing for the presence of a single Wilson coefficient at a
time, the upper bound \refeq{eq:directbound3} implies
\beq \label{eq:BKpboundsmb}
|C_{S,AB} (m_b) | \, < \, 0.8 \,, \qquad |C_{V,AB} (m_b) | \,
< \, 0.8 \,, \qquad |C_{T,A} (m_b) | \, < \, 0.4 \,.
\eeq 

The limits on the Wilson coefficients at the scale $\Lambda = 1 \TeV$
following from $B^+ \to K^+ \tau^+ \tau^-$ are shown in
\reftab{tab:bounds1}.  The quoted numbers have been obtained from
\refeq{eq:BKpboundsmb} by simply applying the relations
\refeq{eq:Bstau2boundsLambda} and \refeq{eq:BXsboundsLambda}. We see
that the exclusive $b \to s \tau^+ \tau^-$ mode provides at present
the most stringent direct constraints on the vector and tensor
operators with flavor content $(\bar s b) (\bar \tau \tau)$, while it
is less restrictive than $B_s \to \tau^+ \tau^-$ for what
concerns the scalar contributions. Notice also that the resummation of
large leading logarithms of the form $\alpha_s^n \ln^n
(\Lambda^2/m_b^2)$ makes the bound on $|C_{S,AB} (\Lambda)|$
($|C_{T,AB} (\Lambda)|$) relative to the limit that applies to
$|C_{S,AB} (m_b)|$ ($|C_{T,AB} (m_b)|$) stronger (weaker).  The
running effects are however moderate in both cases if new physics
enters at the TeV scale, as they change the results by a factor of
around 1.6 and 0.9 only. Numerically, we see that the direct
constraints allow for effects that reach almost ${\cal O} (1)$ in all
$|C_i (1 \TeV)|$. Recalling that the dominant SM contribution to
$\Gamma_{12}^s$ arises from the color-singlet current-current operator
$(\bar s \hspace{0.5mm} \gamma^\mu P_L \hspace{0.25mm} c) (\bar c
\hspace{0.5mm} \gamma_\mu P_L \hspace{0.25mm} b)$, which has a Wilson
coefficient of roughly 1 at the weak scale, we expect that the $(\bar
s b) (\bar \tau \tau)$ operators can give a visible correction to the
off-diagonal element of the decay matrix in the $B_s$-meson system.
We however curb our enthusiasm and postpone a detailed numerical
analysis of the new-physics effects in $\Gamma_{12}^s$ to
\refsec{sec:numerics}, to check first that the indirect constraints
associated to operator mixing do not thwart these potentially large
effects.

%
\section{Indirect Bounds on $\bm{ (\bar s b) (\bar \tau \tau)}$ Operators
  \label{sec:indirect}
}

\begin{figure}[!t]
\begin{center}
\mbox{\includegraphics[height=1.5in]{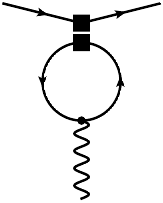}}
\qquad \qquad 
\mbox{\includegraphics[height=1.5in]{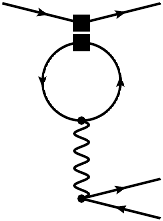}}
\end{center}
\begin{picture}(0,0)(0,0)
\put(150,138){\large $b$}
\put(192,138){\large $s$}
\put(202,90){\large $\tau$}
\put(138,90){\large $\tau$}
\put(185,49){\large $\gamma$}
\put(287,138){\large $b$}
\put(325,138){\large $s$}
\put(334,94){\large $\tau$}
\put(275,94){\large $\tau$}
\put(318,58){\large $\gamma$}
\put(355,46){\large $\ell^-$}
\put(355,25){\large $\ell^+$}
\end{picture}
\vspace{-8mm}
\begin{center}
  \parbox{15.5cm}{\caption{\label{fig:mixing} Diagrams with a penguin
      insertion of a $(\bar s b) (\bar \tau \tau)$ operator (black
      squares) that contribute to the renormalization and the matrix
      element of the electromagnetic dipole (left) and vector-like
      semileptonic (right) operators. The tau loop in both graphs is
      closed.}}
\end{center}
\end{figure}
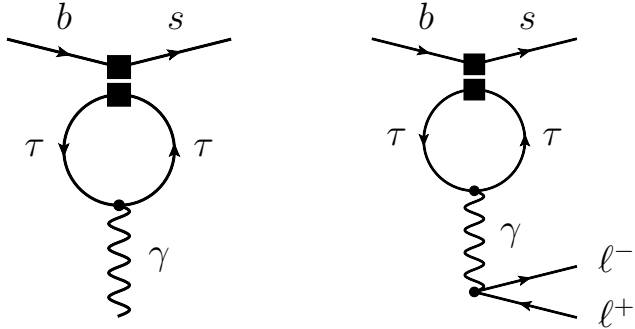

Further constraints on the Wilson coefficients \refeq{eq:CiLambda}
arise indirectly from the experimentally available information on the
$b \to s \gamma$ and $b \to s \ell^+ \ell^-$ ($\ell = e, \mu$)
transitions, because some of the effective operators introduced in
\refeq{eq:Qsbtautau} mix into
\beq \label{eq:penguins}
\Op_{7,A} = \frac{e}{g_s^2} \, m_{\tau} \, (\bar s \, \sigma^{\mu \nu}
P_A \, b) F_{\mu \nu} \,, \qquad \Op_{9,A} = \frac{e^2}{g_s^2} \,
(\bar s \, \gamma^\mu P_A \, b) ( \bar \ell \, \gamma_\mu\, \ell) \,,
\eeq
at the one-loop level. Here $A = L,R$. The corresponding Feynman
diagrams are depicted in \reffig{fig:mixing}. It is important to
realize that due to the flavor structure of the $(\bar s b) (\bar \tau
\tau)$ operators only insertions are possible in which the tau lines
are joined and the photon is emitted from the resulting closed
loop. Our results for the anomalous dimension matrix (ADM) describing
the operator mixing of $\Op_{S,AB}$, $\Op_{V,AB}$, and $\Op_{T,A}$
into $\Op_{7,A}$ and $\Op_{9,A}$ are given in \refapp{app:mixing}. We
find that the ADM is sparse, because most of the relevant penguin
diagrams either contain a vanishing Dirac trace or evaluate to zero
due to current conservation.  In particular, the operators
$\Op_{S,AB}$ mix neither into $\Op_{7,A}$ nor
$\Op_{9,A}$,\footnote{Notice that scalar operators with flavor
  structure $(\bar s b) (\bar q q)$ with $q = s, b$ mix into the
  electromagnetic dipole and vector-like semileptonic operators at the
  one-loop level \cite{Borzumati:1999qt, Hiller:2003js}. The mixing
  arises from graphs constructed by joining two strange or bottom
  quarks belonging to the different disconnected parts of the
  composite operator and attaching the photon to the resulting open
  strange- or bottom-quark loop.} while $\Op_{V,AB}$ ($\Op_{T,A}$)
mixes only into $Q_{9,A}$ ($\Op_{7,A}$). The possible one-loop mixing
of the $(\bar s b) (\bar \tau \tau)$ operators is therefore much more
restricted than what has been claimed in the articles
\cite{Bauer:2010dga, Kim:2010gx, Dutta:2011kg}, which all did not
perform an explicit calculation of the one-loop ADM. We will see in a
moment, that as a result of the particular mixing pattern, the
stringent constraints from the radiative decay $B \to X_s \gamma$ rule
out large contributions to $\Gamma_{12}^s$ only if they arise from the
tensor operators $\Op_{T,A}$.  Similarly, the rare decays $B \to X_s
\ell^+ \ell^-$ and $B \to K^{(\ast)} \ell^+ \ell^-$ primarily limit
contributions stemming from the vector operators $\Op_{V,AB}$.

The new-physics corrections to the partonic $b \to s \gamma$ and $b
\to s \ell^+ \ell^-$ amplitudes involve the low-energy Wilson
coefficients of the effective operators introduced in
\refeq{eq:penguins}. Neglecting the matching corrections to both
$Q_{7,A}$ and $Q_{9,A}$, \ie, setting $C_{7,A} (\Lambda) = C_{9,A}
(\Lambda) = 0$, we find from the analytic solutions to the RGEs as
given in \refapp{app:mixing}, the following expressions
\beq \label{eq:DeltaC7C9}
\begin{split}
  C_{7,A}(m_b) & = \; \eta_6^{4/21} \left (0.6 -
    \eta_6^{-1} \right ) C_{T,A}(\Lambda) \,, \\[2mm]
  C_{9,A} (m_b) & = \left (0.1 - 0.2 \hspace{0.75mm} \eta_6^{-1}
  \right ) \big ( C_{V,AL} (\Lambda) + C_{V,AR} (\Lambda) \big ) \,.
\end{split}
\eeq

Determining the contributions of the operator sets
\refeq{eq:Qsbtautau} and \refeq{eq:penguins} to the radiative and
semileptonic $b \to s$ processes also requires the knowledge of the
corresponding tree-level and one-loop matrix elements. Details on
these calculations are relegated to
\refapp{app:matrixelements}. Following common practice, we include the
effects of the relevant matrix elements by defining so-called
effective Wilson coefficients \cite{Buras:1993xp}. For the SM
operators $\Op_7$ and $\Op_9$ appearing in \refeq{eq:SMbasis}, we
obtain at the scale $\mu = m_b$ the following new-physics corrections
\beq \label{eq:C7C9eff}
\begin{split}
  \Delta C_{7}^{{\rm eff}} (s, m_b) & = \sqrt{x_\tau} \, \big (
  C_{7,R}(m_b) + M_7 (\hat s, x_\tau)
  \hspace{0.5mm} C_{T,R} (m_b) \big ) \,, \\[1mm]
  \Delta C_{9}^{{\rm eff}} (s, m_b) & =C_{9,L} (m_b) + M_9 (\hat s,
  x_\tau) \hspace{0.5mm} \big ( C_{V,LL} (m_b) + C_{V,LR} (m_b) \big )
  \,,
\end{split}
\eeq
with $C_{7,A} (m_b)$ and $C_{9,A} (m_b)$ given in \refeq{eq:DeltaC7C9}
and $C_{T,A} (m_b) = 0.9 \, \eta_6^{4/21} \, C_{T,A} (\Lambda)$.
Furthermore, $\hat s = q^2/m_b^2$ and $x_\tau = m_\tau^2/m_b^2$.  The
analytic expressions for the functions $M_{7} (\hat s, x_\tau)$ and
$M_{9} (\hat s, x_\tau)$ can be found in
\refapp{app:matrixelements}. Analogous formulas hold in the case of
the Wilson coefficients $\Delta C_{7}^{\prime \hspace{0.25mm} {\rm
    eff}} (s, m_b)$ and $\Delta C_{9}^{\prime \hspace{0.25mm} {\rm
    eff}} (s, m_b)$ of the chiral-flipped operators $\Op_{7}^\prime$
and $\Op_{9}^\prime$ after replacing $C_{i,R} (m_b)$ through $C_{i,L}
(m_b)$ $\big ( C_{i,LA} (m_b)$ through $C_{i,RA} (m_b) \big )$ in the
first (second) line of \refeq{eq:C7C9eff}.  Notice that only $\Delta
C_7^{\rm eff} (m_b) = \Delta C_7^{\rm eff} (0,m_b)$ and $\Delta
C_7^{\prime \hspace{0.25mm} {\rm eff}} (m_b) = \Delta C_7^{\prime
  \hspace{0.25mm} {\rm eff}} (0,m_b)$ enter the prediction for $B \to
X_s \gamma$, while all $\Delta C_i^{\rm eff} (s,m_b)$ and their
chirality-flipped partners affect the $b \to s \ell^+ \ell^-$
transitions.

In order to derive a bound on the Wilson coefficients $C_{T,A}$ at the
high scale, we compare the experimental measurement of the $B \to X_s
\gamma$ branching ratio with its SM prediction. For a photon-energy
cut of $E_\gamma > 1.6 \, \GeV$, the experimental world average reads
\cite{Amhis:2012bh}
\beq \label{eq:BRBXsgammaexp}
  {\cal B}(B \to X_s \gamma)_{\rm exp} = \left ( 3.55 \pm 0.24 \pm 0.09
\right ) \cdot 10^{-4} \,,
\eeq
while the SM expectation is given by \cite{Misiak:2006zs,
  Misiak:2006ab}\footnote{Several NNLO QCD corrections (see
  \cite{Asatrian:2006rq, Boughezal:2007ny, Ewerth:2008nv,
    Asatrian:2010rq, Ferroglia:2010xe, Misiak:2010tk} and partly
  \cite{Czakon:2006ss}) that were calculated after the publication of
  \cite{Misiak:2006zs, Misiak:2006ab} are not included in the central
  value of the SM prediction, but remain within the perturbative
  higher-order uncertainty of 3\% that has been estimated in the
  latter two articles.}
\beq \label{eq:BRBXsgammaSM} 
{\cal B} (B \to X_s \gamma)_{\rm SM} = (3.15 \pm 0.23) \cdot 10^{-4}
\, .
\eeq
The good agreement between \refeq{eq:BRBXsgammaexp} and
\refeq{eq:BRBXsgammaSM} puts stringent limits on the corrections
$\Delta C_7^{\rm eff} (m_b)$ and $\Delta C_7^{\prime \hspace{0.25mm}
  {\rm eff}} (m_b)$. Further non-negligible constraints on the
magnitudes and phases of the corrections to the low-energy Wilson
coefficients of the electromagnetic dipole operators arise from the
inclusive \cite{Gambino:2004mv} and exclusive~\cite{Bobeth:2008ij} $b
\to s \ell^+ \ell^-$ transitions as well as the time-dependent
CP-asymmetries and the isospin asymmetry in $B \to K^\ast \gamma$
\cite{DescotesGenon:2011yn}. Including all these constraints and
allowing for the variation of a single effective Wilson coefficient at
a time, we obtain the following 90\% CL upper bounds 
\beq \label{eq:DeltaC7C7p} 
|\Delta C_7^{\rm eff} (m_b)| \, < \, 0.23\,, \qquad |\Delta
C_7^{\prime \hspace{0.25mm} {\rm eff}} (m_b)| \, < \, 0.20\,.
\eeq
Notice that since we are treating the Wilson coefficients as complex,
the upper limit on the magnitude of $\Delta C_7^{\rm eff} (m_b)$ is
weaker by almost a factor of 6 than the bound that holds in the case
of a real coefficient. In contrast, the bound on $|\Delta C_7^{\prime
  \hspace{0.25mm} {\rm eff}} (m_b)|$ is not affected by whether the
correction is treated as complex or real, because $\Op_7^\prime$ does
not interfere with the SM contribution to first order. We also remark
that the $B \to K^{(\ast)} \ell^+ \ell^-$ observables are more efficient
than $B \to K^\ast \gamma$ in restricting the allowed values of the
real and especially the imaginary parts of $\Delta C_7^{\rm eff} (m_b)$.

Utilizing \refeq{eq:DeltaC7C9} and \refeq{eq:C7C9eff}, the limits
\refeq{eq:DeltaC7C7p} translate into
\beq \label{eq:boundsindirect1}
|C_{T,R} (\Lambda)| \, = \, \eta_6^{17/21} \left ( 0.4 + 2.3 \, \eta_6
\right )^{-1} |\Delta C_7^{\rm eff} (m_b) | \, < \, \eta_6^{-4/21}
\left ( \hspace{0.25mm} 10.2 + 1.8 \, \eta_6^{-1} \right )^{-1} \,,
\eeq
and an analog inequality, obtained by replacing $|\Delta C_7^{\rm eff}
(m_b) | $, $10.2$, and $1.8$ by $|\Delta C_7^{\prime \hspace{0.25mm}
  {\rm eff}} (m_b) | $, $11.8$, and $2.1$, holds in the case of
$C_{T,L}$. For a new-physics scale of $\Lambda = 1 \TeV$ the numerical
values of the bounds on $|C_{T,A} (\Lambda)|$ are given in
\reftab{tab:bounds2}. Notice that as a result of the resummation of
large logarithms the bounds on the low-scale Wilson coefficients of
the tensor operators are a factor of around 1.2 stronger than those on
the initial conditions.

The magnitudes of the Wilson coefficients $C_{V,AB}$ can be bounded by
comparing the available data on the inclusive \cite{Aubert:2004it,
  Iwasaki:2005sy, Nakayama:2009zz, Chiang:2010zz} and exclusive
\cite{Aubert:2006vb, Aubert:2008ju, Wei:2009zv, Aaltonen:2011cn,
  LHCbnote38, Aaltonen:2011qs, Aaltonen:2011ja} $b \to s \ell^+
\ell^-$ transitions with the corresponding theoretical predictions
that include the corrections $\Delta C_9^{\rm eff} (s,m_b)$ and
$\Delta C_9^{\prime \hspace{0.25mm} {\rm eff}} (s,m_b)$. Our global
analysis of rare semileptonic $B$ decays relies on the article
\cite{Bobeth:2003at} for what concerns $B \to X_s \ell^+ \ell^-$ and
on the works \cite{Bobeth:2008ij, Bauer:2009cf, Bobeth:2010wg,
  Bobeth:2011gi, EOS} in the case of $B \to K^{({\ast})} \ell^+
\ell^-$. The used data\footnote{Our fit is based on
  \cite{Aubert:2004it, Iwasaki:2005sy}, \cite{Wei:2009zv,
    Aaltonen:2011qs, LHCbnote38, Aaltonen:2011ja}, and
  \cite{Wei:2009zv, Aaltonen:2011qs} for what concerns $B \to X_s
  \ell^+ \ell^-$, $B \to K^\ast \ell^+ \ell^-$, and $B \to K \ell^+
  \ell^-$, respectively.  The latest (but still preliminary) Belle
  measurement of ${\cal B} (B \to X_s \ell^+ \ell^-)$
  \cite{Nakayama:2009zz}, which finds a low-$s$ branching ratio of
  around $2.4 \sigma$ below the SM prediction, is not used, because
  the quoted central values of the branching fractions for electrons
  and muons in the final state differ by a factor of more than 2.  If
  only the LHCb measurements of $B \to K^\ast \ell^+ \ell^-$
  \cite{LHCbnote38} are used, the numbers in \refeq{eq:DeltaC7C7p},
  \refeq{eq:DeltaC9C9p}, and \reftab{tab:bounds2} read $0.29$ and
  $0.20$, $3.1$ and $2.5$, and $1.5$, $1.2$, $0.07$, and $0.13$,
  respectively. While some of these bounds are not as strong as the
  limits based on the combined $B \to K^{(\ast)} \ell^+\ell^-$ data
  \cite{Wei:2009zv, Aaltonen:2011qs, LHCbnote38, Aaltonen:2011ja},
  they clearly show the large impact (and future potential) of LHCb
  measurements on the extraction of low-energy Wilson coefficients. }
includes the branching ratios ${\cal B} (B \to X_s \ell^+ \ell^-)$ and
${\cal B} (B \to K^{(\ast)} \ell^+ \ell^-)$, the forward-backward
asymmetry $A_{\rm FB} (B \to K^\ast \ell^+ \ell^-)$, as well as the
longitudinal $K^\ast$ polarization fraction $F_L (B \to K^\ast \ell^+
\ell^-)$.  In all cases, we consider the low-$s$ region of dilepton
invariant masses $s = [1, 6] \GeV^2$. At high $s$ we consider for
$B\to X_s\ell^+\ell^-$ the region $s > 14.4 \GeV^2$ and for $B\to
K^{(\ast)}\ell^+\ell^-$ the two bins $s = [14.18, 16] \GeV^2$ and $s >
16 \GeV^2$. At 90\% CL our global analysis returns the following upper
bounds
\beq \label{eq:DeltaC9C9p} 
|\Delta C_9^{\rm eff} (s, m_b) | < \, 2.0 \,, \qquad |\Delta
C_9^{\prime \hspace{0.25mm} {\rm eff}} (s, m_b) | \, < \, 1.7 \,,
\eeq 
when the corrections to the effective Wilson coefficients are treated
as additive, $s$-independent contributions and only a single
correction is allowed to float at a time.

\begin{figure}[!t]
\begin{center}
\mbox{\includegraphics[height=1.5in]{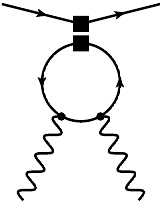}}
\end{center}
\begin{picture}(0,0)(0,0)
\put(214,137.5){\large $b$}
\put(257,137.5){\large $s$}
\put(268,88){\large $\tau$}
\put(203,88){\large $\tau$}
\put(192,35){\large $\gamma$}
\put(278,35){\large $\gamma$}
\end{picture}	
\vspace{-8mm}
\begin{center}
  \parbox{15.5cm}{\caption{\label{fig:mixing:bsgg} 1PI diagram with a
      penguin insertion of a $(\bar s b) (\bar \tau \tau)$ operator
      (black squares) that contribute to the $b \to s \gamma \gamma$
      amplitude. The tau loop in the graph is closed and the diagram
      with interchanged photons is not shown.}}
\end{center}
\end{figure}
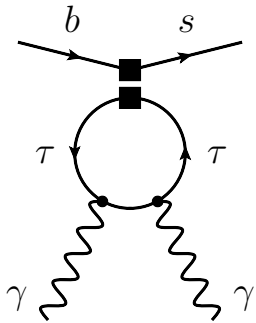

The inequalities \refeq{eq:DeltaC9C9p} can be converted into bounds on
the magnitudes of the high-scale values of $C_{V,AB}$ using
\refeq{eq:DeltaC7C9} and \refeq{eq:C7C9eff}.  Since the expressions
for $\Delta C_9^{\rm eff} (s, m_b)$ and $\Delta C_9^{\prime
  \hspace{0.25mm} {\rm eff}} (s, m_b)$ depend explicitly on $s$
through the function $M_9 (\hat s, m_\tau)$, we extract the limits
directly from our global fit. We find the following exclusion
\beq \label{eq:CVLA}
|C_{V,LA} (\Lambda)| \, = \, \left ( 2.0 -0.2 \, \eta_6^{-1} \right
)^{-1} |\Delta C_9^{\rm eff} (s,m_b)| \, < \,\, 2.0 \left ( 2.0 -0.2
  \, \eta_6^{-1} \right )^{-1} \,.
\eeq
An analog formula holds for $|C_{V,RA} (\Lambda)|$ after replacing
$|\Delta C_9^{\rm eff} (s,m_b)| $ and the factor $2.0$ in front of the
bracket by $|\Delta C_9^{\prime \hspace{0.25mm} \rm eff} (s,m_b)| $
and $1.7$. Assuming a new-physics scale of $\Lambda=1\TeV$, one finds
the numerical values given \reftab{tab:bounds2}.

In contrast to $B \to X_s \gamma$, all $(\bar s b) (\bar \tau \tau)$
operators contribute to the double-radiative $B_{d,s}$-meson decays at
the one-loop level \cite{Gemintern:2004bw, Hiller:2004wc}. In the
following, we will concentrate on the case of $B_s \to \gamma \gamma$,
which turns out to give the strongest constraints on the Wilson
coefficients of the operators under consideration.  At the quark level
the double-radiative $b \to s \gamma \gamma$ decays receive
contributions from one-particle irreducible (1PI) diagrams of the type
shown in \reffig{fig:mixing:bsgg} as well as similar one-particle reducible
(1PR) graphs in which at least one of the two photons is radiated from
an external leg. Since the 1PR diagrams are proportional to the
tree-level matrix elements of $\Op_7^{(\prime)}$, their contributions
are included if the corresponding amplitudes are expressed through the
effective Wilson coefficients $C_7^{(\prime) \rm eff}$. A detailed
discussion of the calculation of the one-loop matrix elements can be
found in \refapp{app:matrixelementsadd}.

Experimentally so far only upper limits on the branching ratio of $B_s
\to \gamma \gamma$ exist. At 90\%~CL the most stringent limit is
\cite{Wicht:2007ni}
\beq \label{eq:BRBsAAexp}
{\cal B} (B_s \to \gamma \gamma)_{\rm exp} < 8.7 \cdot 10^{-6} \,.
\eeq
In order to constrain the Wilson coefficients of the $(\bar s b) (\bar
\tau \tau)$ operators also a value for the branching ratio of $B_s \to
\gamma \gamma$ within the SM is needed. Updating the analysis of
\cite{Bosch:2002bv, Bosch:2002bw}, we obtain
\beq \label{eq:BRBsAASM} 
{\cal B} (B_s \to \gamma \gamma)_{\rm SM} = \left (0.7^{+2.5}_{-0.4}
\right ) \cdot 10^{-6} \,,
\eeq
where the dominant part of the quoted error is due to the hadronic
parameter $\lambda_B$, which parametrizes the first negative moment of
the $B$-meson wave function. Estimates of $\lambda_B$ are very
uncertain, but typically fall in the range between $0.25 \GeV$ and
$0.75 \GeV$ \cite{Grozin:1996pq, Ball:2003fq, Braun:2003wx,
  Lee:2005gza}, and the central value in \refeq{eq:BRBsAASM}
corresponds to $\lambda_B = 0.5 \GeV$.  Numerically subleading errors
in the latter SM prediction arise from the variation of the
renormalization scale and the error on the decay constant $f_{B_s}$.

Assuming the dominance of a single operator, we then find by combining
\refeq{eq:BRBsAAexp} and \refeq{eq:BRBsAASM} the following 90\%~CL
upper bounds on the magnitudes of the Wilson coefficients ($A,B =
L,R$)
\beq \label{eq:BsAAmag}
\begin{split}
  & \hspace{2cm} |\Delta C_7^{ \rm eff} (m_b)| \, < \, 2.2 \,, \quad
  |\Delta C_7^{\prime \hspace{0.25mm} \rm eff} (m_b)| \, < \, 1.9 \,,
  \\[1mm] & |C_{S,AL} (m_b)| \, < \, 3.4 \,, \quad |C_{S,AR} (m_b)| \,
  < \, 2.3 \,, \quad |C_{V,AB} (m_b)| \, < \, 5.9 \,.
\end{split}
\eeq
Clearly, these limits are not competitive with the bounds obtained
earlier from the other tree- and loop-level mediated $B_{d,s}$-meson
decays. The reason for the weakness of these bounds is
threefold. First, as shown in \refapp{app:matrixelementsadd}, the 1PI
contributions of the $(\bar s b) (\bar \tau \tau)$ operators are power
suppressed, second, the experimental bound \refeq{eq:BRBsAAexp} leaves
room for order of magnitude enhancements of ${\cal B} (B_s \to \gamma
\gamma)$ with respect to the SM expectation, and third the theoretical
errors plaguing~\refeq{eq:BRBsAASM} are large.  Notice that the power
suppression of four-fermion operator contributions is a generic
feature of the $B_s \to \gamma \gamma$ decay in the heavy-quark limit
\cite{Bosch:2002bv, Bosch:2002bw}, which is absent in the case of $B
\to K \gamma \gamma$ \cite{Hiller:2004wc}.  The sensitivity of the
latter decay mode to $(\bar s b) (\bar \tau \tau)$ operators is
therefore in principle more pronounced than the one of $B_s \to \gamma
\gamma$.  In practice, however, the possibility to extract
short-distance information from $B \to K \gamma \gamma$ is severely
limited. First, the perturbative part of the $B \to K \gamma \gamma$
spectrum is not accessible behind the large resonance peaks associated
to $B \to K \eta_c \to K \gamma \gamma$ {\it etc.} and, second, not
even an upper limit on ${\cal B} (B \to K \gamma \gamma)$ is currently
available. Hence the bounds \refeq{eq:BsAAmag} represent at present
the most stringent limits on the low-energy Wilson coefficients of
interest that can be derived from double-radiative $B_{d,s}$-meson
decays.

\begin{table}
\begin{center}
\begin{tabular}{|c|c|}
  \hline
  $|C_i (1 \TeV)|$ &
  $b \to s \gamma, s \ell^+ \ell^-, s\gamma \gamma$
  \\
  \hline
  $S,AB$ & --- \\
  $V,LA$ & $1.1$ \\
  $V,RA$ & $1.0$ \\
  $T,L$ & $0.07$ \\
  $T,R$ &$0.08$ \\
  \hline
\end{tabular}

\vspace{4mm}

\parbox{15.5cm}{\caption{\label{tab:bounds2} Indirect upper bounds on
    the high-scale Wilson coefficients at $\Lambda = 1\TeV$.}}
  \end{center}
\end{table}

With a large amount of luminosity collected at a super-flavor factory
it might also be possible to measure the CP asymmetry
\cite{Hewett:2004tv}
\begin{equation} \label{eq:rCP}
  r_{\rm CP} = \frac{|{\cal A} (\bar B_s \to \gamma \gamma)|^2 - 
    |{\cal A} ( B_s \to \gamma \gamma)|^2} {|{\cal A} (\bar B_s \to \gamma
    \gamma)|^2 + |{\cal A} ( B_s \to \gamma \gamma)|^2} \,,
\end{equation}
besides the $B_s \to \gamma \gamma$ branching ratio. The SM
prediction of $r_{\rm CP}$ reads
\begin{equation} \label{eq:rCPSM}
  (r_{\rm CP})_{\rm SM} = (0.5^{+0.6}_{-0.3})\% \,.
\end{equation}
The largest fraction in the quoted error is due to the scale
dependence, followed by the uncertainty coming from $\lambda_B$. The
smallness of \refeq{eq:rCPSM} implies that finding $r_{\rm CP}$ at or
above the $10 \%$ level would constitute an unambiguous signal of the
presence of additional sources of direct CP violation. In fact, the
complex Wilson coefficients of the $(\bar s b) (\bar \tau \tau)$
operators give rise to such an effect and thus allow to change $r_{\rm
  CP}$ dramatically. Employing the 90\%~CL bounds on the low-energy
Wilson coefficients of $|C_{S,AB} (m_b)| < 0.5$, $|C_{V,AB} (m_b)| <
0.8$, $|\Delta C_{7}^{\rm eff} (m_b)| < 0.29$, and $|\Delta
C_{7}^{\prime \hspace{0.25mm} \rm eff} (m_b)| < 0.19$, we find that
the CP asymmetry in $B_s \to \gamma \gamma$ can vary in the ranges
 $(r_{\rm CP})_{S,AL} \in [-95, 95]\%$, $(r_{\rm CP})_{S,AR} \in [-40,
40]\%$, $(r_{\rm CP})_{V,AB} \in [-70, 70]\%$, $(r_{\rm CP})_{T,L} \in
[-20, 20]\%$, and $(r_{\rm CP})_{T,R} \in [-70, 70]\%$, for the
respective Wilson coefficient with an arbitrary phase. In order to
measure $r_{\rm CP}$, SuperB (Belle II) probably needs to collect at
least $75 \, {\rm ab^{-1}}$ ($50 \, {\rm ab^{-1}}$) of data, a goal
that is expected to be achieved in the year 2021. This implies that
the $B_s \to \gamma \gamma$ decay will not be able to provide useful
additional information on the possible size of $(\bar s b) (\bar \tau
\tau)$ contributions in the near future.

By comparing the results displayed in \reftab{tab:bounds1} with those
shown in \reftab{tab:bounds2}, one observes that the bounds on the
high-scale Wilson coefficients $C_i (\Lambda)$ that stem from the
direct constraints, {\it i.e.}, the rare decays $B_s \to \tau^+
\tau^-$, $B \to X_s \tau^+ \tau^-$, and $B^+ \to K^+ \tau^+ \tau^-$,
are currently in 8 out of 10 cases stronger than those that follow
from the indirect constraints, {\it i.e.}, the $b \to s \gamma$, $b
\to s \ell^+ \ell^-$, and $b \to s \gamma \gamma$ transitions. The
only exception are the Wilson coefficients $C_{T,A} (\Lambda)$ of the
tensor operators, which are tightly bound as a result of the one-loop
mixing of $\Op_{T,A}$ into the electromagnetic dipole operators
$\Op_7$ and $\Op_7^\prime$, which give the dominant contribution to
the $B \to X_s \gamma$ rate. On the other hand, sizable effects in
$C_{S,AB} (\Lambda)$ and $C_{V,AB} (\Lambda)$, that can almost compete
with the leading current-current SM interactions in strength, are
allowed, because rare $B_s$ and $B_d$ decays involving a tau pair in
the final state are so poorly constrained.  While the direct
constraints are unlikely to change notable with almost all BaBar and
Belle data analyzed, the indirect bounds on the vector operators are
expected to improve with LHCb precision measurements of $B \to
K^{(\ast)} \ell^+ \ell^-$ under way.

%
\section{Effects of $\bm{(\bar s b) (\bar \tau \tau)}$ Operators in
  $\bm{ \Gamma_{12}^s}$}
\label{sec:numerics}

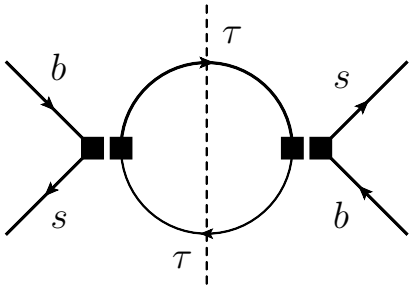
\begin{figure}[!t]
\begin{center}
\mbox{\includegraphics[height=1.5in]{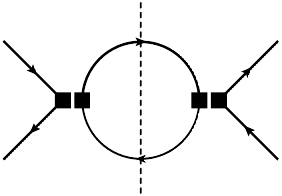}}
\end{center}
\begin{picture}(0,0)(0,0)
\put(180,106){\large $b$}
\put(180,50){\large $s$}
\put(287,50){\large $b$}
\put(287,103){\large $s$}
\put(245,120){\large $\tau$}
\put(226,35){\large $\tau$}
\end{picture}
\vspace{-8mm}
\begin{center}
  \parbox{15.5cm}{\caption{\label{fig:gamma12} Diagram with a double
      insertion of a $(\bar s b) (\bar \tau \tau)$ operator (black
      squares) that contributes to $\Gamma_{12}^s$ at LO.  The tau
      loop in the diagram is closed and the cut (dashed line)
      indicates that only the imaginary part of the graph furnishes a
      correction to the off-diagonal element of the decay-width
      matrix.}}
\end{center}
\end{figure}

The off-diagonal element of the decay-width matrix is related via the
optical theorem to the absorptive part of the forward-scattering
amplitude which converts a $\bar B_s$ into a $B_s$ meson. Working to
LO in the strong coupling constant and $\Lambda_{\rm QCD}/m_b$, the
contributions from the complete set of operators \refeq{eq:Qsbtautau}
to $\Gamma_{12}^s$ is found by computing the matrix elements between
quark states depicted in \reffig{fig:gamma12}. Details on the actual
calculation are given in \refapp{app:amplitude}. Considering again
only the self-interference of a single $(\bar s b) (\bar \tau \tau)$
operator at a time, we obtain in terms of the hadronic matrix elements
$\langle \Op_i \rangle = \langle B_s | \Op_i |\bar B_s \rangle$ of the
following $\Delta B = 2$ operators
\beq \label{eq:DBOp}
\begin{split}
  \Op_S^A & = \left ( \bar s \, P_A \, b \right) \left ( \bar s \, P_A
    \, b \right) \,, \\
  \tilde \Op_S^A & = \left ( \bar s_\alpha \, P_A \, b_\beta \right)
  \left ( \bar s_\beta \, P_A \, b_\alpha \right) \,, \\
  \Op_V^A & = \left ( \bar s \, \gamma^\mu P_A \, b \right) \left (
    \bar s \, \gamma_\mu P_A \, b \right) \,,
\end{split}
\eeq
the results
\beq \label{eq:Gamma12s}
\begin{split}
  \left (\Gamma_{12}^s \right )_{S,LA} & = {\cal N}_{\Gamma_{12}^s} \,
  3 \hspace{0.25mm} \hspace{0.25mm} x_\tau \hspace{0.25mm} \beta_\tau
  \left \langle
    \Op_S^L \right \rangle (C_{S,LA})^2 \,, \\
  \left (\Gamma_{12}^s \right )_{V,LA} & = {\cal N}_{\Gamma_{12}^s} \,
  \beta_\tau\, \Big [ (1 - x_\tau) \left \langle \Op_V^L \right
  \rangle + (1 + 2 x_\tau) \left \langle \Op^R_S \right \rangle \Big ]
  \, (C_{V,LA})^2 \,, \\ 
  \left (\Gamma_{12}^s \right )_{T,L} & = -{\cal N}_{\Gamma_{12}^s} \,
  12 \hspace{0.25mm} \hspace{0.25mm} x_\tau \hspace{0.25mm} \beta_\tau
  \, \Big [ \hspace{0.25mm} 4 \left \langle  \Op_S^L \right \rangle + 8  \hspace{0.75mm}
  \big \langle \tilde \Op_S^L \big \rangle \hspace{0.5mm} \Big ] \,
  (C_{T,L})^2 \,.
\end{split}
\eeq
Here $\alpha, \beta$ are color indices and we have introduced
$\beta_\tau = \sqrt{1-4x_\tau}$ as well as
\beq \label{eq:NGamma}
{\cal N}_{\Gamma_{12}^s} = -\frac{G_F^2 \hspace{0.25mm} m_b^2}{6
  \hspace{0.25mm} \pi M_{B_s}} \, (V_{ts}^\ast V_{tb})^2 \,.
\eeq
The corresponding expressions for $\left (\Gamma_{12}^s \right
)_{S,RA} $, $\left (\Gamma_{12}^s \right )_{V,RA}$, and $\left
  (\Gamma_{12}^s \right )_{T,R}$ are obtained from \refeq{eq:Gamma12s}
by exchanging the labels $L$ and $R$ everywhere.  Following common
practice, we express the matrix elements in terms of hadronic
parameters $B_i$. In our analysis, we employ
\beq \label{eq:DBOpME}
\left \langle \Op^A_S \right \rangle = -\frac{5}{12} \, f_{B_s}^2
M_{B_s}^2 B_S \,, \qquad \big \langle \tilde \Op^{A}_S \big \rangle =
\frac{1}{12} \, f_{B_s}^2 M_{B_s}^2 \tilde B_S \,, \qquad \left
  \langle \Op^A_V \right \rangle = \frac{2}{3} \, f_{B_s}^2 M_{B_s}^2
B_V \,,
\eeq
with $f_{B_s} = 231 \MeV$ \cite{Lenz:2010gu}, $B_S \approx 1.3$,
$\tilde B_S \approx 1.4$ \cite{Becirevic:2001xt}, $B_V \approx 0.84$
\cite{Lenz:2010gu}, and the meson states normalized as $\langle B_s |
B_s \rangle = \langle \bar B_s | \bar B_s \rangle = 2 M_{B_s}$. Notice
that the latter value of $f_{B_s}$ has been used to obtain the SM
predictions $(\Delta \Gamma_s)_{\rm SM}$ and $(a_{fs}^s)_{\rm SM}$
presented in \refeq{eq:phiDGSM} and \refeq{eq:afsSM}, respectively.

\begin{figure}[!t]
\begin{center}
\mbox{\includegraphics[height=3in]{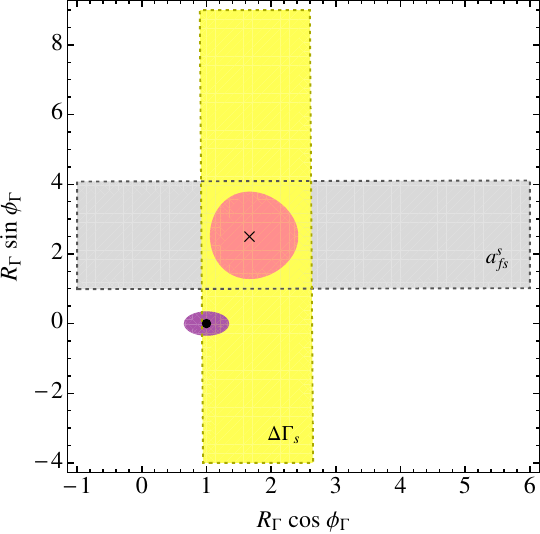}}
\qquad
\mbox{\includegraphics[height=3in]{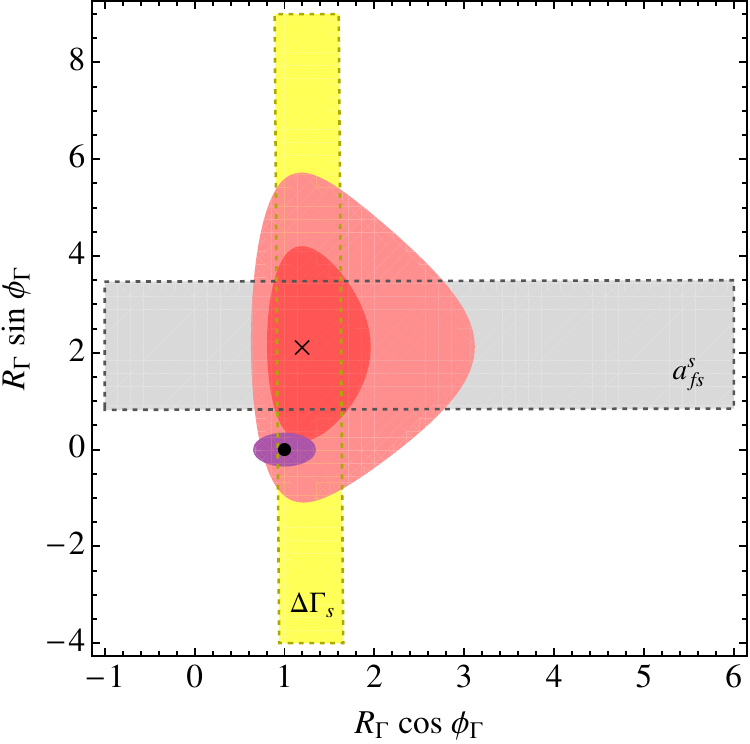}}

\vspace{4mm}
\mbox{\includegraphics[height=3in]{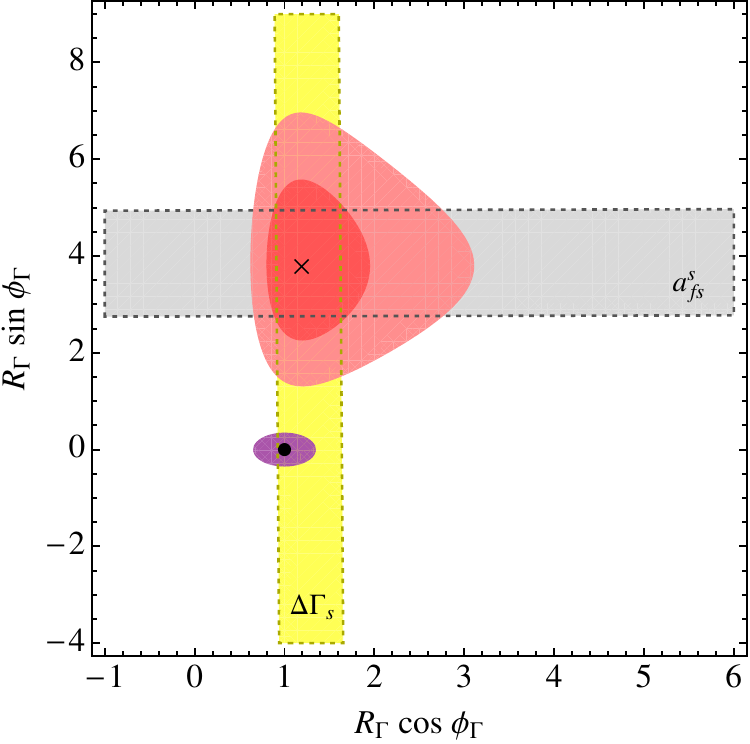}}
\end{center}
\vspace{-8mm}
\begin{center}
  \parbox{15.5cm}{\caption{\label{fig:sbtautau}
      Upper left (upper right, lower):
      Constraints from $\Delta \Gamma_s$ and $a_{fs}^s$ on the
      parameter $R_\Gamma$ and $\phi_\Gamma$ in scenario S2 employing
      the data set D1 (D2, D3). The yellow and gray bands represent
      $68\%$~CL regions (${\rm dofs} = 1$), while the purple ellipses
      illustrate the parameter space that is accessible for $|C_{V,AB}
      (m_b)| < 0.8$ and $C_{S,AB} (m_b) = C_{T,A} (m_b) = 0$. For
      comparison, the 68\% (95\%) probability region of the combined
      fit to the $B_s$-meson mixing data (${\rm dofs} = 2$) is shown
      in red (light red). In all panels
      the SM prediction (best-fit solution) is indicated by a dot 
      (cross).}}
\end{center}
\end{figure}

The direct and indirect constraints discussed in
Sections~\ref{sec:direct} and \ref{sec:indirect} restrict the
magnitude of the low-energy Wilson coefficients but not their relative
weak phase.  We can thus determine only the maximal\footnote{Of
  course, there is also a model-independent lower limit on
  $(R_\Gamma)_i$. It corresponds to full destructive interference and
  is obtained from \refeq{eq:RGammai} and \refeq{eq:RGammaSVT} by
  replacing the plus with minus signs.} allowed value of the parameter
$R_\Gamma$ introduced in \refeq{eq:M12G12paraNP} in a
model-independent fashion. Allowing the new-physics contribution of
one operator at a time to be non-zero, the corresponding inequality is
given in terms of the off-diagonal elements of the decay matrix
\refeq{eq:Gamma12s} by
\beq \label{eq:RGammai}
(R_\Gamma)_i < 1 + \frac{2 \hspace{0.5mm} |(\Gamma_{12}^s)_i|}{(\Delta
  \Gamma_s)_{\rm SM}} \, \cos \hspace{0.5mm} ( \phi_{J/\psi
  \phi}^s)_{\rm SM} \approx 1+ \frac{2 \hspace{0.5mm}
  |(\Gamma_{12}^s)_i|}{(\Delta \Gamma_s)_{\rm SM}}\,.
\eeq
Numerically, we find for the individual contributions
\beq \label{eq:RGammaSVT}
\begin{split} 
  (R_\Gamma)_{S,AB} & < 1 + \left ( 0.4 \pm 0.1 \right )
  |C_{S,AB} (m_b)|^2 \,, \\[2mm]
  (R_\Gamma)_{V,AB} & < 1+ \left ( 0.4 \pm 0.1 \right )
  |C_{V,AB} (m_b)|^2 \,,\\[2mm] 
  (R_\Gamma)_{T,A} &
 < 1+ \left ( 3.6 \pm 0.9 \right ) |C_{T,A}
  (m_b)|^2 \,,
\end{split}
\eeq
where the quoted uncertainties are due to the error on $(\Delta
\Gamma_s)_{\rm SM}$ as given in \refeq{eq:phiDGSM}. Employing now the
90\% CL bounds on the low-energy Wilson coefficients derived in the
previous two sections, \ie, $|C_{S,AB} (m_b)| < 0.5$, $|C_{V,AB}
(m_b)| < 0.8$, $|C_{T,L} (m_b)| < 0.06$, and $|C_{T,R} (m_b)| < 0.09$,
it follows that
\beq \label{eq:RGammaSVTbounds} 
(R_\Gamma)_{S,AB} < 1.15 \,,  \qquad (R_\Gamma)_{V,AB} < 1.35 \,, \qquad
(R_\Gamma)_{T,L} < 1.02 \,, \qquad (R_\Gamma)_{T,R} < 1.04 \,.
\eeq
These numbers imply that $(\bar s b) (\bar \tau \tau)$ operators of
scalar (vector) type can lead to enhancements of $|\Gamma_{12}^s|$
over its SM value by 15\% (35\%) without violating any existing
constraint. In contrast, contributions from tensor operators can alter
$|\Gamma_{12}^s|$ by at most $4\%$, because larger corrections would
be at variance with $B \to X_s \gamma$.

In \reffig{fig:sbtautau} we compare the parameter space in the
$R_\Gamma \hspace{0.25mm} \cos \phi_\Gamma$--$R_\Gamma \hspace{0.25mm}
\sin \phi_\Gamma$ plane that can be populated in scenario S2 with
$|C_{V,AB} (m_b)| < 0.8$ and $C_{S,AB} (m_b) = C_{T,A} (m_b) = 0$
(purple ellipses) to the constraints imposed by the measurement of
$\Delta \Gamma_s$ and $a_{fs}^s$ (yellow and gray bands). The upper
left and right panel correspond to the data set D1 and D2, respectively,
while the lower panel illustrates the situation in the case of the D3 data set.
We see that the contribution of a single  vector operator with flavor
structure $(\bar s b) (\bar \tau \tau)$ can lead to an improvement of the
fit. For example, in the case of the new data set D2 an agreement with
the data at 68\%~CL (red area) can be achieved. It is however also
evident that the best-fit solution (cross) of both data sets cannot be
accommodated.  The allowed parameter space is further 
reduced for scalar and tensor operators, so that we do not explicitly display 
these cases in the figures.  This demonstrates that absorptive new physics in 
$\Gamma_{12}^s$ in form of $(\bar s b)(\bar \tau \tau)$ operators cannot provide 
a satisfactory explanation of the anomaly in the dimuon charge asymmetry data 
observed by the D\O \ collaboration.

%
\section{New-Physics Models}
\label{sec:newphysics}

So far our discussion has been completely general since we
parametrized the new-physics effects in terms of effective couplings
of higher-dimensional operators. We now will consider explicit
scenarios of new physics that can give rise to $(\bar s b) (\bar \tau
\tau)$ operators with large Wilson coefficients.  Discussions similar
to ours have been presented previously in \cite{Dighe:2010nj,
  Alok:2010ij, Kim:2010gx, Dighe:2007gt}, where it was found that new-physics 
  models containing leptoquarks or $Z^\prime$ boson may explain the observed 
  tensions in the $B_s$-meson data. In the following we will show that this is not
   the case. Our work hence clarifies and extends these existing studies.

We first consider scenarios with leptoquarks (LQ), {\ie},
color-triplet or -antitriplet particles carrying both lepton ($L$) and
baryon ($B$) number.  Such new degrees of freedom are encountered in
various extensions of the SM, such as grand unified theories,
technicolor, and composite models (see \cite{Buchmuller:1986zs,
  Leurer:1993em, Davidson:1993qk, Hewett:1997ce} for topical reviews).
At low energies, the interactions between the SM particles and the LQs
can be captured by writing down the most general Lagrangian compatible
with renormalizability, $L$ and $B$ conservation, and invariance under
the SM gauge group. These general requirements allow for scalar and
vector LQs interacting directly with the SM fermions as well as for LQ
couplings to the Higgs boson.

In the following, we will elaborate in detail only on the case of
$SU(2)$ singlet scalar LQs. From our discussion it should however
become clear how the given results have to be adapted to cover the
other cases.  The relevant interaction terms involving a charged
lepton, a down-type quark, and a $SU(2)$ singlet scalar LQ are given
by
\beq \label{eq:LLQ}
{\cal L}_{\rm LQ} \, \supset \,( \lambda_{R \tilde S_0} )_{ij}
\hspace{0.5mm} \bar d^c_j \hspace{0.25mm} P_R \hspace{0.5mm} e_i
\hspace{0.5mm} \tilde S_0+ {\rm h.c.} \,.
\eeq
Here $\lambda_{R \tilde S_0}$ is a complex $3 \times 3$ matrix in the
lepton and quark flavor spaces, $d_i$ and $e_i$ are $SU(2)$ SM
singlets, and the subscript of $\tilde S_0$ indicates the $SU(2)$
transformation property of the LQ.  After integrating out the LQ and
performing a Fierz rearrangement, the interactions \refeq{eq:LLQ} give
rise to a $\Delta B = 1$ $(\bar s b) (\bar \tau \tau)$ operator of
vector type. Explicitly, one finds
\beq \label{eq:LeffLQ}
{\cal L}_{\rm eff} \, \supset \, -\frac{( \lambda^\ast_{R \tilde S_0}
  )_{32} (\lambda_{R \tilde S_0})_{33}}{2 M_{\tilde S_0}^2} \,
\Op_{V,RR} \,,
\eeq
where $M_{\tilde S_0}$ denotes the mass of $\tilde S_0$. The
coefficient multiplying the operator $\Op_{V,RR}$ is bound by the
requirement $|C_{V,AB} (\mu)| < 0.8$, which holds for any scale
$\mu$. Numerically, we find
\beq \label{eq:LQbound}
\frac{| (\lambda^\ast_{R \tilde S_0} )_{32} (\lambda_{R \tilde
    S_0})_{33}|}{M_{\tilde S_0}^2} \, < \, 2.1 \, {\rm TeV}^{-2} \,.
\eeq
The same bound also applies to the possible $SU(2)$ doublet scalar LQs
(commonly denoted by $S_2$ and $\tilde S_{2}$ in the literature),
which generate $\Op_{V,LR}$ operators, after an appropriate
replacement of couplings and masses.

\begin{figure}[!t]
\begin{center}
\mbox{\includegraphics[width=1.9in]{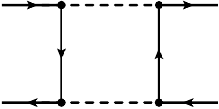}}
\begin{picture}(0,0)(0,0)
\put(-124,76){\large $b$}
\put(-124,-15){\large $s$}
\put(-26,-15){\large $b$}
\put(-26,76){\large $s$}
\put(-120,30){\large $\tau$}
\put(-32,30){\large $\tau$}
\put(-78,78){\large $\tilde S_0$}
\put(-78,-20){\large $\tilde S_0$}
\end{picture} \qquad \quad
\mbox{\raisebox{2.25mm}{\includegraphics[width=1.9in]{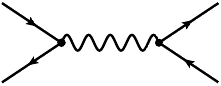}}}
\begin{picture}(0,0)(0,0)
\put(-124,56){\large $b$}
\put(-124,4){\large $s$}
\put(-29,4){\large $b$}
\put(-29,56){\large $s$}
\put(-78,47.5){\large $Z^\prime$}
\end{picture}
\end{center}
\vspace{2mm}
\begin{center}
  \parbox{15.5cm}{\caption{\label{fig:M12} Left: One-loop contribution
      to $M_{12}^s$ arising from a box diagram involving tau leptons
      and $SU(2)$ singlet scalar LQs. Right: Tree-level contribution
      to $M_{12}^s$ associated to the exchange of a $Z^\prime$
      boson.}}
\end{center}
\end{figure}
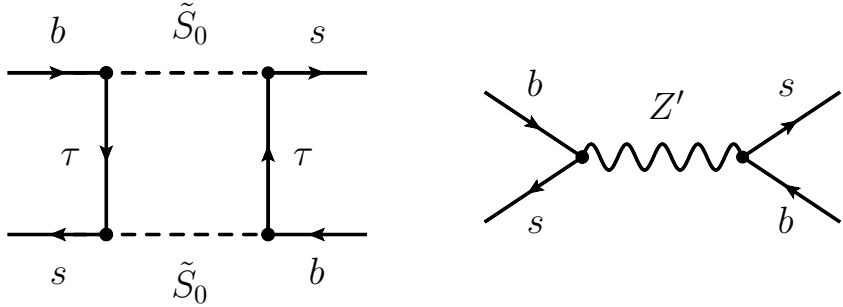

Importantly, the interactions \refeq{eq:LLQ} give rise to corrections
to both $\Gamma_{12}^s$ and $M_{12}^s$.  The contribution to the
former quantity arise from a double insertion of
\refeq{eq:LeffLQ}. Applying the general formulas \refeq{eq:Gamma12s}
to the specific case of a scalar LQ, we obtain
\beq \label{eq:G12LQ}
(\Gamma_{12}^s)_{\rm LQ} = -\frac{1}{ 288 \pi} \, \frac{\big (
  (\lambda^\ast_{R \tilde S_0} )_{32} (\lambda_{R \tilde S_0})_{33}
  \big )^2}{M_{\tilde S_0}^4} \, \beta_\tau \left [ (1-x_\tau ) B_V -
  \frac{5}{8} \hspace{0.5mm} (1+ 2x_\tau ) B_S \right ] m_b^2
\hspace{0.25mm} f_{B_s}^2 M_{B_s} \,.
\eeq
This formula disagrees with the analytic expression first given in
\cite{Dighe:2007gt}, which does not contain terms proportional to
$B_S$ at all.  Numerically, the difference amounts to a factor of
around $-0.4$, with our result being smaller in magnitude.  The
correction to $M_{12}^s$ originates from the box diagram displayed on
the left-hand side of \reffig{fig:M12}. In agreement with
\cite{Dighe:2007gt}, we find for this contribution\footnote{We
  integrate out the top quark and the LQ simultaneously and
  consequently neglect small RG effects in the six-flavor theory.}
\beq \label{eq:M12LQ}
(M_{12}^s)_{\rm LQ} = \frac{1}{384 \pi^2} \, \frac{\big (
  (\lambda^\ast_{R \tilde S_0} )_{32} (\lambda_{R \tilde S_0})_{33}
  \big )^2}{M_{\tilde S_0}^2} \, \hat \eta_{B_s} I ( a_\tau )
f_{B_s}^2 M_{B_s} B_V \,,
\eeq
where $\hat \eta_{B_s} = \eta_5^{6/23} \approx 0.8$ encodes the QCD
corrections due to RG running and the Inami-Lim function $I(a)$ reads
\beq \label{eq:SSa}
I (a) = \frac{1+a}{(1-a)^2}+\frac{2 a}{(1-a)^3} \, \ln a \,.
\eeq
Notice that the above function approaches 1 for $a \to 0$, which is
the relevant limit, since $a_\tau = m_\tau^2/M_{\tilde S_0}^2 \ll 1$
for LQ masses at or above the electroweak scale.

Combining now \refeq{eq:G12LQ} and \refeq{eq:M12LQ}, one obtains
numerically
\beq \label{eq:M12oG12LQ}
r_{\rm LQ} = \frac{(M_{12}^s)_{\rm LQ}}{(\Gamma_{12}^s)_{\rm LQ}} =
2084 \left ( \frac{M_{\tilde S_0}}{250 \GeV} \right )^2 = 2084 \,
z_{\tilde S_0} \,,
\eeq
which is real and positive. In contrast, $r_{\rm SM} = (M_{12}^s)_{\rm
  SM}/(\Gamma_{12}^s)_{\rm SM} \approx -200 \hspace{0.25mm} \exp \left
  (i \phi_{\rm SM}^s \right ) \approx -200$. Notice also that
\refeq{eq:M12oG12LQ} scales as $M_{\tilde S_0}^2$ which reflects the
fact that $(M_{12}^s)_{\rm LQ}$ arises from a dimension-six correction
(single operator insertion), whereas $(\Gamma_{12}^s)_{\rm LQ}$ stems
from a dimension-eight contribution (operator double insertion).

In fact, given that $\Delta M_s$ (or equivalently $R_M$) is bounded
experimentally, the sign and magnitude of $r_{\rm LQ}$ completely
determine the ranges of possible $R_\Gamma$, $\Delta\Gamma_s$, and
$a_{fs}^s$ values in the LQ scenario \refeq{eq:LLQ}. In
\refapp{app:bounds} we show that this is a feature of all new-physics
scenarios with real $(M_{12}^s)_{\rm NP}/(\Gamma_{12}^s)_{\rm NP}$.
Applying the general formula \refeq{eq:RGminmax1} to the case at hand,
we find at 90\% CL
\beq \label{eq:RGminmaxLQ}
0.96 = 1 - \frac{0.03}{z_{\tilde S_0}} < (R_\Gamma)_{\rm LQ} < 1 +
\frac{0.2}{z_{\tilde S_0}} = 1.31 \,.
\eeq
Here the bound on $R_M$ given in \refeq{eq:D1CL68} and
\refeq{eq:D2CL68} has been used and the final numerical values have
been obtained for $M_{\tilde S_0} = 210 \GeV$, which represents the
current 95\%~CL lower bound on third-generation scalar LQs decaying to
a $b \tau$ final state~\cite{Abazov:2008jp}. Comparing
\refeq{eq:RGminmaxLQ} to the relevant model-independent upper bound,
\ie, the inequality in \refeq{eq:RGammaSVTbounds} involving
$(R_\Gamma)_{V,AB}$, we see that at present both limits are
(accidentally) almost the same. Since the bound on $M_{\tilde S_0}$
will improve with the LHC collecting more and more luminosity as we
speak, the model-dependent upper limit will, however, soon become
stronger than the model-independent one.

\begin{figure}[!t]
\begin{center}
\mbox{\includegraphics[height=3in]{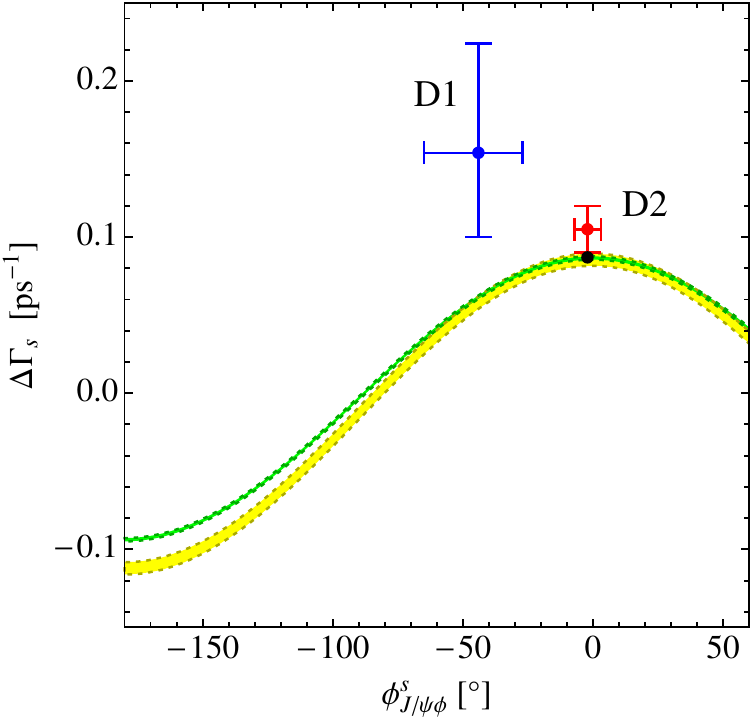}} \quad 
\mbox{\includegraphics[height=3in]{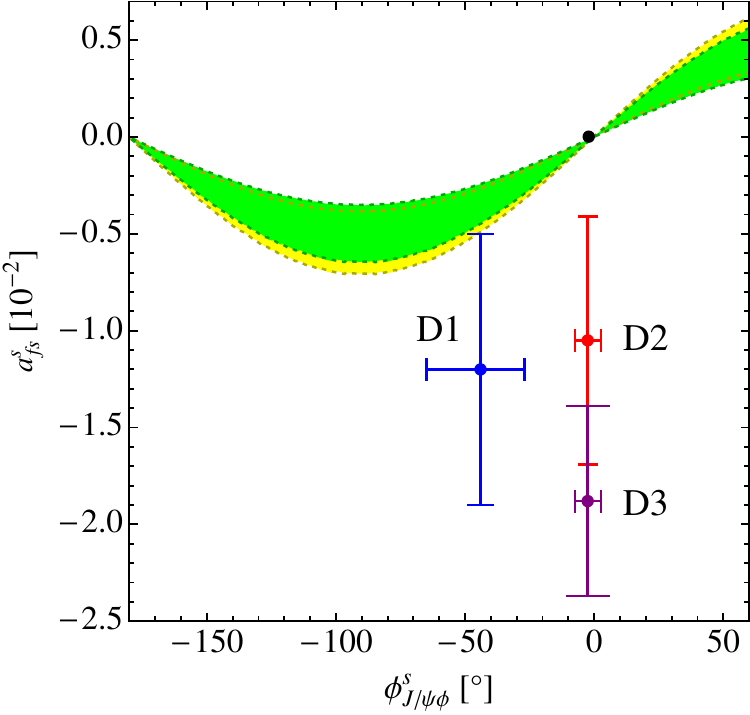}} 
\end{center}
\vspace{-6mm}
\begin{center}
  \parbox{15.5cm}{\caption{\label{fig:LQ} Predictions for $\Delta
      \Gamma_s$ (left) and $a_{fs}^s$ (right) as a function of
      $\phi_{J/\psi \phi}^s$ in the presence of a singlet scalar LQ
      with $M_{\tilde S_0} = 210 \GeV$ (yellow areas) and $M_{\tilde
        S_0} = 400 \GeV$ (green areas). Since the left (right) panel
      is symmetric (antisymmetric) under $\phi_{J/\psi \phi}^s
      \leftrightarrow -\phi_{J/\psi \phi}^s$ only positive values of
      $\phi_{J/\psi \phi}^s$ up to $50^\circ$ are shown.  In both
      panels, the blue (red) error bar corresponds to the data set D1
      (D2), while the SM point is indicated as a black point. The
      purple error bar in the right panel indicates the D3 data set.
      See text for details.}}
\end{center}
\end{figure}

Similarly, one finds from the general formulas \refeq{eq:DGminmax1}
and \refeq{eq:afssminmax1}, the following two double inequalities
\begin{gather}
  -0.11 \ {\rm ps}^{-1} = -\left (1 + \frac{0.2}{z_{\tilde S_0}}
  \right )(\Delta \Gamma_s)_{\rm SM} < (\Delta\Gamma_s)_{\rm LQ} <
  \left (1 +\frac{0.02}{z_{\tilde S_0}} \right ) (\Delta
  \Gamma_s)_{\rm SM} = 0.09 \, {\rm ps}^{-1} \,, \;\;\;\; \nonumber
  \\[-3mm]  \label{eq:DGafssminmaxLQ} \\[-3mm]
  -0.7 \cdot 10^{-2} = - 331 \left (1 + \frac{0.1}{z_{\tilde S_0}}
  \right ) (a_{fs}^s)_{\rm SM} < (a_{fs}^s)_{\rm LQ} < 331 \left (1
    +\frac{0.1}{z_{\tilde S_0}} \right ) (a_{fs}^s)_{\rm SM} = 0.7
  \cdot 10^{-2} \,. \;\;\;\; \nonumber
\end{gather}
These numbers imply that light singlet scalar LQs can change the
predictions for $\Delta \Gamma_s$ and $a_{fs}^s$ with respect to the
SM expectations \refeq{eq:phiDGSM} and \refeq{eq:afsSM} by a factor of
more than $-1.3$ and $\pm 375$, respectively. In contrast, a strong
enhancement of $\Delta \Gamma_s$ is not possible.\footnote{In 
  \cite{Dighe:2011du} significant violations of  $\Delta \Gamma_s \leq 
  (\Delta \Gamma_s)_{\rm SM}$ were reported. We are unable to reproduce 
  this finding.} The maximal effects do, however, not occur simultaneously, 
  since the observables in question are strongly correlated, as can be seen from
\reffig{fig:LQ}. In both panels the parameter space that is accessible
at 90\%~CL, assuming a singlet scalar LQ mass of $210 \GeV$ ($400
\GeV$), is indicated in yellow (green).  We see that $\Delta \Gamma_s$
becomes maximal (minimal) for $\phi_{J/\psi \phi}^s = 0^\circ$
($\phi_{J/\psi \phi}^s = -180^\circ$) and that the LQ predictions form
a band with a cosine-like shape (left panel). The predictions for
$a_{fs}^s$, on the other hand, are minimal for $\phi_{J/\psi \phi}^s =
0^\circ, -180^\circ$ and maximal for $\phi_{J/\psi \phi}^s =
-90^\circ$, and the borders of the accessible parameter space in the
$\phi_{J/\psi \phi}^s\hspace{0.25mm}$--$\hspace{0.5mm} a_{fs}^s$ plane
are sine curves (right panel). Notice that the cosine- and sine-like
behavior of the predictions is an immediate consequence of
\refeq{eq:observablesNP}. By comparing the accessible parameter space
to the experimentally preferred regions, we see from the yellow
colored regions that at 68\%~CL none of the considered data sets, \ie,
D1 (blue error bars), D2 (red error bars), and D3 (purple error bar),
can be accommodated by a light singlet scalar LQ.  Increasing the LQ mass,
reduces the allowed parameter space, as can be seen from the green areas,
further limiting the possible improvement of the tension in the $B_s$-meson
sector.\footnote{While this work was being completed, the preprint
  \cite{Dorsner:2011ai} appeared, which performs a dedicated global
  fit of scalar LQ couplings, utilizing an assortment of tree- and
  loop-level observables in the charged lepton and quark sector. This
  article finds that the tension in the $B_s$-meson system cannot be
  cured by LQs, if one wants to explain the anomaly of the anomalous
  magnetic moment of the muon, $(g-2)_\mu$, and simultaneously fulfill
  the bounds on lepton-flavor violating processes. Notice, however,
  that if the former requirement is dropped, no bound on $|
  (\lambda^\ast_{R \tilde S_0} )_{32} (\lambda_{R \tilde
    S_0})_{33}|/M_{\tilde S_0}^2$ can be derived from the rare decays
  considered in the latter paper.}

The above general considerations are readily applied to the case of
vector-like LQs. Considering for example LQ interactions of the form
${\cal L}_{\rm LQ} \supset ( \lambda_{R V_0} )_{ij} \hspace{0.5mm}
\bar d_j \hspace{0.25mm} \gamma_\mu P_R \hspace{0.5mm} e_i
\hspace{0.5mm} V_0^\mu+ {\rm h.c.}$ will lead to an effective
Lagrangian ${\cal L}_{\rm eff} \supset -( \lambda^\ast_{R V_0} )_{32}
(\lambda_{R V_0})_{33}/M_{V_0}^2 \, \Op_{V,RR}$ with $M_{V_0}$
denoting the mass of the vector-like LQ. It follows that the
model-independent bound on such a LQ is twice as strong as the limit
derived in \refeq{eq:LQbound}. Furthermore, the ratio between
$(M_{12}^s)_{\rm LQ}$ and $(\Gamma_{12}^s)_{\rm LQ}$ is found to be
equal to the result given in \refeq{eq:M12oG12LQ}. This tells us that
(for $M_{V_0} = M_{\tilde S_0}$) the possible effects in $\Delta
\Gamma_s$ and $a_{fs}^s$ due to $V_0$ are identical to those
originating from $\tilde S_0$. Vector-like LQs thus fail to reproduce
the data sets D1 and D2 at 68\%~CL as well.  Notice finally, that the
situation is utterly hopeless, if one considers LQ couplings to the
Higgs boson, since after a Fierz rearrangement the corresponding
effective scalar interactions with flavor structure 
$(\bar{s} \tau)(\bar{\tau}b)$ contain tensor operators, which are most
tightly constrained by the model-independent bounds
\refeq{eq:RGammaSVTbounds}.
 
We now switch gear and consider SM extensions where the gauge group
contains an additional $U(1)^\prime$ factor and the resulting
$Z^\prime$ boson possess family non-universal, flavor-changing
couplings \cite{Langacker:2000ju}.  Such new degrees of freedom arise
in many well-motivated new-physics models, such as grand unified
theories and/or string constructions (see \cite{Nakamura:2010zzi,
  Langacker:2008yv} for comprehensive reviews).  In the physical
basis, the flavor-changing neutral currents generically appear at tree
level in both the left- and right-handed sectors.  The interaction
Lagrangian relevant for the further discussion can be written as
\beq \label{eq:LZprime}
{\cal L}_{Z^\prime} \, \supset \, \frac{g}{\cos \theta_W} \, \Big [
\left ( \kappa_{sb}^L \, \bar s \, \gamma^\mu P_L \, b + \kappa_{sb}^R
  \, \bar s \, \gamma^\mu P_R\, b + {\rm h.c.} \right ) +
\kappa_{\tau\tau}^L \, \bar \tau \, \gamma^\mu P_L \, \tau +
\kappa_{\tau\tau}^R \, \bar \tau \, \gamma^\mu P_R\, \tau
\hspace{0.25mm} \Big] \, Z_\mu^\prime \,,
\eeq
where $g$ is the $SU(2)$ coupling of the SM and $\cos \theta_W$
denotes the cosine of the weak mixing angle. The $Z^\prime$-boson
coupling constant does not appear explicitly in \refeq{eq:LZprime},
because it has been absorbed into the $\kappa_{ij}^{L,R}$ factors.
Notice that the flavor off-diagonal couplings are in general complex,
while the diagonal ones have to be real due to the hermiticity of the
Lagrangian.  Integrating out the $Z^\prime$ boson leads to the
following $\Delta B = 1$ four-fermion interactions
\beq \label{eq:LZprimeeff}
{\cal L}_{\rm eff} \, \supset \, -\frac{8 \hspace{0.125mm}
  G_F}{\sqrt{2}} \, \frac{M_Z^2}{M_{Z^\prime}^2} \, \kappa_{sb}^{A}
\hspace{0.5mm} \kappa_{\tau \tau}^B \; \Op_{V,AB} \,.
\eeq 
Here $M_{Z^\prime}$ denotes the mass of the extra $U(1)^\prime$ gauge
boson. From the numbers collected in \reftab{tab:bounds1}, it follows
that
\beq \label{eq:LZprimebound}
\frac{| \kappa_{sb}^{A} \hspace{0.25mm} \hspace{0.25mm} \kappa_{\tau
    \tau}^B| }{M_{Z^\prime}^2} < 1.9 \, {\rm TeV}^{-2} \,.
\eeq

\begin{figure}[!t]
\begin{center}
\mbox{\includegraphics[height=3in]{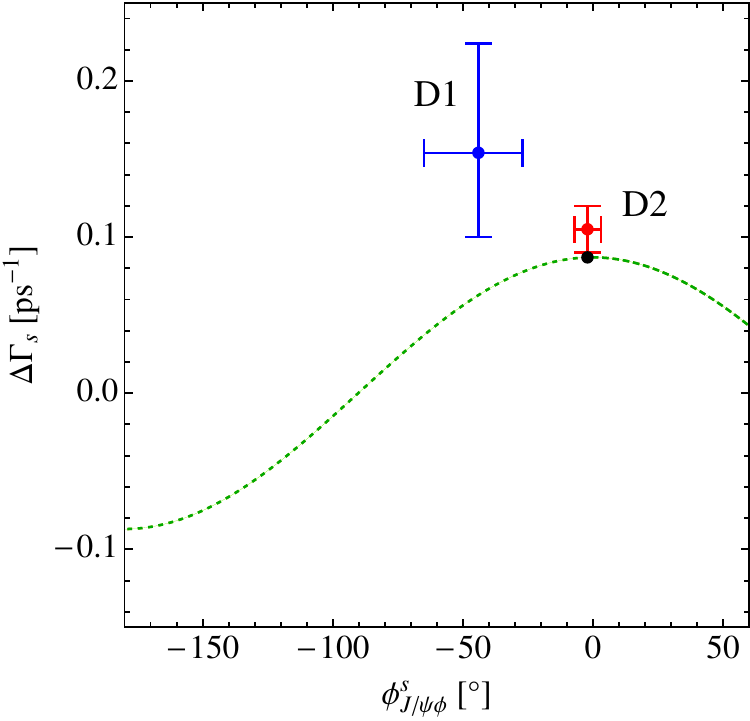}} \quad 
\mbox{\includegraphics[height=3in]{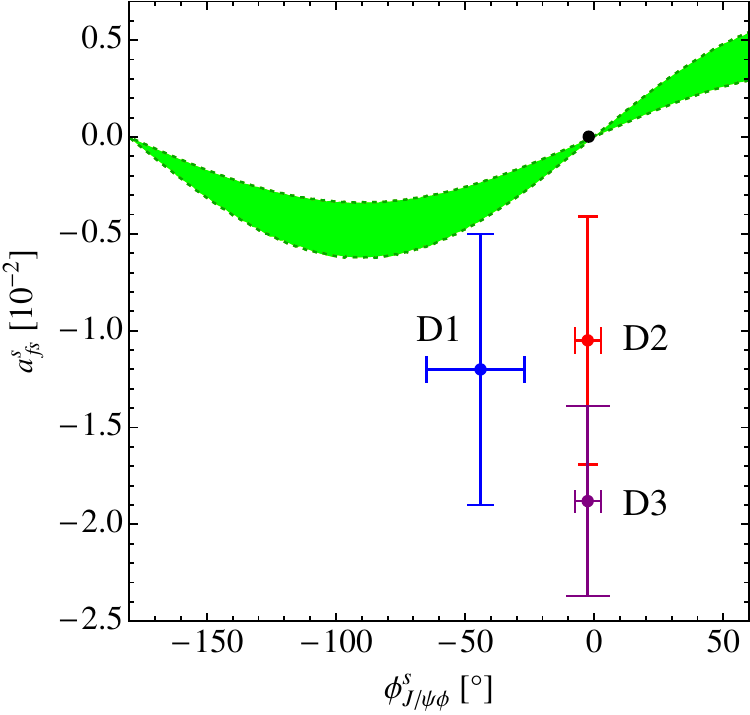}} 
\end{center}
\vspace{-6mm}
\begin{center}
  \parbox{15.5cm}{\caption{\label{fig:Zprime} Predictions for $\Delta
      \Gamma_s$ (left) and $a_{fs}^s$ (right) as a function of
      $\phi_{J/\psi \phi}^s$ in the presence of a $Z^\prime$ with
      $M_{Z^\prime} = 400 \GeV$ (yellow areas) and $M_{Z^\prime} =
      1000 \GeV$ (green areas), which both overlap completely. Since
      the left (right) panel is symmetric (antisymmetric) under 
      $\phi_{J/\psi \phi}^s \leftrightarrow -\phi_{J/\psi \phi}^s$ 
      only positive values of $\phi_{J/\psi \phi}^s$ up to $50^\circ$ 
      are shown. In both panels, the blue (red) error bar corresponds
      to the data set D1 (D2), while the SM point is indicated as a black
      point. The purple error bar in the right panel indicates the
      D3 data set. See text for details.}}
\end{center}
\end{figure}

After integrating out the $Z^\prime$ boson, the interactions \refeq{eq:LZprime} 
give rise to one-loop corrections to $\Gamma_{12}^s$ as well as tree-level
corrections to $M_{12}^s$. In the following,
we restrict ourselves to the case of purely left-handed
interactions. The inclusion of right-handed currents is however
straightforward.  For $\Gamma_{12}^s$, we find in accordance with
\cite{Alok:2010ij}
\beq \label{eq:G12Zprime}
(\Gamma_{12}^s)_{Z^\prime} = -\frac{4 \hspace{0.25mm} G_F^2}{9 \pi}
\left ( \frac{M_Z^2}{M_{Z^\prime}^2} \right )^2 \big ( \kappa_{sb}^{L}
\hspace{0.25mm} \kappa_{\tau \tau}^L \big )^2 \, \beta_\tau \left [
  (1-x_\tau ) B_V - \frac{5}{8} \hspace{0.5mm} (1+ 2x_\tau ) B_S
\right ] m_b^2 \hspace{0.25mm} f_{B_s}^2 M_{B_s} \,.
\eeq
The element $M_{12}^s$ receives contributions from the graph shown on
the right-hand side of \reffig{fig:M12}. A simple tree-level
calculation leads to \cite{He:2006bk}
\beq \label{eq:M12Zprime}
(M_{12}^s)_{Z^\prime} = \frac{4 \hspace{0.125mm} G_F}{3\sqrt{2}} \,
\frac{M_Z^2}{M_{Z^\prime}^2} \, \big ( \kappa_{sb}^{L} \big )^2 \,
\hat \eta_{B_s} f_{B_s}^2 M_{B_s} B_V \,.
\eeq

The ratio of \refeq{eq:M12Zprime} and \refeq{eq:G12Zprime} is given in
semi-numerical form by
\beq \label{eq:rZprime}
r_{Z^\prime} = 6.0 \cdot 10^{5} \; \, \left ( \frac{M_{Z^\prime}}{250
    \GeV} \, \frac{1}{\kappa_{\tau \tau}^L} \right )^2 = 6.0 \cdot
10^{5} \; y_{Z^\prime} \,.
\eeq
It is real and positive. Notice that the numerical coefficient in
\refeq{eq:rZprime} is enhanced relative to \refeq{eq:M12oG12LQ} by a
loop factor and thus very large.  From the discussion in
\refapp{app:bounds} it hence follows that the effects of a $Z^\prime$
boson in $R_\Gamma$, $\Delta \Gamma_s$, and $a_{fs}^s$ are severely
constrained. Explicitly, we find
\begin{gather}
  1 - \frac{1.0 \cdot 10^{-4}}{y_{Z^\prime}} < (R_\Gamma)_{Z^\prime} <
  1 + \frac{7.7 \cdot
    10^{-4}}{y_{Z^\prime}}  \,, \nonumber \\
  -\left (1 + \frac{7.7 \cdot 10^{-4}}{y_{Z^\prime}} \right )(\Delta
  \Gamma_s)_{\rm SM} < (\Delta\Gamma_s)_{Z^\prime} < \left (1
    +\frac{7.0 \cdot 10^{-5}}{y_{Z^\prime}}
  \right ) (\Delta \Gamma_s)_{\rm SM} \,, \;\;\;\;  
  \label{eq:DGafssminmaxZprime} \\[1mm]
  - 331 \left (1 + \frac{3.3 \cdot 10^{-4}}{y_{Z^\prime}} \right )
  (a_{fs}^s)_{\rm SM} < (a_{fs}^s)_{Z^\prime} < 331 \left (1
    +\frac{3.3 \cdot 10^{-4}}{y_{Z^\prime}} \right ) (a_{fs}^s)_{\rm
    SM} \,. \;\;\;\; \nonumber
\end{gather}

The $Z^\prime$-boson mass $M_{Z^\prime}$ is constrained by direct
searches for resonant production of tau lepton pairs.  At present the
best bound is provided by the CDF measurement \cite{Acosta:2005ij},
which rules out a $Z^\prime$ with SM-like couplings to $q \bar q$ and
$\tau^+ \tau^-$ and a mass below $399 \GeV$ at 95\%~CL. Although this
bound is model-dependent,\footnote{The mass $M_{Z^\prime}$ and the
  couplings $\kappa_{\tau \tau}^{L,R}$ are constrained by the LEP
  measurement of the $Z$-boson couplings to taus
  \cite{LEPEWWG:2005ema}.  Applying the results of the one-loop $Z \to
  f \bar f$ form factors given in \cite{Haisch:2011up}, we find that
  even for couplings close to the non-perturbative limit the resulting
  bound is however with $M_{Z^\prime} \gtrsim 25 \GeV$ rather weak.
  We will not entertain the possibility of a very light
  $Z^\prime$-boson in the following. } we will use it as a guideline
and consider a relatively light $Z^\prime$-boson of $M_{Z^\prime} =
400 \GeV$ as well as a heavier one with $M_{Z^\prime} = 1000
\GeV$. The $Z^\prime$-boson predictions in the $\phi_{J/\psi
  \phi}^s\hspace{0.25mm}$--$\hspace{0.5mm} \Delta \Gamma_s$ $\big
(\phi_{J/\psi \phi}^s\hspace{0.25mm}$--$\hspace{0.5mm} a_{fs}^s \big)$
plane are shown in the left (right) panel of \reffig{fig:Zprime}. The
parameter space of 90\% probability is depicted in yellow (green) for
$M_{Z^\prime} = 400 \GeV$ ($M_{Z^\prime} = 1000 \GeV$), which both 
overlap completely.  While the
shapes of the colored areas resemble the ones of \reffig{fig:LQ}, one
observes that compared to the case of LQs the allowed parameter space
is even more reduced for a $U(1)^\prime$ gauge boson. In particular, a
$Z^\prime$ boson with mass at or above the electroweak scale and
purely left-handed perturbative couplings clearly fails to describe
the $B_s$-meson data within 68\% CL. Because the parameter
$r_{Z^\prime}$ as defined in \refeq{eq:rZprime} is of the order
$G_F^{-1} \hspace{0.25mm} m_b^{-2} \hspace{0.25mm}
M_{Z^\prime}^2/M_Z^2$ and therefore generically large, the latter
statement is also true if the $Z^\prime$ boson couples only
right-handed or both left- and right-handed.

%
\section{Summary}
\label{sec:summary}

Motivated by the observation that the anomalously large dimuon charge
asymmetry measured by the D\O \ collaboration, can be fully explained
only if new physics contributes to the absorptive part of the
$B_s$--$\bar B_s$ mixing amplitude, we have presented a
model-independent study of the contributions to $\Gamma_{12}^s$
arising from the complete set of dimension-six operators with flavor
content $(\bar s b) (\bar \tau \tau)$.  Taking into account the direct
bounds from $B_s \to \tau^+ \tau^-$, $B \to X_s \tau^+ \tau^-$, and
$B^+ \to K^+ \tau^+ \tau^-$ as well as the indirect constraints from
$b \to s \gamma$, $b \to s \ell^+ \ell^-$ ($\ell = e, \mu$), and $b
\to s \gamma \gamma$, we have demonstrated that only the Wilson
coefficients of the tensor operators are severely constrained by data,
while those of the scalar and vector operators can be sizable and
almost reach the size of the Wilson coefficient of the leading
current-current operator in the standard model. The model-independent
90\%~CL limits on the magnitudes of the Wilson coefficients are
summarized in \reftab{tab:bounds3}. Given these loose bounds, it
follows that the presence of a single $(\bar s b) (\bar \tau \tau)$
operator can lead to an enhancement of $\Gamma_{12}^s$ of at most 35\%
compared to its standard model value. Since a resolution of 
the tension in the $B_{s}$-meson sector would require the effects to be 
of the order of 300\% (or larger), the allowed shifts are by far too small to 
provide a satisfactory explanation of the issue. We emphasize that after 
minor modifications, our general results can be applied to other 
dimension-six operators involving quarks and leptons. For example, 
as a result of the 90\%~CL limit ${\cal B} (B^+ \to K^+ \tau^\pm \mu^\mp) 
< 7.7 \cdot 10^{-5}$~\cite{Aubert:2007rn}, the direct bounds on the Wilson
coefficients of the set of $(\bar s b) (\bar \tau \mu)$ operators turn
out to be roughly a factor of $7$ stronger than those in the $(\bar s
b) (\bar \tau \tau)$ case.  Possible effects of $(\bar s b) (\bar \tau
\mu)$ operators are therefore generically too small to lead to a
notable improvement of the tension present in the current $B_{d,s}$-meson data. 
 Similarly, a contribution from $(\bar d b) (\bar \tau \tau)$ operators to 
$\Gamma_{12}^d$ large enough to explain the $A_{\rm SL}^b$ data is 
excluded by  the 90\% CL bound ${\cal B} (B \to \tau^+ \tau^-) < 4.1 \cdot 
10^{-3}$~\cite{Aubert:2005qw}. Naively, also $(\bar b d) (\bar c c)$ operators are 
heavily constrained (meaning that their Wilson coefficients should be 
smaller than those of the QCD/electroweak penguins in the standard model) by 
the plethora of exclusive $B$ decays. A dedicated analysis of the latter class of 
contributions is however not available in the literature.

\begin{table}
\begin{center}
\begin{tabular}{|c|c|c|c|}
  \hline
  Operator & Bound on $\Lambda$ ($C_i^\Lambda = 1$) & 
  Bound on $C_i^\Lambda$ ($\Lambda = 1 \TeV$)  & Observable \\
  \hline
  $(\bar s \, P_A \, b) (\bar \tau \, P_B \, \tau)$ & $2.0 \TeV$ & 
 $4.2 \cdot 10^{-1}$ & $B_s \to  \tau^+ \tau^-$ \\
  $(\bar s \, \gamma^\mu P_A \, b) (\bar \tau \, \gamma_\mu P_B \, 
  \tau)$ & $1.0 \TeV$ & 
  $1.1$ & $B^+ \to K^+ \tau^+ \tau^-$ \\
  $(\bar s \, \sigma^{\mu \nu} P_L \, b) (\bar \tau \, \sigma_{\mu \nu} 
  P_L \, \tau)$ & 
  $3.2 \TeV$ & $9.2 \cdot 10^{-2}$ & $B \to X_s \gamma$ \\
  $(\bar s \, \sigma^{\mu \nu} P_R \, b) (\bar \tau \, \sigma_{\mu \nu} 
  P_R \, \tau)$ & $2.8 \TeV$ & $1.1 \cdot 10^{-1}$ & $B \to X_s \gamma$ \\
  \hline
\end{tabular}

\vspace{4mm}

  \parbox{15.5cm}{
    \caption{\label{tab:bounds3} Model-independent limits on the
      Wilson coefficients of all possible four-fermion operators with
      flavor structure $(\bar s b) (\bar \tau \tau)$. The second
      column shows the bound on the suppression scale $\Lambda$
      assuming an effective coupling strength of $C_i^\Lambda = 1$,
      while the third column gives the value for $C_i^\Lambda$ for a
      new-physics scale of $\Lambda = 1\TeV$. Notice that only the
      bounds on the tensor operators depend on the specific chirality
      $A,B = L,R$.}}
  \end{center}
\end{table}

Our model-independent study of non-standard effects in $\Gamma_{12}^s$
is complemented by an analysis of $B_s$--$\bar B_s$ mixing in two
explicit standard model extensions. We consider in detail the
corrections to $M_{12}^s$ and $\Gamma_{12}^s$ due to a $SU(2)$ singlet
scalar leptoquark and a left-handed $Z^\prime$ boson with a mass at or
above the electroweak scale. In both cases, we find that depletions of
order one in $\Delta \Gamma_s$ and changes by more than two orders of
magnitude in $a_{fs}^s$ can occur, while a notable enhancement of the
former quantity with respect to its standard model value is not
possible. The large effects do, however, not occur simultaneously,
since the non-standard contributions to $M_{12}^s$ and $\Gamma_{12}^s$
are strongly correlated in the considered new-physics scenarios. While
$\Delta \Gamma_s$ becomes maximal (minimal) for $\phi_{J/\psi \phi}^s
= 0^\circ$ ($\phi_{J/\psi \phi}^s = -180^\circ$) and the predictions
form a band with a cosine-like shape in the $\phi_{J/\psi
  \phi}^s\hspace{0.25mm}$--$\hspace{0.5mm} \Delta \Gamma_s$ plane,
$a_{fs}^s$ is minimal for $\phi_{J/\psi \phi}^s = 0^\circ, -180^\circ$
and maximal for $\phi_{J/\psi \phi}^s = -90^\circ$, and the accessible
parameter space in the $\phi_{J/\psi
  \phi}^s\hspace{0.25mm}$--$\hspace{0.5mm} a_{fs}^s$ plane is bounded
by sine curves. In turn, neither a $SU(2)$ singlet scalar leptoquark
nor a left-handed $Z^\prime$ boson renders a significant better description of the
current data than the standard model.  The same conclusion also applies to vector-like
leptoquarks and $Z^\prime$ bosons with right-handed or both left- and
right-handed couplings.  In fact, we have shown that the pattern of
deviations found in the case of leptoquarks and $Z^\prime$ bosons is a
feature of all new-physics model with real $(M_{12}^s)_{\rm
  NP}/(\Gamma_{12}^s)_{\rm NP}$ and that in this class of models the
measurement of $\Delta M_s$ generically puts stringent constraints on
both $\Delta \Gamma_s$ and $a_{fs}^s$. As it turns out, these bounds
are weakest if the ratio $r_{\rm NP} = (M_{12}^s)_{\rm NP}/
(\Gamma_{12}^s)_{\rm NP}$ is positive and as small as possible. Since
on dimensional grounds $r_{\rm NP}$ scales as the square of the
new-physics scale, this general observation implies that standard
model extensions that aim at a good description of the Tevatron
data should have new degrees of freedom below the electroweak scale
and/or be equipped with a mechanism
that renders the contribution to $M_{12}^s$ small. Furthermore, models
in which $(M_{12}^s)_{\rm NP}$ is generated beyond Born level seem
more promising, since in such a case $r_{\rm NP}$ is suppressed by a
loop factor with respect to the case where $(M_{12}^s)_{\rm NP}$
arises already at tree level.

The above discussion implies that  a full explanation of the observed discrepancies is not even
possible for the most general case $(M_{12}^s)_{\rm NP} \neq 0$ and
$(\Gamma_{12}^s)_{\rm NP} \neq 0$. Numerically, we find 
that the addition of a single $(\bar s b) (\bar \tau \tau)$ vector
operator giving $(R_\Gamma)_{V,AB} = 1.35$ on top of dispersive new physics with $(M_{12}^s)_{\rm NP} \neq
0$, can only improve the quality of the fit to the latest set of
measurements (\ie, \refeq{eq:DMsexp}, \refeq{eq:phiDGnew}, and
\refeq{eq:afssnew}) to $\chi^2 = 1.2$ compared to $\chi^2 = 3.2$
within the standard model. This might indicate that the high central value of
$A_{\rm SL}^b$ observed by the D\O \ collaboration is (partly) due to
a statistical fluctuation.
Future improvements in the measurement of
the CP phase $\phi_{J/\psi \phi}^s$ and, in particular, a first
determination of the difference $a_{fs}^s-a_{fs}^d$ between the $B_s$
and $B_d$ semileptonic asymmetries by LHCb, are soon expected to shed
light on this issue. Will these measurements continue to make a case
for new physics in $\Gamma_{12}^s$? Time will tell!

%
\subsubsection*{Acknowledgments}

We are grateful to Yossi~Nir for his critical questions that allowed 
us to spot mistakes in~(\ref{eq:Gamma12s}), (\ref{eq:RGammaSVT}), 
(\ref{eq:RGammaSVTbounds}), and (\ref{eq:fierzid}).
A big ``thank you'' to Sacha Davidson, Florian Goertz, Alex Kagan,
Jernej Kamenik, Alex~Lenz, and Quim Matias for useful discussions
concerning leptoquarks, $B_s$--$\bar B_s$ mixing, charm counting, and
radiative and semileptonic $b \to s$ decays. We furthermore thank
Gudrun Hiller for reminding us of the role of $(\bar s b) (\bar \tau
\tau)$ operators in $b \to s \gamma \gamma$ transitions. We are also
grateful to Kevin Flood and Malachi Schram for helpful informations
about experimental aspects of $B^+ \to K^+ \tau^+ \tau^-$. Stimulating
discussions with Christian Bauer and Bogdan Dobrescu in June 2010 at
the Aspen Workshop ``Forefront QCD and LHC Discoveries'', which
triggered the present research, are acknowledged. UH thanks finally
the Aspen Center for Physics for hospitality and the NSF (Grant
No. 1066293) for partial support.

%
\begin{appendix}

\section{Operator Mixing}
\label{app:mixing}

\renewcommand{\theequation}{A\arabic{equation}}
\setcounter{equation}{0}

In the following we present the analytic expressions that are needed
to evolve the Wilson coefficients from the high-energy scale $\Lambda$
down to the low-energy scale $m_b$. In the basis $(\Op_{S,AL},
\Op_{V,AL}, \Op_{T,A}, \Op_{7,A}, \Op_{9,A})$ with $A = L, R$ the LO
ADM, which describes the mixing of these operators, reads
\renewcommand{\arraystretch}{2}
\setlength{\arraycolsep}{5pt}
\beq \label{eq:gammastau}
 \gamma_{s} = \frac{\alpha_s}{4\pi} \, \left ( \begin{array}{ccccc}
      -8 & 0 & 0 & 0 & 0 \\[0mm] 
      0 & 0 & 0 & 0 & \displaystyle \frac{4}{3}  \\[0mm] 
      0 & 0 & \displaystyle \frac{8}{3} & 8  & 0 \\[0mm] 
      0 & 0& 0 & \displaystyle \frac{8}{3} -2 \beta_s & 0 \\[0mm] 
      0 & 0 & 0 & 0 & -2 \beta_s
    \end{array} \right )
  \,, 
\eeq
where $\beta_s = 11 - 2/3 \hspace{0.25mm} N_f$ and $N_f$ denotes the
number active quark flavors. Notice that the self-mixing of
$\Op_{S,AL}$ is, up to an overall sign, equal to the one of the quark
mass. This has to be the case, because a composite operator built out
of a scalar current and a quark mass is conserved and thus has zero
anomalous dimension. Furthermore, the self-mixings of $\Op_{T,A}$ and
$\Op_{7,A}$ are the same, up to the factor $-2\beta_s$ (which is due
to the normalization of $\Op_{7,A}$), since the tau mass does not run
in QCD. Of course, the result \refeq{eq:gammastau}, which after a
suitable replacement of color and charge factors agrees with
\cite{Borzumati:1999qt, Hiller:2003js}, also holds for the operators
$(\Op_{S,AR}, \Op_{V,AR}, \Op_{T,A}, \Op_{7,A}, \Op_{9,A})$ given that
gluon interactions conserve chirality.

Solving the RGEs, we find that the Wilson coefficients at the scale
$\mu_1 < \mu_0$ are given in terms of their initial conditions
evaluated at $\mu_0$ by
\beq \label{eq:RGEsolution}
\begin{split}
  C_{S,AL} (\mu_1) & = \eta^{-4/\beta_s} \, C_{S,AL} (\mu_0) \,,
  \\[1.5mm]
  C_{V,AL} (\mu_1) & = C_{V,AL} (\mu_0) \,, \\[1.5mm]
  C_{T,A} (\mu_1) & = \eta^{4/(3\,\beta_s)} \, C_{T,A} (\mu_0) \,,
  \\
  C_{7,A} (\mu_1) & = \eta^{4/(3\,\beta_s) - 1}\, C_{7,A} (\mu_0) +
  \frac{4}{\beta_s} \left( 1 - \eta^{-1} \right) \eta^{4/(3\,\beta_s)}
  \, C_{T,A} (\mu_0) \,, \\
  C_{9,A} (\mu_1) & = \eta^{-1} \, C_{9,A} (\mu_0) +
  \frac{2}{3\hspace{0.25mm} \beta_s} \left( 1 - \eta^{-1} \right) \,
  \big (C_{V,AL} (\mu_0) + C_{V,AR} (\mu_0) \big ) \,,
\end{split}
\eeq
where $\eta = \alpha_s (\mu_0)/\alpha_s (\mu_1)$.  These results hold
if one does not cross a quark threshold in between $\mu_0$ and
$\mu_1$. Generalizing them to include threshold effects is however
straightforward, since the evolution factorizes and finite matching
corrections are not present at LO.

%
\section{Kinematic Functions}
\label{app:kinfun}

\renewcommand{\theequation}{B\arabic{equation}}
\setcounter{equation}{0}

In this appendix, we collect the analytic results for the functions
$M_i^{X_s} (s)$ and $M_i^{K^+} (s)$, which appear in the formulas
\refeq{eq:dBrXsdq2} and \refeq{eq:dBrKdq2} for the differential
branching ratios of the $B \to X_s \tau^+ \tau^-$ and $B \to K^+
\tau^+ \tau^-$ decays.
 
In the case of the inclusive $b \to s \tau^+ \tau^-$ transition, the
relevant kinematic functions are given by the following expressions
\cite{Fukae:1998qy}
\bea \label{eq:MXsanalytic}
\begin{split}
  M_{S}^{X_s} (s) &= 2 s \left(1 - \frac{2 \hspace{0.25mm}
      m_\tau^2}{s}\right)
  \left(m_b^2+m_s^2-s\right) g_{X_s} (s) \,, \\[1mm]
  M_{V}^{X_s} (s) &= 2
  \left[\left(m_b^2-m_s^2\right)^2-s^2-\frac{g^2_{X_s}(s)}{3}\right]
  g_{X_s} (s) \,, \\[1mm]
  M_{T}^{X_s} (s) &= \frac{32}{3} \left(1 + \frac{2 \hspace{0.25mm}
      m_\tau^2}{s} \right) \left[ \hspace{0.25mm} 2
    \left(m_b^2-m_s^2\right)^2- s \left(m_b^2+m_s^2+s\right) \right]
  \, g_{X_s}(s) \,,
\end{split}
\eea
where 
\beq \label{eq:gXs} 
g_{X_s}(s) = \beta \, \sqrt{\lambda (s, m_b, m_s)} \,, 
\eeq 
and 
\beq \label{eq:betalambda}
\beta = \sqrt{1 -\frac{4 \hspace{0.125mm} m_\tau^2}{s}} \,, \qquad
\lambda (a, b, c) = \left (a - (b + c)^2 \right ) \left (a - (b - c)^2
\right ) \,.
\eeq

In the case of the exclusive decay, the relevant expressions take
instead the form \cite{Bobeth:2007dw}
\beq \label{eq:MiKpanalytic} 
\begin{split}
  M_S^{K^+}(s) & = \frac{3}{16} \left ( \frac{M_{B^+}^2 -
      M_{K^+}^2}{m_b - m_s} \right )^2 s \left ( 1- \frac{2
      \hspace{0.25mm} m_\tau^2}{s} \right )
  f_0^2 (s)\; g_{K^+} (s) \,, \\[1mm]
  M_V^{K^+}(s) & = \frac{1}{8 \hspace{0.25mm} s} \, \bigg[
  \hspace{0.25mm} 3 \hspace{0.25mm} m_\tau^2 \! \hspace{0.25mm} \left
    (M_{B^+}^2 - M_{K^+}^2 \right )^2 f_0^2 (s) +
  \lambda(s,M_{B^+},M_{K^+}) \, (s - m_\tau^2 ) \,
  f_+^2 (s) \bigg] \, g_{K^+} (s) \,,  \\[1mm]
  M_T^{K^+}(s) & = \frac{s}{\left ( M_{B^+} + M_{K^+} \right )^2}
  \left (1 + \frac{2 \hspace{0.25mm} m_\tau^2}{s} \right )
  \lambda(s,M_{B^+},M_{K^+}) \, f_T^2 (s) \, g_{K^+} (s) \,,
\end{split}
\eeq
with 
\beq \label{eq:gKp} 
g_{K^+}(s) = \beta \, \sqrt{\lambda (s, M_{B^+}, M_{K^+})} \,,
\eeq 
and $M_{K^+} = 493.677 \MeV$ \cite{Nakamura:2010zzi}.  The
$s$-dependence of the $B^+ \to K^+$ form factors $f_{i}(s)$ entering
\refeq{eq:MiKpanalytic} is modeled using the results obtained in
\cite{Khodjamirian:2010vf}. Explicitly, we employ ($\bar s = s/{\rm
  GeV}^2$)
\beq \label{eq:BPformfactors}
\begin{split}
  f_+ (s) &= \left (-5.6314 + 6.9718 \, \sqrt{33.3258 - \bar s} +
    0.6206\, \bar s \right )
  d_1(\bar s) \, d_2(\bar s) \,, \\[0.5mm]
  f_T (s) &= \left (-7.7322 + 8.2175 \, \sqrt{33.3258 - \bar s} +
    0.7643\, \bar s \right )
  d_1(\bar s)\, d_2(\bar s) \,, \\[0.5mm]
  f_0 (s) &= \left (-30.040 + 11.200 \, \sqrt{33.3258 - \bar s} +
    1.6269\, \bar s \right ) d_2(\bar s) \,,
\end{split}
\eeq
with $d_1(\bar s) = (1 - \bar s/M^2_{B_s^\ast(1^-)})^{-1}$, $d_2(\bar
s) = \left ( 4.3173 + \sqrt{33.3258 - \bar s} \right )^{-2}$, and
$M_{B_s^\ast(1^-)} = 5.412 \GeV$.

\section{Matrix Elements for $\bm {b \to s \ell^+ \ell^-}$}
\label{app:matrixelements}

%
\renewcommand{\theequation}{C\arabic{equation}}
\setcounter{equation}{0}

This appendix contains the analytic results for the $b \to s \ell^+
\ell^-$ matrix elements of the complete set of $(\bar s b) (\bar \tau
\tau)$ operators
\beq \label{eq:Qiapp}
Q_i = (\bar s \, \Gamma^i_{\bar sb} \, b) (\bar \tau \, \Gamma^i_{\bar
  \tau \tau}\hspace{0.25mm} \tau) \,,
\eeq
with $\Gamma_{\bar sb}^i \otimes \Gamma_{\bar \tau\tau}^i = \{P_A
\otimes P_B, \gamma^\mu P_A \otimes \gamma_\mu P_B, \sigma^{\mu \nu}
P_A \otimes \sigma_{\mu \nu} P_A \}$.  The diagram with the closed tau
loop shown on the right-hand side in \reffig{fig:mixing} yields the
following amplitude for the operator insertion
\beq \label{eq:amplitude}
\begin{split}
  {\cal A} = i \hspace{0.5mm} {\cal N} \hspace{0.25mm} Q_\tau \,
  (\bar{s} \, \Gamma_{\bar sb}^i \, b) ( \bar{\ell} \, \gamma^\alpha
  \, \ell ) & \; \Bigg\{ \frac{m_\tau}{q^2}\left [ \Delta + \ln x_\mu
    -\frac{1}{4} \hspace{0.25mm} M_7(\hat s, x_\tau) \right ] {\rm Tr}
  \left [\gamma_\alpha \hspace{0.25mm} q\!\!\!/ \hspace{0.25mm}
    \Gamma_{\bar \tau \tau}^i \right ] \\ & \, + \left[ \, \frac{1}{3}
    \, (\Delta + \ln x_\mu ) - \frac{1}{2} \hspace{0.25mm} M_9 (\hat
    s, x_\tau) \right] {\rm Tr} \left [\gamma_\alpha \Gamma_{\bar \tau
      \tau}^i \right ] \Bigg\} \; C_i \,.
\end{split}  
\eeq
Here $C_i$ is the relevant Wilson coefficient and 
\beq \label{eq:definitions}
{\cal N} = \frac{4 \hspace{0.25mm} G_F}{\sqrt{2}} \, \frac{\alpha}{4
  \pi} \, V_{ts}^\ast V_{tb} \,, \qquad \Delta = \frac{1}{\varepsilon}
- \gamma_E + \ln\left (4\pi \right ) \,, \qquad x_\mu =
\frac{\mu^2}{m_b^2}\,,
\eeq
and $\gamma_E \approx 0.577216$ denotes the Euler-Mascheroni
constant. The functions $M_7 (\hat s, x_\tau)$ and $M_9 (\hat s,
x_\tau)$ entering \refeq{eq:amplitude} are given by
\beq \label{eq:M7}
\begin{split}
  M_7 (\hat{s}, x_\tau) & =
  -8 + 4 \hspace{0.25mm} \ln x_\tau + 4 \hspace{0.5mm} g(y) \,,\\
  M_7(0,x_\tau) & = 4 \hspace{0.25mm} \ln x_\tau \,,
\end{split}  
\eeq
and 
\beq \label{eq:M9}
\begin{split}
  M_9(\hat{s}, x_\tau) & = - \frac{2}{9} \, (5 + 3 y) + \frac{2}{3}
  \hspace{0.25mm} \ln x_\tau +\frac{1}{3}
  \, (2 + y ) \hspace{0.25mm} g(y) \,, \\
  M_9(\hat s,0) & = \frac{2}{3} \, (\ln\hat{s} - i\pi) - \frac{10}{9}
  \,,
\end{split}  
\eeq
where $y = 4\hspace{0.25mm} m_\tau^2/s$ and 
\beq \label{eq:gy}
g(y) = \sqrt{|1 - y|} \, 
\begin{cases} \, \left[\displaystyle \ln\left(\frac{1 + \sqrt{1 -
          y}}{1 - \sqrt{1 - y}} \right) - i\pi \right], &
  y<1\,,\\[4mm] \, 2 \tan^{-1} \left (\displaystyle \frac{1}{\sqrt{y -
        1}} \right ) , & y>1 \,.
  \end{cases} 
\eeq

The terms ${\rm Tr} \left [\gamma_\alpha \hspace{0.25mm} q\!\!\!/
  \hspace{0.25mm} \Gamma_{\bar \tau \tau}^i \right ] $ and ${\rm Tr}
\left [\gamma_\alpha \Gamma_{\bar \tau \tau}^i \right ] $ in
\refeq{eq:amplitude} give rise to the contributions proportional to
the electromagnetic dipole operators $\Op_{7,A}$ and the vector-like
semileptonic operators $\Op_{9,A}$, respectively. The relevant
non-zero Dirac traces are
\beq \label{eq:diractraces}
{\rm Tr} \left [ \gamma_\alpha \hspace{0.25mm} q \!\!\!  /
  \hspace{0.25mm} P_A \right ] = 2 \hspace{0.25mm} q_\alpha \,, \quad
{\rm Tr} \left [ \gamma_\alpha \hspace{0.25mm} \gamma_\mu P_A \right ]
= 2 \hspace{0.24mm} g_{\alpha \mu} \,, \quad {\rm Tr} \left [
  \gamma_\alpha \hspace{0.25mm} q \!\!\!  / \hspace{0.25mm}
  \sigma_{\mu \nu} P_A \right ] = 4 \hspace{0.25mm} i \left ( q_\mu
  \hspace{0.25mm} g_{\alpha \nu} - q_\nu \hspace{0.25mm} g_{\alpha
    \mu} \right) \,.
\eeq
Finally, the tree-level $b \to s \ell^+ \ell^-$ matrix elements
involving the operators \refeq{eq:penguins} read
\beq \label{eq:treeME}
\begin{split}
  \langle \Op_{7,A} \rangle_{\rm tree} & = i \hspace{0.5mm} {\cal N}
  \, 2 \, \frac{m_\tau}{q^2} (\bar{s}\, \gamma^\mu q\!\!\!/ P_A \, b )
  (\bar{\ell} \, \gamma_\mu \, \ell ) \,, \\
  \langle \Op_{9,A} \rangle_{\rm tree} & = i \hspace{0.5mm} {\cal N}
  \, (\bar{s} \, \gamma^\mu P_A \, b) (\bar{\ell} \, \gamma_\mu \ell)
  \,.
\end{split}
\eeq
By combining \refeq{eq:amplitude}, \refeq{eq:diractraces}, and
\refeq{eq:treeME} it is a matter of simple algebra to derive both the
expressions given in \refeq{eq:C7C9eff} as well as the relevant 
entries of the LO ADM in \refeq{eq:gammastau}.

\section{Matrix Elements for $\bm {b \to s \gamma \gamma}$}
\label{app:matrixelementsadd}

%
\renewcommand{\theequation}{D\arabic{equation}}
\setcounter{equation}{0}

In the following, we present the analytic results for the one-loop
matrix elements involving an insertion of a $(\bar s b) (\bar \tau
\tau)$ operator, which enter the LO prediction for $B_s \to \gamma
\gamma$. The relevant decay amplitude has the following general
structure
\begin{equation} \label{eq:ABsAA}
  \begin{split}
    {\cal A} & = \frac{\cal N}{3} \, \frac{f_{B_s}}{2} \, \big \langle
    \gamma \gamma | A_+ \, F_{\mu \nu} F^{\mu \nu} - i \hspace{0.25mm}
    A_- \, F_{\mu \nu} \tilde F^{\mu \nu} | 0 \big \rangle \\[1mm] & =
    \frac{\cal N}{3} \, f_{B_s} \, \Big \{ A_+ \left ( 2 \hspace{0.25mm}
      k_1 \cdot \varepsilon_2 \, k_2 \cdot \varepsilon_1 - M_{B_s}^2 \,
      \varepsilon_1\cdot \varepsilon_2 \right ) - 2 \hspace{0.25mm} i
    \hspace{0.25mm} A_- \, \epsilon (k_1, k_2, \varepsilon_1,
    \varepsilon_2 ) \Big \} 
\end{split}
\end{equation}
where ${\cal N}$ has been defined in \refeq{eq:definitions}, $F_{\mu
  \nu}$ and $\tilde F_{\mu \nu} = \epsilon_{\mu \nu \lambda \rho}/2 \,
F^{\lambda \rho}$ are the photon field strength tensor and its dual,
$k_{1,2}$ and $\varepsilon_{1,2}$ denote the momenta and the
polarization vectors of the two photons, and we have introduced
$\epsilon (a,b,c,d) =\epsilon_{\mu \nu \lambda \rho} \hspace{0.25mm}
a^\mu \hspace{0.25mm} b^\nu \hspace{0.25mm} c^\lambda \hspace{0.25mm}
d^\rho$. The subscripts $\pm$ on the coefficients $A_\pm$, denote the
CP properties of the corresponding two-photon final state with $A_+$
($A_-$) being proportional to the parallel (perpendicular) spin
polarization $\vec \varepsilon_1 \cdot \vec \varepsilon_2$ ($\vec
\varepsilon_1 \times \vec \varepsilon_2$).

We find that the non-vanishing matrix elements $\langle \Op_i \rangle
= \langle \gamma \gamma |\Op_i | \bar B_s \rangle$ arising from
operator insertions into the Feynman diagram of \reffig{fig:mixing:bsgg},
are given by ($y_\tau = m_\tau^2/M_{B_s}^2$)
\begin{equation} \label{eq:MEbsAA}
\begin{split}
  \left \langle Q_{S,LA} \right \rangle & = \frac{ \alpha}{8
    \hspace{0.25mm} \pi} \, f_{B_s} \, \frac{M_{B_s}}{m_\tau} \,
  Q_\tau^2 \, \Bigg \{ \hspace{0.5mm} \frac{1}{2} \, \big[ 2 + \left (
    1- 4 \hspace{0.25mm} y_\tau \right) h(y_\tau) \big ] \left ( 2
    \hspace{0.25mm} k_1 \cdot \varepsilon_2 \, k_2 \cdot \varepsilon_1
    - M_{B_s}^2 \, \varepsilon_1\cdot \varepsilon_2 \right ) \\ &
  \hspace{3.75cm} \mp \big [ 2+ h(y_\tau) \big] \, i \hspace{0.5mm}
  \epsilon (k_1, k_2, \varepsilon_1, \varepsilon_2) \Bigg \} \,, \\[1mm]
  \left \langle Q_{V,LA} \right \rangle & = \frac{\alpha}{4
    \hspace{0.25mm} \pi} \, f_{B_s} \, Q_\tau^2 \, \Bigg \{ \pm
  h(y_\tau) \, i \hspace{0.5mm} \epsilon (k_1, k_2, \varepsilon_1,
  \varepsilon_2) \Bigg \} \,,
\end{split}
\end{equation}
with 
\begin{equation} \label{eq:hz}
  h (y) = -2 \left [ 1 + y \ln^2 \left (
      \frac{\sqrt{1-4\hspace{0.25mm}y} - 1} {\sqrt{1-4\hspace{0.25mm}y}+1}
    \right ) \right ] \,.
\end{equation}
Here $\alpha = 1/137.036$ and the upper (lower) signs hold in the case
$A=L$ ($A=R$). The expressions \refeq{eq:MEbsAA} can be shown to
resemble the results of \cite{Hiller:2004wc, Bosch:2002bw}.  The
formulas that apply in the case of $\Op_{S,RA}$ and $\Op_{V,RA}$ are
obtained from \refeq{eq:MEbsAA} by reversing the overall sign. Notice
that although the 1PI diagrams with an insertion of a tensor operator
with flavor content $(\bar s b) (\bar \tau \tau)$ evaluate to zero,
the operators $\Op_{T,A}$ contribute to the amplitude
\refeq{eq:ABsAA}, because the tensor operators lead to corrections
$\Delta C_7^{(\prime) \rm eff}$ to the effective Wilson coefficients
of $\Op_7^{(\prime)}$.  We finally mention that $\langle \Op_{S,AB}
\rangle$ scales like the mass of the charged lepton in the loop. This
implies that $\langle \Op_{S,AB} \rangle \to 0$ for $y \to 0$, since
$h(y) \to -2$ in this limit. The matrix elements of scalar operators
involving electrons or muons are therefore to first approximation
equal to zero. In the case of tau leptons the chiral suppression is
compensated by the large logarithms $\ln y$ and the numerical factors
appearing in $h(y) = -2 + \left (-2 \ln^2 y + 2 \pi^2 -4
  \hspace{0.25mm} i \hspace{0.25mm} \pi \ln y \right ) y +{\cal O}
(y^2)$.

Combining \refeq{eq:ABsAA} and \refeq{eq:MEbsAA} it follows that the
new-physics corrections to the coefficients $A_\pm$ take the form
\beq \label{eq:DApm}
\begin{split}
  \Delta A_+ &= \frac{M_{B_s}}{\lambda_B} \left ( \Delta C_7^{\rm eff}
    - \Delta C_7^{ \prime \hspace{0.25mm} \rm eff} \right ) -
  \frac{3}{4} \, \frac{M_{B_s}}{m_\tau} \, \big[ 2 + \left (1-4
    \hspace{0.25mm} y_\tau \right ) h(y_\tau) \big ] \hspace{0.25mm}
  F_S^+ \,, \\[1mm] \Delta A_- & = \frac{M_{B_s}}{\lambda_B} \left (
    \Delta C_7^{\rm eff} + \Delta C_7^{ \prime \hspace{0.25mm} \rm
      eff} \right ) - \frac{3}{4} \, \frac{M_{B_s}}{m_\tau} \, \big[ 2
  + h(y_\tau) \big ] \hspace{0.25mm} F_S^- + \frac{3}{2} \, h(y_\tau)
  \hspace{0.25mm} F_V^- \,,
\end{split}
\eeq
where 
\beq \label{eq:Fipm}
F_i^\pm = C_{i,LL} \pm C_{i,LR} - C_{i,RL} \mp  C_{i,RR} \,.
\eeq
Notice that only the contributions associated with $\Op_7^{(\prime)}$
contribute to the $B_s \to \gamma \gamma$ amplitudes at leading power
in $\Lambda_{\rm QCD}/m_b$, while the effects arising from the $(\bar
s b) (\bar \tau \tau)$ insertions are relative to the leading terms
suppressed by one power of $\lambda_B/M_{B_s} = {\cal O} (\Lambda_{\rm
  QCD}/m_b)$.

After summing over the photon polarizations the branching ratio of
$B_s \to \gamma \gamma$ becomes
\beq \label{eq:BRBsAA}
{\cal B} (B_s \to \gamma \gamma) = \frac{{\cal N}^2 \hspace{0.125mm}
  M_{B_s}^3 \hspace{0.25mm} f_{B_s}^2 \hspace{0.25mm} \tau_{B_s}} {144
  \hspace{0.25mm} \pi} \, \Big ( \left |(A_+)_{\rm SM} + \Delta A_+
\right |^2 + \left |(A_-)_{\rm SM} + \Delta A_- \right |^2 \Big ) \,,
\eeq
where $(A_\pm)_{\rm SM}$ denotes SM contributions to $A_\pm$. Explicit
formulas for the SM coefficients, including leading as well as
subleading corrections in $\Lambda_{\rm QCD}/m_b$, can be found in
\cite{Bosch:2002bv, Bosch:2002bw}.

%
\section{Forward-Scattering Amplitudes}
\label{app:amplitude}

\renewcommand{\theequation}{E\arabic{equation}}
\setcounter{equation}{0}

In this appendix, we give some details on the calculation of the
contributions from the complete set of operators \refeq{eq:Qsbtautau}
to $\Gamma_{12}^s$. Since the decay width is related to the absorptive
part of the forward scattering amplitude, the off-diagonal element of
the decay-width matrix may be written as
\beq \label{eq:Gamma12sdef}
\Gamma_{12}^s = \frac{1}{2 M_{B_s}} \, \langle \bar B_s | {\cal T} |
B_s \rangle \,,
\eeq
with the transition operator $\cal T$ given by 
\beq \label{eq:calT}
{\cal T} = 8 \hspace{0.25mm} G_F^2 \left ( V_{ts}^\ast V_{tb} \right
)^2 \hspace{0.5mm} \sum_{i,j} \hspace{0.5mm} {\rm Im} \left \{ i \int
  \! d^4x \; T \, \big [ C_i \hspace{0.25mm} Q_i (x) \, C_j
  \hspace{0.25mm} Q_j (0) \big ] \right \}\,.
\eeq
Here $T$ denotes time ordering, the Wilson coefficients $C_i$ are
normalized as in \refeq{eq:Leff} and \refeq{eq:CiLambda}.

At leading power in $\Lambda_{\rm QCD}/m_b$ the contribution to
$\Gamma_{12}^s$ is found by computing the matrix elements of ${\cal
  T}$ between quark states. The corresponding Feynman diagram
involving an insertion of a pair $(i, j )$ of operators
\refeq{eq:Qiapp} is depicted in \reffig{fig:gamma12}. Treating the
external quarks as on-shell and neglecting the strange-quark mass, we
obtain from the discontinuity of the graph
\beq \label{eq:Gamma12ij}
\begin{split}
  (\Gamma_{12}^s)_{ij} & = -\frac{3}{2} \;{\cal N}_{\Gamma_{12}^s} \,
  C_i \, C_j \, \beta_\tau \, \bigg \{ \, x_\tau \hspace{0.25mm} {\rm
    Tr} \left [ \Gamma^i_{\bar \tau \tau} \, \Gamma^j_{\bar \tau \tau}
  \right ] + \frac{\sqrt{x_\tau}}{2 \hspace{0.25mm} m_b} \, {\rm Tr}
  \left [ \Gamma^i_{\bar \tau \tau} \left [ \Gamma^j_{\bar \tau \tau}
      , {p\!\!\!/}_b \right ] \right ]
  \\[1mm]
  & \hspace{2cm} -\frac{1 - 4 \hspace{0.25mm} x_\tau}{12} \, {\rm Tr}
  \left [ \Gamma^i_{\bar \tau \tau} \hspace{0.25mm} \gamma^\alpha
    \hspace{0.25mm} \Gamma^j_{\bar \tau \tau} \hspace{0.25mm}
    \gamma_\alpha \right ] -\frac{1 + 2 \hspace{0.25mm} x_\tau}{6
    \hspace{0.25mm} m_b^2} \, {\rm Tr} \left [ \Gamma^i_{\bar \tau
      \tau} \hspace{0.25mm} {p\!\!\!/}_b \hspace{0.25mm}
    \Gamma^j_{\bar \tau \tau} \hspace{0.25mm} {p\!\!\!/}_b \right ]
  \bigg \} \, \langle \Gamma_{\bar s b}^i \, \Gamma_{\bar s b}^j
  \rangle \,,
\end{split}
\eeq
with ${\cal N}_{\Gamma_{12}^s}$ and $\beta_\tau$ given in and before
\refeq{eq:NGamma}, respectively.  Notice that the prefactor
$\beta_\tau$ is related to the imaginary part of the Passarino-Veltman
two-point integral $B_0 (m_b^2, m_\tau^2, m_\tau^2)$. Explicitly, one
has $\pi \hspace{0.25mm} \beta_\tau = {\rm Im} \left [B_0 (m_b^2,
  m_\tau^2, m_\tau^2) \right ]$.  The result \refeq{eq:Gamma12ij} can be 
  shown to agree with the general findings of \cite{Golowich:2006gq, 
  Chen:2007dg, Badin:2007bv}.

The general result \refeq{eq:Gamma12ij} simplifies if one considers
the self-interference of operators only.  Due to the cyclicity of the
trace one has ${\rm Tr} \left [ \Gamma^i_{\bar \tau \tau} \left [
    \Gamma^i_{\bar \tau \tau} , {p\!\!\!/}_b \right ] \right ] = 0$
for all $i = S,AB, \hspace{0.5mm} V,AB, \hspace{0.5mm} T,A$.  In the
case of $i = S,AB, \hspace{0.5mm} T,A$ the only non-vanishing Dirac
traces read
\beq \label{eq:STTr}
{\rm Tr} \left [P_{L,R} \, P_{L,R} \right ] = 2\,, \qquad {\rm Tr}
\left [\sigma^{\mu \nu} P_{L,R} \, \sigma^{\lambda \rho} P_{L,R}
\right ] = 2 \left (g^{\mu \lambda} g^{\nu \rho} - g^{\mu \rho} g^{\nu
    \lambda} \mp i \epsilon^{\mu \nu \lambda \rho} \right ) \,,
\eeq
where the Levi-Civita Tensor $\epsilon^{\mu \nu \lambda \rho}$ is
related to $\gamma_5$ via $\gamma_5 = -i/4! \, \epsilon^{\mu \nu
  \lambda \rho} \, \gamma_\mu \gamma_\nu \gamma_\lambda
\gamma_\rho$. For $i = V,AB$ one needs instead
\beq \label{eq:VTr}
{\rm Tr} \left [\gamma^\mu P_{L,R} \, \gamma^\alpha \, \gamma^\nu
  P_{L,R} \, \gamma_\alpha \right ] = -4\hspace{0.25mm} g^{\mu \nu}
\,, \qquad {\rm Tr} \left [\gamma^\mu P_{L,R} \, {p\!\!\!/}_b \,
  \gamma^\nu P_{L,R} \, {p\!\!\!/}_b \right ] = 2 \left ( 2
  \hspace{0.5mm} p_b^\mu \hspace{0.25mm} p_b^\nu - m_b^2
  \hspace{0.25mm} g^{\mu \nu} \right ) \,.
\eeq
In order to derive the expressions given in \refeq{eq:Gamma12s}, one
still has to take into account that
\beq \label{eq:epsid}
i \epsilon^{\mu \nu \lambda \rho} \left (\bar s \, \sigma_{\mu \nu}
  P_{L,R} \, b \right ) \left (\bar s \, \sigma_{\lambda \rho} P_{L,R}
  \, b \right ) = \mp \hspace{0.25mm} 2 \left (\bar s \, \sigma^{\mu
    \nu} P_{L,R} \, b \right ) \left (\bar s \, \sigma_{\mu \nu}
  P_{L,R} \, b \right ) \,,
\eeq
and apply the Fierz identity  
\beq \label{eq:fierzid}
\left (\bar s \, \sigma^{\mu \nu} P_{L,R} \, b \right ) \left (\bar s
  \, \sigma_{\mu \nu} P_{L,R} \, b \right ) = -4 \left (\bar s \, P_{L,R}
  \, b \right ) \left (\bar s \, P_{L,R} \, b \right ) - 8 \left (\bar
  s_\alpha \, P_{L,R} \, b_\beta \right ) \left (\bar s_\beta \,
  P_{L,R} \, b_\alpha \right ) \,.
\eeq

\section{Bounds on $\bm {\Delta \Gamma_s}$ and $\bm{a_{fs}^s}$ from
  $\bm{\Delta M_s}$ }
\label{app:bounds}

%
\renewcommand{\theequation}{F\arabic{equation}}
\setcounter{equation}{0}

In this appendix, we show that in the class of SM extensions with real
$(M_{12}^s)_{\rm NP}/(\Gamma_{12}^s)_{\rm NP}$, a bound on $\Delta
M_s$ necessarily results in a limit on both $\Delta \Gamma_s$ and
$a_{fs}^s$. We start our derivation by noting that under the
assumption that $\arg \left ( (M_{12}^s)_{\rm NP}
  /(\Gamma_{12}^s)_{\rm NP} \right )$ is either equal to $0^\circ$ or
$180^\circ$, the parameters $R_{M,\Gamma}$ introduced in
\refeq{eq:M12G12paraNP} can be expressed in terms of the three ratios
\beq \label{eq:rs}
r_M = \frac{(M_{12}^s)_{\rm NP}}{(M_{12}^s)_{\rm SM}} \,, \qquad 
r_{\rm SM} = \frac{(M_{12}^s)_{\rm SM}}{(\Gamma_{12}^s)_{\rm SM}}\,, \qquad 
r_{\rm NP} =  \frac{(M_{12}^s)_{\rm NP}}{(\Gamma_{12}^s)_{\rm NP}}  \,,
\eeq
and $\phi^s_{\rm SM}$, as follows
\beq \label{eq:RMRG}
R_M = | 1 + r_M | \, , \qquad R_\Gamma = \left | 1 - {\rm sgn} \left(
    r_{\rm NP} \right) \frac{| r_{\rm SM}|}{|r_{\rm NP}|} \, r_M \,
  e^{i \phi^s_{\rm SM}} \right | \approx \left | 1 - {\rm sgn} \left(
    r_{\rm NP} \right) \frac{| r_{\rm SM}|}{|r_{\rm NP}|} \, r_M
\right | \,.
\eeq
Because $\phi^s_{\rm SM}$ is only a fraction of $1^\circ$, the
approximation made to obtain the final expression for $R_\Gamma$ is
sufficient for all practical purposes.

In order to simplify the further discussion, we consider the two
possible signs of $r_{\rm NP}$ separately.  We begin with ${\rm sgn}
\left (r_{\rm NP} \right ) = +1$. From \refeq{eq:RMRG} it is readily
seen that in this case, the value of $R_\Gamma$ becomes maximal
(minimal) if $r_M$ is real, negative (positive), and as large as
possible in magnitude. A two-sided bound on $\Delta M_s$ or
equivalently $R_M$,
\beq \label{eq:RMminmax}
R_M^{\rm min}  < R_M < R_M^{\rm max}  \,,
\eeq 
with $R_M^{\rm min} < 1$ and $R_M^{\rm max} > 1$,\footnote{This is the
  relevant case in view of the good agreement between $(\Delta
  M_s)_{\rm SM}$ and measured value of the mass difference.  It is
  straightforward to extend the given formulas to the other possible
  cases.} thus constrains the possible values of $R_\Gamma$ to lie in
the interval
\beq \label{eq:RGminmax1}
{\rm max} \left ( 0, 1- \big ( R_M^{\rm max} -1 \big ) \,
  \frac{|r_{\rm SM}|}{|r_{\rm NP}|} \right )< R_\Gamma < 1 + \big (
R_M^{\rm max} + 1 \big ) \, \frac{|r_{\rm SM}|}{|r_{\rm NP}|} \,.
\eeq
According to \refeq{eq:observablesNP} a bound on $R_\Gamma$ also
restricts the allowed range for $\Delta \Gamma_s$. Since one has
$\phi_M = 180^\circ$ and $\phi_\Gamma = 0^\circ$ if $R_\Gamma$ becomes
maximal, the upper limit in \refeq{eq:RGminmax1} leads to a lower
bound on $\Delta \Gamma_s$. On the other hand, the upper limit on
$\Delta \Gamma_s$ arises if $\phi_M = \phi_\Gamma = 0^\circ$ and $r_M
= R_M^{\rm min} -1 < R_M^{\rm max} -1$. One therefore has
\beq \label{eq:DGminmax1}
-\left [ 1 + (R^{\rm max}_M + 1) \, \frac{|r_{\rm SM}|}{|r_{\rm NP}|}
\right ] (\Delta \Gamma_s)_{\rm SM} < (\Delta \Gamma_s)_{\rm NP} <
\left [ 1 - (R^{\rm min}_M - 1) \, \frac{|r_{\rm SM}|}{|r_{\rm NP}|}
\right ] (\Delta \Gamma_s)_{\rm SM} \,.
\eeq
Given that both $R_M$ and $R_\Gamma$ are bounded it is not surprising
that also $a_{fs}^s$ satisfies a double inequality. Because the
expression \refeq{eq:observablesNP} for $a_{fs}^s$ is inversely
proportional to $R_M$, the CP asymmetry becomes extremal for $R_M =
R_M^{\rm min}$. It is then easy to convince oneself, that for this
value of $R_M$ the combination of $R_\Gamma \hspace{0.25mm} \sin \left
  (\phi_M - \phi_\Gamma \right )$ is maximized/minimized if $r_M$ is
purely imaginary, corresponding to $\phi_M = \pm 90^\circ$. After some
algebraic simplifications, one arrives at
\beq \label{eq:afssminmax1}
-\frac{1}{R_M^{\rm min}} \left [ 1 + \frac{|r_{\rm SM}|}{|r_{\rm NP}|}
\right ] \frac{(a_{fs}^s)_{\rm SM}}{\phi_{\rm SM}^s} < (a_{fs}^s)_{\rm
  NP} < \frac{1}{R_M^{\rm min}} \left [ 1 + \frac{|r_{\rm
      SM}|}{|r_{\rm NP}|} \right ] \frac{(a_{fs}^s)_{\rm
    SM}}{\phi_{\rm SM}^s} \,.
\eeq

Using the same line of reasoning as above, it is also not too
difficult to derive the constraints for ${\rm sgn} \left (r_{\rm NP}
\right ) = -1$ that follow from \refeq{eq:RMminmax} with $R_M^{\rm
  min} < 1$ and $R_M^{\rm max} > 1$.  One has to differentiate between
the case $|r_{\rm SM}|/|r_{\rm NP}| < 1$ and $|r_{\rm SM}|/|r_{\rm
  NP}| > 1$.  In the former case, we obtain
\begin{gather}
  {\rm max} \left ( 0, 1- \big ( R_M^{\rm max} +1 \big ) \,
    \frac{|r_{\rm SM}|}{|r_{\rm NP}|} \right )< R_\Gamma < 1 + \big (
  R_M^{\rm max} - 1 \big ) \,
  \frac{|r_{\rm SM}|}{|r_{\rm NP}|} \,, \;\; \nonumber \\[-3mm]
  \label{eq:RG2} \\[-3mm]
  \left [ -1 + (R^{\rm min}_M + 1) \, \frac{|r_{\rm SM}|}{|r_{\rm
        NP}|} \right ] (\Delta \Gamma_s)_{\rm SM} < (\Delta
  \Gamma_s)_{\rm NP} < \left [ 1 + (R^{\rm max}_M - 1) \,
    \frac{|r_{\rm SM}|}{|r_{\rm NP}|} \right ] (\Delta \Gamma_s)_{\rm
    SM} \,, \;\; \nonumber
\end{gather}
while in the latter case the relations
\begin{gather}
  {\rm max} \left ( 0, 1+ \big ( R_M^{\rm min} -1 \big ) \,
    \frac{|r_{\rm SM}|}{|r_{\rm NP}|} \right ) < R_\Gamma < -1 + \big
  ( R_M^{\rm max} + 1 \big ) \, \frac{|r_{\rm SM}|}{|r_{\rm NP}|}
  \,, \nonumber \\ \label{eq:RG3}
  \left [ 1 + (R^{\rm min}_M - 1) \, \frac{|r_{\rm SM}|}{|r_{\rm NP}|}
  \right ] (\Delta \Gamma_s)_{\rm SM} < (\Delta \Gamma_s)_{\rm NP} <
  \left [ -1 + (R^{\rm max}_M + 1) \, \frac{|r_{\rm SM}|}{|r_{\rm
        NP}|} \right ] (\Delta \Gamma_s)_{\rm SM} \,,
\end{gather}
apply. The constraint on the CP asymmetry is independent of whether
$|r_{\rm SM}|/|r_{\rm NP}|$ is smaller or bigger than 1. It takes the
simple form
\beq \label{eq:afss2}
-\frac{1}{R_M^{\rm min}} \left | 1 - \frac{|r_{\rm SM}|}{|r_{\rm NP}|}
\right | \frac{(a_{fs}^s)_{\rm SM}}{\phi_{\rm SM}^s} < (a_{fs}^s)_{\rm
  NP} < \frac{1}{R_M^{\rm min}} \left |1 - \frac{|r_{\rm SM}|}{|r_{\rm
      NP}|} \right | \frac{(a_{fs}^s)_{\rm SM}}{\phi_{\rm SM}^s} \,.
\eeq
The formulas \refeq{eq:RGminmax1} to \refeq{eq:afss2} have a couple of
features that are worth mentioning. First, the possible variations
increase with decreasing $|r_{\rm SM}|/|r_{\rm NP}| $.  Second, the
limits on $\Delta \Gamma_s$ and $a_{fs}^s$ are less stringent if ${\rm
  sgn} \left (r_{\rm NP} \right ) = +1$. The only exception is the
upper bound on $\Delta \Gamma_s$, which becomes weakest for ${\rm sgn}
\left (r_{\rm NP} \right ) = -1$ and $|r_{\rm SM}|/|r_{\rm NP}| > 1$.
In new-physics scenarios with real $(M_{12}^s)_{\rm
  NP}/(\Gamma_{12}^s)_{\rm NP}$, non-standard effects in $\Delta
\Gamma_s$ and $a_{fs}^s$ are hence the least constrained by the
measurement of $\Delta M_s$, if the ratio $r_{\rm SM}/r_{\rm NP}$ is
positive and as small as possible.

\end{appendix}

%


\begin{thebibliography}{99}

\bibitem{Abulencia:2006ze}
  A.~Abulencia {\it et al.}  [CDF Collaboration],
  Phys.\ Rev.\ Lett.\  {\bf 97}, 242003 (2006)
  [arXiv:hep-ex/0609040].
  
\bibitem{LHCbnote50}
LHCb Collaboration, LHCb-CONF-2011-050.
  
\bibitem{Lenz:2011ti}
  A.~Lenz and U.~Nierste,
  arXiv:1102.4274 [hep-ph].
  
\bibitem{Aaltonen:2007he}
  T.~Aaltonen {\it et al.}  [CDF Collaboration],
  Phys.\ Rev.\ Lett.\  {\bf 100}, 161802 (2008)
  [arXiv:0712.2397 [hep-ex]].
  
\bibitem{Aaltonen:2007gf}
  T.~Aaltonen {\it et al.}  [CDF collaboration],
  Phys.\ Rev.\ Lett.\  {\bf 100}, 121803 (2008)
  [arXiv:0712.2348 [hep-ex]].
  
\bibitem{D0noteB58}  
D\O \ Collaboration, Conference Note 5933-CONF, May 28, 2009, \href{http://www-d0.fnal.gov/Run2Physics/WWW/results/prelim/B/B58/B58.pdf}{http://www-d0.fnal.gov/Run2Physics/WWW/results/prelim/B/B58/B58.pdf}

\bibitem{Asner:2010qj}
  D.~Asner {\it et al.}  [Heavy Flavor Averaging Group],
  arXiv:1010.1589 [hep-ex], updated results available at
  \href{http://www.slac.stanford.edu/xorg/hfag/}{\tt http://www.slac.stanford.edu/xorg/hfag/}

\bibitem{Aaij:2012eq}
  R.~Aaij {\it et al.}  [LHCb Collaboration],
  Phys.\ Rev.\ Lett.\  {\bf 108}, 241801 (2012)
  [arXiv:1202.4717 [hep-ex]].
   
\bibitem{Lenz:2010gu}
  A.~Lenz {\it et al.},
  Phys.\ Rev.\  D {\bf 83}, 036004 (2011)
  [arXiv:1008.1593 [hep-ph]].

\bibitem{Lenz:2011zz}
  A.~Lenz,
  Phys.\ Rev.\  D {\bf 84}, 031501 (2011)
  [arXiv:1106.3200 [hep-ph]].

\bibitem{Amhis:2012bh} 
  Y.~Amhis {\it et al.}  [Heavy Flavor Averaging Group Collaboration],
  arXiv:1207.1158 [hep-ex].

\bibitem{:2012ie} 
  [CDF Collaboration],
  Phys.\ Rev.\ Lett.\  {\bf 109}, 171802 (2012)
  [arXiv:1208.2967 [hep-ex]].

\bibitem{Abazov:2011ry} 
  V.~M.~Abazov {\it et al.}  [D\O \ Collaboration],
  Phys.\ Rev.\ D {\bf 85}, 032006 (2012)
  [arXiv:1109.3166 [hep-ex]].

\bibitem{LHCbnote49}
   LHCb Collaboration, LHCb note CERN-LHCb-CONF-2012-002, March 5, 2012, 
   \href{http://cdsweb.cern.ch/record/1423592}{http://cdsweb.cern.ch/record/1423592}

\bibitem{LHCb:2012ad}
  R.~Aaij {\it et al.}  [LHCb Collaboration],
  Phys.\ Lett.\ B {\bf 713}, 378 (2012)
  [arXiv:1204.5675 [hep-ex]].

\bibitem{:2012fu} 
  G.~Aad {\it et al.}  [ATLAS Collaboration],
  arXiv:1208.0572 [hep-ex].
  
\bibitem{:2012uia} 
  V.~M.~Abazov {\it et al.}  [D{\O} Collaboration],
  Phys.\ Rev.\ D {\bf 86}, 072009 (2012)
  [arXiv:1208.5813 [hep-ex]].

\bibitem{CDFnote9015}
CDF Collaboration, CDF Note 9015, October 16, 2007,
\href{http://www-cdf.fnal.gov/physics/new/bottom/070816.blessed-acp-bsemil/public-acp-bsemil.ps}{http://www-cdf.fnal.gov/physics/new/bottom/070816.blessed-acp-bsemil/public-acp-bsemil.ps}

\bibitem{Abazov:2010hv}
  V.~M.~Abazov {\it et al.}  [D\O \ Collaboration],
  Phys.\ Rev.\  D {\bf 82}, 032001 (2010)
  [arXiv:1005.2757 [hep-ex]].
  
\bibitem{Abazov:2010hj}
  V.~M.~Abazov {\it et al.}  [D\O \ Collaboration],
  Phys.\ Rev.\ Lett.\  {\bf 105}, 081801 (2010)
  [arXiv:1007.0395 [hep-ex]].
  
\bibitem{Abazov:2011yk}
  V.~M.~Abazov {\it et al.}  [D\O \ Collaboration],
  Phys.\ Rev.\ D {\bf 84}, 052007 (2011)
  [arXiv:1106.6308 [hep-ex]].
 
\bibitem{LHCb:CONF-2012-022}
  LHCb Collaboration, LHCb-CONF-2012-022.
   
\bibitem{Dighe:2010nj}
  A.~Dighe, A.~Kundu and S.~Nandi,
  Phys.\ Rev.\  D {\bf 82}, 031502 (2010)
  [arXiv:1005.4051 [hep-ph]].
  
\bibitem{Dobrescu:2010rh}
  B.~A.~Dobrescu, P.~J.~Fox and A.~Martin,
  Phys.\ Rev.\ Lett.\  {\bf 105}, 041801 (2010)
  [arXiv:1005.4238 [hep-ph]].
  
\bibitem{Ligeti:2010ia}
  Z.~Ligeti, M.~Papucci, G.~Perez and J.~Zupan,
  Phys.\ Rev.\ Lett.\  {\bf 105}, 131601 (2010)
  [arXiv:1006.0432 [hep-ph]].
  
\bibitem{Bauer:2010dga}
  C.~W.~Bauer and N.~D.~Dunn,
  Phys.\ Lett.\  B {\bf 696}, 362 (2011)
  [arXiv:1006.1629 [hep-ph]].
    
\bibitem{Bai:2010kf}
  Y.~Bai and A.~E.~Nelson,
  Phys.\ Rev.\  D {\bf 82}, 114027 (2010)
  [arXiv:1007.0596 [hep-ph]].
  
\bibitem{Oh:2010vc}
  S.~Oh and J.~Tandean,
  Phys.\ Lett.\  B {\bf 697}, 41 (2011)
  [arXiv:1008.2153 [hep-ph]].
 
\bibitem{Alok:2010ij}
  A.~K.~Alok, S.~Baek and D.~London,
  JHEP {\bf 1107}, 111 (2011)
  [arXiv:1010.1333 [hep-ph]].
 
\bibitem{Kim:2010gx}
  J.~E.~Kim, M.~S.~Seo and S.~Shin,
  Phys.\ Rev.\  D {\bf 83}, 036003 (2011)
  [arXiv:1010.5123 [hep-ph]].
  
\bibitem{Oh:2011nb}
  S.~Oh and J.~Tandean,
  Phys.\ Rev.\  D {\bf 83}, 095006 (2011)
  [arXiv:1102.1680 [hep-ph]].
 
\bibitem{Dutta:2011kg} 
  B.~Dutta, S.~Khalil, Y.~Mimura and Q.~Shafi,
  JHEP {\bf 1205}, 131 (2012)
  [arXiv:1104.5209 [hep-ph]].
   
\bibitem{Dighe:2011du}
  A.~Dighe, D.~Ghosh, A.~Kundu and S.~K.~Patra,
  Phys.\ Rev.\ D {\bf 84}, 056008 (2011)
  [arXiv:1105.0970 [hep-ph]].
    
\bibitem{Goertz:2011nx}
  F.~Goertz and T.~Pfoh, 
  Phys.\ Rev.\ D {\bf 84}, 095016 (2011)
  [arXiv:1105.1507 [hep-ph]].

\bibitem{Dighe:2007gt}
  A.~Dighe, A.~Kundu and S.~Nandi,
  Phys.\ Rev.\  D {\bf 76}, 054005 (2007)
  [arXiv:0705.4547 [hep-ph]].
  
\bibitem{Carpentier:2010ue}
  M.~Carpentier and S.~Davidson,
  Eur.\ Phys.\ J.\  C {\bf 70}, 1071 (2010)
  [arXiv:1008.0280 [hep-ph]].

\bibitem{Grossman:1996qj}
  Y.~Grossman, Z.~Ligeti and E.~Nardi,
  Phys.\ Rev.\  D {\bf 55}, 2768 (1997)
  [arXiv:hep-ph/9607473].

\bibitem{Nakamura:2010zzi}
  K.~Nakamura {\it et al.}  [Particle Data Group],
  J.\ Phys.\ G {\bf 37}, 075021 (2010), updated results available at
  \href{http://pdglive.lbl.gov}{\tt http://pdglive.lbl.gov}
  
\bibitem{Lenz:1997aa}
  A.~Lenz, U.~Nierste and G.~Ostermaier,
  Phys.\ Rev.\  D {\bf 56}, 7228 (1997)
  [arXiv:hep-ph/9706501].
  
  \bibitem{Bagan:1994qw} 
  E.~Bagan, P.~Ball, V.~M.~Braun and P.~Gosdzinsky,
  Phys.\ Lett.\ B {\bf 342}, 362 (1995)
  [Erratum-ibid.\ B {\bf 374}, 363 (1996)]
  [hep-ph/9409440].
  
  \bibitem{Neubert:1996we} 
  M.~Neubert and C.~T.~Sachrajda,
  Nucl.\ Phys.\ B {\bf 483}, 339 (1997)
  [hep-ph/9603202].

\bibitem{Kagan:1997qna}
  A.~L.~Kagan and J.~Rathsman,
  arXiv:hep-ph/9701300.

\bibitem{Kagan:1997sg}
  A.~Kagan,
  arXiv:hep-ph/9806266.
  
\bibitem{Aubert:2006mp}
  B.~Aubert {\it et al.}  [BaBar Collaboration],
  Phys.\ Rev.\  D {\bf 75}, 072002 (2007)
  [arXiv:hep-ex/0606026].

\bibitem{Abreu:1998xb}
  P.~Abreu {\it et al.} [DELPHI Collaboration],
  Phys.\ Lett.\  {\bf B426}, 193-206 (1998).

\bibitem{Flood:2010zz}
  K.~Flood  [BaBar Collaboration],
  PoS {\bf ICHEP2010}, 234 (2010).

\bibitem{Bobeth:2002ch}
  C.~Bobeth, T.~Ewerth, F.~Kr\"uger and J.~Urban,
  Phys.\ Rev.\  D {\bf 66}, 074021 (2002)
  [arXiv:hep-ph/0204225].

\bibitem{Laiho:2009eu}
  J.~Laiho, E.~Lunghi and R.~S.~Van de Water,
  Phys.\ Rev.\  D {\bf 81}, 034503 (2010)
  [arXiv:0910.2928 [hep-ph]].

\bibitem{Charles:2004jd}
  J.~Charles {\it et al.}  [CKMfitter Group],
  Eur.\ Phys.\ J.\ C {\bf 41}, 1 (2005) [arXiv:hep-ph/0406184],
  updated results available at
  \href{http://ckmfitter.in2p3.fr}{\tt http://ckmfitter.in2p3.fr}

\bibitem{Bobeth:2003at}
  C.~Bobeth, P.~Gambino, M.~Gorbahn and U.~Haisch,
  JHEP {\bf 0404}, 071 (2004)
  [arXiv:hep-ph/0312090].

\bibitem{Fukae:1998qy}
  S.~Fukae, C.~S.~Kim, T.~Morozumi and T.~Yoshikawa,
  Phys.\ Rev.\  D {\bf 59}, 074013 (1999)
  [arXiv:hep-ph/9807254].

\bibitem{Nir:1989rm}
  Y.~Nir,
  Phys.\ Lett.\  B {\bf 221}, 184 (1989).

\bibitem{Hewett:1995dk}
  J.~L.~Hewett,
  Phys.\ Rev.\  D {\bf 53}, 4964 (1996)
  [arXiv:hep-ph/9506289].
  
\bibitem{Du:1993sh}
  D.~S.~Du, C.~Liu and D.~X.~Zhang,
  Phys.\ Lett.\  B {\bf 317}, 179 (1993).

\bibitem{Bobeth:2007dw}
  C.~Bobeth, G.~Hiller and G.~Piranishvili,
  JHEP {\bf 0712}, 040 (2007)
  [arXiv:0709.4174 [hep-ph]].
  
\bibitem{Borzumati:1999qt}
  F.~Borzumati, C.~Greub, T.~Hurth and D.~Wyler,
  Phys.\ Rev.\  D {\bf 62}, 075005 (2000)
  [arXiv:hep-ph/9911245].
  
\bibitem{Hiller:2003js}
  G.~Hiller and F.~Kr\"uger,
  Phys.\ Rev.\  D {\bf 69}, 074020 (2004)
  [arXiv:hep-ph/0310219].
  
\bibitem{Buras:1993xp}
  A.~J.~Buras, M.~Misiak, M.~M\"unz and S.~Pokorski,
  Nucl.\ Phys.\  B {\bf 424}, 374 (1994)
  [arXiv:hep-ph/9311345].
  
\bibitem{Misiak:2006zs}
  M.~Misiak {\it et al.},
  Phys.\ Rev.\ Lett.\  {\bf 98}, 022002 (2007) [arXiv:hep-ph/0609232].

\bibitem{Misiak:2006ab}
  M.~Misiak and M.~Steinhauser,
  Nucl.\ Phys.\  B {\bf 764}, 62 (2007) [arXiv:hep-ph/0609241].
  
  \bibitem{Asatrian:2006rq}
  H.~M.~Asatrian,T.~Ewerth, H.~Gabrielyan and C.~Greub,
  Phys.\ Lett.\  B {\bf 647}, 173 (2007) [arXiv:hep-ph/0611123].
  
\bibitem{Boughezal:2007ny}
  R.~Boughezal, M.~Czakon and T.~Schutzmeier,
  JHEP {\bf 0709}, 072 (2007) [arXiv:0707.3090 [hep-ph]].
  
\bibitem{Ewerth:2008nv}
  T.~Ewerth,
  Phys.\ Lett.\  B {\bf 669}, 167 (2008)
  [arXiv:0805.3911 [hep-ph]].
  
\bibitem{Asatrian:2010rq}
  H.~M.~Asatrian, T.~Ewerth, A.~Ferroglia, C.~Greub and G.~Ossola,
  Phys.\ Rev.\  D {\bf 82}, 074006 (2010)
  [arXiv:1005.5587 [hep-ph]].
  
\bibitem{Ferroglia:2010xe}
  A.~Ferroglia and U.~Haisch,
  Phys.\ Rev.\  D {\bf 82}, 094012 (2010)
  [arXiv:1009.2144 [hep-ph]].
  
\bibitem{Misiak:2010tk}
  M.~Misiak and M.~Poradzinski,
  Phys.\ Rev.\  D {\bf 83}, 014024 (2011)
  [arXiv:1009.5685 [hep-ph]].
  
\bibitem{Czakon:2006ss}
  M.~Czakon, U.~Haisch and M.~Misiak,
  JHEP {\bf 0703}, 008 (2007) [arXiv:hep-ph/0612329].
  
\bibitem{Gambino:2004mv}
  P.~Gambino, U.~Haisch and M.~Misiak,
  Phys.\ Rev.\ Lett.\  {\bf 94}, 061803 (2005)
  [arXiv:hep-ph/0410155].
  
\bibitem{Bobeth:2008ij}
  C.~Bobeth, G.~Hiller and G.~Piranishvili,
  JHEP {\bf 0807}, 106 (2008)
  [arXiv:0805.2525 [hep-ph]].
  
\bibitem{DescotesGenon:2011yn}
  S.~Descotes-Genon, D.~Ghosh, J.~Matias and M.~Ramon,
  JHEP {\bf 1106}, 099 (2011)
  [arXiv:1104.3342 [hep-ph]].

\bibitem{Aubert:2004it}
  B.~Aubert {\it et al.}  [BaBar Collaboration],
  Phys.\ Rev.\ Lett.\  {\bf 93}, 081802 (2004)
  [arXiv:hep-ex/0404006].
  
\bibitem{Iwasaki:2005sy}
  M.~Iwasaki {\it et al.}  [Belle Collaboration],
  Phys.\ Rev.\  D {\bf 72}, 092005 (2005)
  [arXiv:hep-ex/0503044].

\bibitem{Nakayama:2009zz}
  H.~Nakayama,
  ``Precision measurement of the electroweak flavor-changing neutral current
  decays of $B$ mesons,'', Ph. D. thesis, University of Tokyo,
  \href{http://belle.kek.jp/belle/theses/doctor/2009/Nakayama.pdf}{http://belle.kek.jp/belle/theses/doctor/2009/Nakayama.pdf}
  
\bibitem{Chiang:2010zz}
  C.~C.~Chiang  [Belle Collaboration],
  PoS {\bf ICHEP2010}, 231 (2010).
  
\bibitem{Aubert:2006vb}
  B.~Aubert {\it et al.}  [BaBar Collaboration],
  Phys.\ Rev.\  D {\bf 73}, 092001 (2006)
  [arXiv:hep-ex/0604007].
  
\bibitem{Aubert:2008ju}
  B.~Aubert {\it et al.}  [BaBar Collaboration],
  Phys.\ Rev.\  D {\bf 79}, 031102 (2009)
  [arXiv:0804.4412 [hep-ex]].
  
\bibitem{Wei:2009zv}
  J.~T.~Wei {\it et al.}  [Belle Collaboration],
  Phys.\ Rev.\ Lett.\  {\bf 103}, 171801 (2009)
  [arXiv:0904.0770 [hep-ex]].
  
\bibitem{Aaltonen:2011cn}
  T.~Aaltonen {\it et al.}  [CDF Collaboration],
  Phys.\ Rev.\ Lett.\  {\bf 106}, 161801 (2011)
  [arXiv:1101.1028 [hep-ex]].
  
\bibitem{Aaltonen:2011qs} 
  T.~Aaltonen {\it et al.}  [CDF Collaboration],
  Phys.\ Rev.\ Lett.\  {\bf 107}, 201802 (2011)
  [arXiv:1107.3753 [hep-ex]].
  
\bibitem{LHCbnote38}  
LHCb Collaboration, LHCb-CONF-2011-038, August 10, 2011,
\href{http://cdsweb.cern.ch/record/1367849/files/LHCb-CONF-2011-038.pdf}
{http://cdsweb.cern.ch/record/1367849/files/LHCb-CONF-2011-038.pdf}
 
 \bibitem{Aaltonen:2011ja} 
  T.~Aaltonen {\it et al.}  [CDF Collaboration],
  Phys.\ Rev.\ Lett.\  {\bf 108}, 081807 (2012)
  [arXiv:1108.0695 [hep-ex]].

\bibitem{Bauer:2009cf}
  M.~Bauer, S.~Casagrande, U.~Haisch and M.~Neubert,
  JHEP {\bf 1009}, 017 (2010)
  [arXiv:0912.1625 [hep-ph]].
  
\bibitem{Bobeth:2010wg}
  C.~Bobeth, G.~Hiller and D.~van Dyk,
  JHEP {\bf 1007}, 098 (2010)
  [arXiv:1006.5013 [hep-ph]].

\bibitem{Bobeth:2011gi}
  C.~Bobeth, G.~Hiller and D.~van Dyk,
  JHEP {\bf 1107}, 067 (2011)
  [arXiv:1105.0376 [hep-ph]].
    
\bibitem{EOS}  
 EOS Collaboration, "EOS: A HEP Program for Flavor Observables",
 \href{http://project.het.physik.tu-dortmund.de/eos/}{http://project.het.physik.tu-dortmund.de/eos/}
 
 \bibitem{Gemintern:2004bw}
  A.~Gemintern, S.~Bar-Shalom and G.~Eilam,
  Phys.\ Rev.\  D {\bf 70}, 035008 (2004)
  [arXiv:hep-ph/0404152].

\bibitem{Hiller:2004wc}
  G.~Hiller and A.~S.~Safir,
  JHEP {\bf 0502}, 011 (2005)
  [arXiv:hep-ph/0411344v4].
  
\bibitem{Wicht:2007ni}
  J.~Wicht {\it et al.} [Belle Collaboration],
  Phys.\ Rev.\ Lett.\  {\bf 100}, 121801 (2008)
  [arXiv:0712.2659 [hep-ex]].
  
\bibitem{Bosch:2002bv}
  S.~W.~Bosch and G.~Buchalla,
  JHEP {\bf 0208}, 054 (2002)
  [hep-ph/0208202].      
 
\bibitem{Bosch:2002bw}
  S.~W.~Bosch,
  [hep-ph/0208203].
  
\bibitem{Grozin:1996pq}
  A.~G.~Grozin and M.~Neubert,
  Phys.\ Rev.\  {\bf D55}, 272-290 (1997)
  [hep-ph/9607366].
  
\bibitem{Ball:2003fq}
  P.~Ball and  E.~Kou,
  JHEP {\bf 0304}, 029 (2003)
  [hep-ph/0301135].
  
\bibitem{Braun:2003wx}
  V.~M.~Braun, D.~Y.~Ivanov and G.~P.~Korchemsky,
  Phys.\ Rev.\  {\bf D69}, 034014 (2004)
  [hep-ph/0309330].      
  
\bibitem{Lee:2005gza}
  S.~J.~Lee and  M.~Neubert,
  Phys.\ Rev.\  {\bf D72}, 094028 (2005)
  [hep-ph/0509350].  
  
\bibitem{Hewett:2004tv} 
  T.~Abe {\it et al.},
  [hep-ph/0503261].  
 
 \bibitem{Becirevic:2001xt}
  D.~Becirevic, V.~Gimenez, G.~Martinelli, M.~Papinutto and J.~Reyes,
  JHEP {\bf 0204}, 025 (2002)
  [arXiv:hep-lat/0110091].
  
\bibitem{Buchmuller:1986zs}
  W.~Buchm\"uller, R.~R\"uckl and D.~Wyler,
  Phys.\ Lett.\  B {\bf 191}, 442 (1987)
  [Erratum-ibid.\  B {\bf 448}, 320 (1999)].
  
\bibitem{Leurer:1993em}
  M.~Leurer,
  Phys.\ Rev.\  D {\bf 49}, 333 (1994)
  [arXiv:hep-ph/9309266].
  
\bibitem{Davidson:1993qk}
  S.~Davidson, D.~C.~Bailey and B.~A.~Campbell,
  Z.\ Phys.\  C {\bf 61}, 613 (1994)
  [arXiv:hep-ph/9309310].
  
\bibitem{Hewett:1997ce}
  J.~L.~Hewett and T.~G.~Rizzo,
  Phys.\ Rev.\  D {\bf 56}, 5709 (1997)
  [arXiv:hep-ph/9703337].
  
  \bibitem{Abazov:2008jp}
  V.~M.~Abazov {\it et al.}  [D\O \ Collaboration],
  Phys.\ Rev.\ Lett.\  {\bf 101}, 241802 (2008)
  [arXiv:0806.3527 [hep-ex]].
  
\bibitem{Dorsner:2011ai}
  I.~Dorsner, J.~Drobnak, S.~Fajfer, J.~F.~Kamenik and N.~Kosnik,
  arXiv:1107.5393 [hep-ph].
  
\bibitem{Langacker:2000ju}
  P.~Langacker and M.~Pl\"umacher,
  Phys.\ Rev.\  D {\bf 62}, 013006 (2000)
  [arXiv:hep-ph/0001204].
  
\bibitem{Langacker:2008yv}
  P.~Langacker,
  Rev.\ Mod.\ Phys.\  {\bf 81}, 1199 (2009)
  [arXiv:0801.1345 [hep-ph]].
  
\bibitem{He:2006bk}
  X.~G.~He and G.~Valencia,
  Phys.\ Rev.\  D {\bf 74}, 013011 (2006)
  [arXiv:hep-ph/0605202].
  
\bibitem{Acosta:2005ij}
  D.~E.~Acosta {\it et al.}  [CDF Collaboration],
  Phys.\ Rev.\ Lett.\  {\bf 95}, 131801 (2005)
  [arXiv:hep-ex/0506034].
  
\bibitem{LEPEWWG:2005ema}
  S. Schael {\it et al.} \ [ALEPH~Collaboration],
  Phys.\ Rept.\ {\bf 427}, 257 (2006)
  [arXiv:hep-ex/0509008].
  
\bibitem{Haisch:2011up}
  U.~Haisch and S.~Westhoff,
  JHEP {\bf 1108}, 088 (2011)
  [arXiv:1106.0529 [hep-ph]].
  
\bibitem{Aubert:2007rn}
  B.~Aubert {\it et al.}  [BaBar Collaboration],
  Phys.\ Rev.\ Lett.\  {\bf 99}, 201801 (2007)
  [arXiv:0708.1303 [hep-ex]].
  
\bibitem{Aubert:2005qw} 
  B.~Aubert {\it et al.}  [BABAR Collaboration],
  Phys.\ Rev.\ Lett.\  {\bf 96}, 241802 (2006)
  [hep-ex/0511015].

\bibitem{Khodjamirian:2010vf}
  A.~Khodjamirian, T.~Mannel, A.~A.~Pivovarov and Y.~M.~Wang,
  JHEP {\bf 1009}, 089 (2010)
  [arXiv:1006.4945 [hep-ph]].
  
\bibitem{Golowich:2006gq}
  E.~Golowich, S.~Pakvasa and A.~A.~Petrov,
  Phys.\ Rev.\ Lett.\  {\bf 98}, 181801 (2007)
  [arXiv:hep-ph/0610039].
  
\bibitem{Chen:2007dg}
  S.~L.~Chen, X.~G.~He, A.~Hovhannisyan and H.~C.~Tsai,
  JHEP {\bf 0709}, 044 (2007)
  [arXiv:0706.1100 [hep-ph]].
  
\bibitem{Badin:2007bv}
  A.~Badin, F.~Gabbiani and A.~A.~Petrov,
  Phys.\ Lett.\  B {\bf 653}, 230 (2007)
  [arXiv:0707.0294 [hep-ph]].

\end{thebibliography}
\end{document}